%% file: arXiv-topological-descriptors-main.tex
\documentclass{egpubl-noname}
\usepackage{eurovis2021-noname}     

\usepackage[T1]{fontenc}
\usepackage{dfadobe}  

\usepackage{cite}  
\BibtexOrBiblatex
\electronicVersion
\PrintedOrElectronic
\ifpdf \usepackage[pdftex]{graphicx} \pdfcompresslevel=9
\else \usepackage[dvips]{graphicx} \fi

\usepackage{egweblnk} 
\usepackage{amsmath}
\usepackage{amssymb}

\usepackage{multirow}
\usepackage{rotating}

\usepackage{bm}

\graphicspath{{./figs/}}

\newtheorem{theorem}{Theorem}
\newtheorem{definition}{Definition}

\newcommand {\mm}[1] {\ifmmode{#1}\else{\mbox{\(#1\)}}\fi}

\usepackage[nameinlink,capitalize]{cleveref}

\newcommand{\Mspace}        {\mm{\mathbb{M}}}
\newcommand{\Nspace}        {\mm{\mathbb{N}}}

\newcommand{\Rspace}        {\mm{\mathbb{R}}}
\newcommand{\Xspace}        {\mm{\mathbb{X}}}
\newcommand{\Yspace} {\mm{\mathbb{Y}}}

\newcommand{\Tspace}        {\mm{\mathbb{T}}}
\newcommand{\Sspace}        {\mm{\mathbb{S}}}

\newcommand{\Hgroup}        {\mm{\mathsf H}}

\newcommand{\Fcal}        {\mm{\mathcal F}}

\newcommand{\Ncal}        {\mm{\mathcal N}}
\newcommand{\Ucal}        {\mm{\mathcal U}}
\newcommand{\Vcal}        {\mm{\mathcal V}}

\newcommand{\D}        {\mm{\mathcal{D}}}

\newcommand{\G}        {\mm{\mathcal G}}
\newcommand{\T}        {\mm{\mathcal T}}
\newcommand{\A}        {\mm{\mathcal A}}
\newcommand{\M}        {\mm{\mathcal M}}

\usepackage{mathtools}

\newcommand{\Hcal}        {\mm{{\mathcal H}}}
\newcommand{\Dcal}        {\mm{{\mathcal D}}}

\usepackage[dvipsnames]{xcolor}

\usepackage{enumitem}

\newcommand{\para}[1]        {\vspace{1mm}\noindent{\textbf{#1}}}

\newcommand{\grad}[1]     {{\nabla {#1}}}

\newcommand{\etal}{\textit{et al.}}
\newcommand{\eg}{\textit{e.g.}}
\newcommand{\wrt}{\textit{w.r.t.}}
\newcommand{\cf}{\textit{cf.}}
\newcommand{\mywlog}{\textit{w.l.o.g.}}
\newcommand{\ie}{\textit{i.e.}}

\definecolor{myred}{rgb}{0.96, 0.76, 0.76}
\definecolor{mygreen}{rgb}{0.66, 0.89, 0.63}
\definecolor{myteal}{rgb}{0.63, 0.79, 0.95}
\definecolor{mypurple}{rgb}{0.82, 0.62, 0.91}

\definecolor{celadon}{rgb}{0.67, 0.88, 0.69}

\definecolor{etonblue}{rgb}{0.59, 0.78, 0.64}
\definecolor{grannysmithapple}{rgb}{0.66, 0.89, 0.63}
\definecolor{junebud}{rgb}{0.74, 0.85, 0.34}
\definecolor{lightgreen}{rgb}{0.56, 0.93, 0.56}
\definecolor{magicmint}{rgb}{0.67, 0.94, 0.82}
\usepackage{booktabs}

\usepackage{makecell}

\usepackage{xcolor,colortbl}
\definecolor{green}{rgb}{0.1,0.1,0.1}

\definecolor{amethyst}{rgb}{0.6, 0.4, 0.8}
\definecolor{aliceblue}{rgb}{0.94, 0.97, 1.0}
\definecolor{apricot}{rgb}{0.98, 0.81, 0.69}
\definecolor{aquamarine}{rgb}{0.5, 1.0, 0.83}	
\definecolor{ashgrey}{rgb}{0.7, 0.75, 0.71}
\definecolor{asparagus}{rgb}{0.53, 0.66, 0.42}
\definecolor{babyblue}{rgb}{0.54, 0.81, 0.94}
\definecolor{babypink}{rgb}{0.96, 0.76, 0.76}
\definecolor{burlywood}{rgb}{0.87, 0.72, 0.53}
\definecolor{brightlavender}{rgb}{0.75, 0.58, 0.89}
\definecolor{darkpink}{rgb}{0.91, 0.33, 0.5}
\definecolor{azure}{rgb}{0.0, 0.5, 1.0}

\usepackage{lipsum,multicol}
\usepackage{tabularx}

\title[Scalar Field Comparison with Topological Descriptors]
      {Scalar Field Comparison with Topological Descriptors: \\
      Properties and Applications for Scientific Visualization}
      
\author[Lin Yan et al.]
{\parbox{\textwidth}{\centering 
        Lin Yan$^{1}$\orcid{0000-0001-7017-0329}, 
        Talha Bin Masood$^{2}$\orcid{0000-0001-5352-1086}, 
        Raghavendra Sridharamurthy$^3$\orcid{0000-0001-8463-0488},
        Farhan Rasheed$^{2}$\orcid{0000-0003-0632-1545},\\
        Vijay Natarajan$^{3}$\orcid{0000-0002-7956-1470}, 
        Ingrid Hotz$^{2}$\orcid{0000-0001-7285-0483},       
        Bei Wang$^{1}$\orcid{0000-0002-9240-0700}
        }
        \\
{\parbox{\textwidth}{\centering 
$^1$Scientific Computing and Imaging Institute, University of Utah, USA\\
$^2$Department of Science and Technology (ITN), Link\"oping University, Norrk\"oping, Sweden\\
$^3$Department of Computer Science and Automation, Indian Institute of Science Bangalore, India}
}
}
\begin{document}
\maketitle
%-------------------------------------------------------------------------
\begin{abstract}
\input{sec-abstract}
\end{abstract}  

%-------------------------------------------------------------------------
\input{sec-introduction}

\input{sec-classify}

\input{sec-background}

\input{sec-comparative-measures}

\input{sec-navigation}
\input{sec-classify-single-fields}
\input{sec-classify-time-varying}

\input{sec-classify-ensembles}

\input{sec-math-properties}

\input{sec-future}

\input{sec-conclusion}

%-------------------------------------------------------------------------
\section*{Acknowledgement}
This work is partially supported by the United States Department of Energy (DOE) grant DE-SC0021015 and National Science Foundation (NSF) grant IIS-1910733. 
This work is also partially supported by an Indo-Swedish joint network project: DST/INT/SWD/VR/P-02/2019 and Swedish Research Council (VR) grant 2018-07085, and the VR grant 2019-05487. 
RS and VN are partially supported by a scholarship from MHRD, Swarnajayanti Fellowship from the Department of Science and Technology, India (DST/SJF/ETA-02/2015-16), and a Mindtree Chair research grant.

%-------------------------------------------------------------------------
\input{STAR-MT-refs.bbl}

%-------------------------------------------------------------------------

\input{sec-bio}

%-------------------------------------------------------------------------

\end{document}

%% file: sec-abstract.tex
In topological data analysis and visualization, topological descriptors such as persistence diagrams, merge trees, contour trees, Reeb graphs, and Morse--Smale complexes play an essential role in capturing the shape of scalar field data. 
We present a state-of-the-art report on scalar field comparison using topological descriptors. 
We provide a taxonomy of existing approaches based on visualization tasks associated with three categories of data: single fields, time-varying fields, and ensembles. 
These tasks include symmetry detection, periodicity detection, key event/feature detection, feature tracking, clustering, and structure statistics. Our main contributions include the formulation of a set of desirable mathematical and computational properties of comparative measures,  and the classification of visualization tasks and applications that are enabled by these measures. 

%% file: sec-introduction.tex
\section{Introduction}
\label{sec:introduction}

Topological data analysis (TDA) provides fundamental tools for scientific visualization in terms of abstraction and summarization. 
These tools have great potential for data comparison, feature tracking, and  ensemble analysis.
For these purposes, a large variety of comparative measures have been proposed targeting different topological descriptors and employed in a variety of visualization applications, which are the focus of this survey.
This state-of-the-art report aims to categorize, summarize, and analyze existing approaches that utilize comparative measures and identify open problems and opportunities for future work.
We thus are interested in both the mathematical foundations and properties of comparative measures and their use in real-world visualization applications. 

Popular topological descriptors for scalar field data, considered in this survey, can be classified into three categories: set-based such as persistence diagrams~\cite{EdelsbrunnerLetscherZomorodian2002} and barcodes~\cite{Ghrist2008, CarlssonZomorodianCollins2004}; graph-based such as  merge trees~\cite{BeketayevYeliussizovMorozov2014}, contour trees~\cite{CarrSnoeyinkAxen2003}, and Reeb graphs~\cite{Reeb1946}; and complex-based such as Morse and Morse-Smale complexes~\cite{GerberPotter2012,EdelsbrunnerHarerZomorodian2001,EdelsbrunnerHarerNatarajan2003}, respectively. 
Our work is especially motivated by the following questions:
\begin{itemize}
\item Which role do topological methods in comparative analysis and visualization play and what are the typical applications?
\item What comparative measures have been proposed and where are they applied? 
\item What are the desirable properties of a comparative measure for topological descriptors?
\end{itemize}

Our contributions include:
\begin{itemize}
\item We provide a classification of approaches in TDA and visualization relevant to the comparative study of scalar fields.
\item We collect a set of desirable properties of a comparative measure  concerning metricity, stability, discriminativity, and computational complexity;
\item We analyze existing approaches with respect to these properties.
\item We derive a list of opportunities and challenges for future work.  
\end{itemize}

We provide three navigation aids that help the reader. Two tables provide an overview of different visualization tasks that are supported by comparative measures over topological descriptors~(\autoref{table:navigate-descriptor-by-task}) and desirable properties for the various published comparative measures~(\autoref{table:properties}). 
Further, we complement our survey with a visual literature browser (\url{https://git.io/Jt2Hq}) developed with the SurVis~\cite{BeckKochWeiskopf2015} framework for an interactive navigation of the state of the art.

\para{Existing surveys.}
An organized classification of the literature related to scalar field comparison is a valuable addition that complements existing state of the art on topology-based tool sets. 
The survey by Heine \etal~\cite{HeineLeitteHlawitschka2016} provides an overview of topology-based visualization, including a classification of such models for scalar fields, vector fields, tensor fields, and multi-fields. 
Specifically for scalar fields, the survey discusses topological descriptors, including their computation, simplification, visualization, and application. 
Our paper is different from the above survey because it focuses on comparative measures of topological descriptors for scalar fields, a topic not covered by previous surveys.  

\para{Review methodology.} 
To complete the survey, we first gathered papers from a number of visualization, computational topology, and TDA venues whose title and/or abstract contain keywords relevant to topological descriptors and their comparative measures, such as ``persistent homology'', ``merge trees'', ``scalar field comparison'', etc. 
The list of venues includes but is not limited to: journals such as IEEE Transactions on Visualization and Computer Graphics (TVCG), Computer Graphics Forum (CGF), Journal of Applied and Computational Topology, Journal of Computational Geometry (JoCG), as well as conferences/workshops such as IEEE Visualization Conference (VIS) and its associated events (\textit{e.g.} IEEE Large Scale Data Analysis and Visualization or LDAV), EG/VGTC Conference on Visualization (EuroVis), IEEE Pacific Visualization Symposium (PacificVis),  International Symposium on Computational Geometry (SoCG), etc. 
Since our primary focus is on topological descriptors and their applications in visualization, we did not survey papers from machine learning venues. 
We created a virtual index card summarizing each paper under topics such as ``summary'', ``contributions'', ``topological descriptions proposed/used'', ``comparative measures proposed/used/parameters'', ``applications'',  ``properties'', ``future directions mentioned in the paper'', and additional ``tags'' and ``notes''. 
These index cards were then used during the categorization process and each card was checked by two authors. 
In total, this process resulted in approximately 200 papers that passed the initial screening, $\sim$100 of which were deemed most relevant and included in this survey. 

\para{Overview.}
This report is organized as follows:
after introducing the basic classification categories and the desired properties in \autoref{sec:classify}, some technical background on scalar field topology is summarized in \autoref{sec:background}.
Comparative measures developed for topological descriptors with mathematical definition (if applicable) are summarized in \autoref{sec:comparative-measures}.
\autoref{sec:navigation} serves as a reminder of the structure of the survey and provides navigation for the following sections, which  focus on visualization application structured by visualization tasks for single fields~(\autoref{sec:single-fields}), time-dependent fields~(\autoref{sec:time-varying-fields}), and ensembles~(\autoref{sec:ensembles}), respectively. 
A detailed discussion of desirable mathematical and computational properties and a systematic analysis of the surveyed comparative measures can be found in \autoref{sec:properties}.
The report ends with an outlook on future work and opportunities in~\autoref{sec:future} and a conclusion in~\autoref{sec:conclusion}.

%% file: sec-classify.tex
\section{Literature Research Procedure and Classification}
\label{sec:classify}

We review representative papers in the field of computational topology, TDA, and visualization that develop or utilize topological descriptors for the comparative analysis and visualization of scalar fields.  
The annotation of each paper is guided primarily by a set of visualization tasks that are associated with three categories of data, and secondarily by a set of desirable mathematical and computational properties. 
Our primary categories are loosely inspired by an existing survey~\cite{HeineLeitteHlawitschka2016} that classifies papers based on the complexity of data types; and our secondary categorization is untreated in previous works.

\subsection{Primary Categories Based on Visualization Tasks}
During our literature review, we observed that comparative measures were  developed with a focus either on a specific topological descriptor or a specific visualization task and application.  
We therefore identified three categories of data where topological comparison was applied: \textit{single fields}, \textit{time-varying fields}, and \textit{ensembles}. 
 
A single field $f$ is a scalar-valued field defined on a 2D, 3D, or higher-dimensional domain $\Xspace$, $f: \Xspace \to \Rspace$. 
A time-varying field $F$ is a dynamically changing field, and is defined over the Cartesian product of a spatial domain $\Xspace$ and a time axis $\Rspace$, $F: \Xspace \times \Rspace \to \Rspace$. 
Time-dependent data is typically available as a discrete set of temporal snapshots.
An ensemble refers to a collection $\Fcal$ of fields that are indexed by a collection of parameters, $\Fcal = \{f_i : i \in I\}$ (where $I$ is an index set). 

\para{Single fields.} Comparative measures help extract, visualize, and highlight similar structures within a single field -- broadly referred to as the symmetry detection problem in scalar fields. These measures also enable the comparison of two or more single fields for shape matching and retrieval ({\eg},~\cite{ThomasNatarajan2011, ThomasNatarajan2013, SaikiaSeidelWeinkauf2014}).

\para{Time-varying fields.} For time-varying fields, comparative measures between successive time steps have been used to detect periodic behavior, key events, or outliers~\cite{NarayananThomasNatarajan2015, SaikiaWeinkauf2017, LukasczykWeberMaciejewski2017, SridharamurthyMasoodKamakshidasan2020}.  
Comparative measures also drive explicit feature tracking in time-varying data. 

\para{Ensembles.} For ensembles, comparative measures help identify similar or dissimilar behavior between members. 
They help identify clusters of members, outliers, or unique members of the ensemble. 
More recently, they have been used to compute structure statistics that describe the distribution of the ensemble members in the parameter space~\cite{SolerPlainchaultConche2018,YanWangMunch2020,AthawaleMaljovecYan2020}.

\subsection{Secondary Categories Based on Desirable Properties}
\label{sec:classify-properties}

We discuss desirable properties of a comparative measure $d = d(\A_1, \A_2)$ between a pair of topological descriptors (of the same type), $\A_1$ and $\A_2$. 
We focus on four types of properties surrounding \emph{metricity}, \emph{stability}, \emph{discriminativity}, and \emph{computational complexity}. 
These properties have been studied across scattered literature in TDA and visualization. 
We systematically investigate these properties and their relations to existing approaches in \autoref{sec:properties}.

\para{Metricity and Pseudometricity.}
Requiring $d$ to be a metric is desirable. 
That is, $d$ satisfies the following metric properties:  
\begin{itemize}
\item[1.] Non-negativity: $d(\A_1, \A_2) \geq 0$;
\item[2.] Identity: $d(\A_1, \A_2)=0$ iff $\A_1 = \A_2$;
\item[3.] Symmetry: $d(\A_1, \A_2) = d(\A_2, \A_1)$;
\item[4.] Triangle inequality: $d(\A_1, \A_2) \leq d(\A_1, \A_3) + d(\A_2, \A_3).$
\end{itemize}
If the triangle inequality (item 4) above is not required, $d$ becomes a \emph{dissimilarity measure} instead. 
If the identity is not required, $d$ becomes a \emph{pseudometric}, replacing item 2 above by:
\begin{itemize}
\item $d(\A_1, \A_1)=0$ (but possibly $d(\A_1,\A_2) = 0$ for some distinct $\A_1 \neq \A_2$).
\end{itemize}

\para{Stability.}
Many definitions of stability for a distance metric $d$ with respect to the underlying scalar field have been proposed.
Stability can refer to whether $d$ is stable with respect to simplification or perturbation of the underlying function. 
For example, given two scalar fields $f_1$ and $f_2:\Xspace \to \Rspace$ that give rise to a pair of topological descriptors $\A_1$ and $\A_2$, $d$ is $L^{\infty}$-stable if for some constant $C>0$, 
$$d(\A_1, \A_2) \leq C\cdot ||f_1-f_2||_\infty.$$

\para{Discriminativity.}
Discriminativity also has various definitions. For instance, using a comparative measure $d_0$ as a baseline, $d$ is considered to be more discriminative than $d_0$ if for some constant $c>0$, 
$$d_0(\A_1, \A_2) \leq c \cdot d(\A_1,\A_2)$$ and there exists no constant $c'>0$ such that $d_0 = c' \cdot d$ (that is, $d$ is not a scaled version of $d_0$).

\para{Computational complexity.}
We investigate the computational complexity of $d$ in terms of the time and space complexity, scalability, and parallel computing. 
We investigate whether $d$ is \emph{easily implementable}, referring to whether an algorithmic solution has been proposed which affects its practicality. 
 
The above properties are particularly desirable for analysis and visualization tasks that are supported by a comparative measure. They lead to theoretically sound, interpretable, robust, reliable, and practical methods for comparative visualization.

%% file: sec-background.tex
\section{Technical Foundations on Scalar Field Topology}
\label{sec:background}

In this section, we review the technical foundations for scalar field topology, including the definitions of Morse functions and topological descriptors;  see~\cite{Zomorodian2005,EdelsbrunnerHarer2010} for computation-oriented and ~\cite{Tierny2017} for visualization-oriented introduction to scalar field topology. 
We review set-based (persistence diagrams, barcodes), graph-based (merge trees, contour trees, Reeb graphs), and complex-based (Morse and Morse-Smale complexes) topological descriptors and their variants. 
The graph-based descriptors are largely based on contours of a function, whereas complex-based ones are primarily based on its gradient. 
We also briefly mention relevant topological descriptors for multivariate functions (Jacobi sets, Reeb spaces, joint contour nets), as they are natural extensions of their univariate counterparts. 

\subsection{Morse Functions and Morse Theory}
Most of the topological descriptors described in this section are rooted in Morse theory~\cite{Milnor1963}.
We give a high-level review here; see~\cite{Matsumoto2002} for a friendly introduction and~\cite{Milnor1963} for the original treatment.   

\begin{figure}[!ht]
    \centering
    \includegraphics[width=0.98\columnwidth]{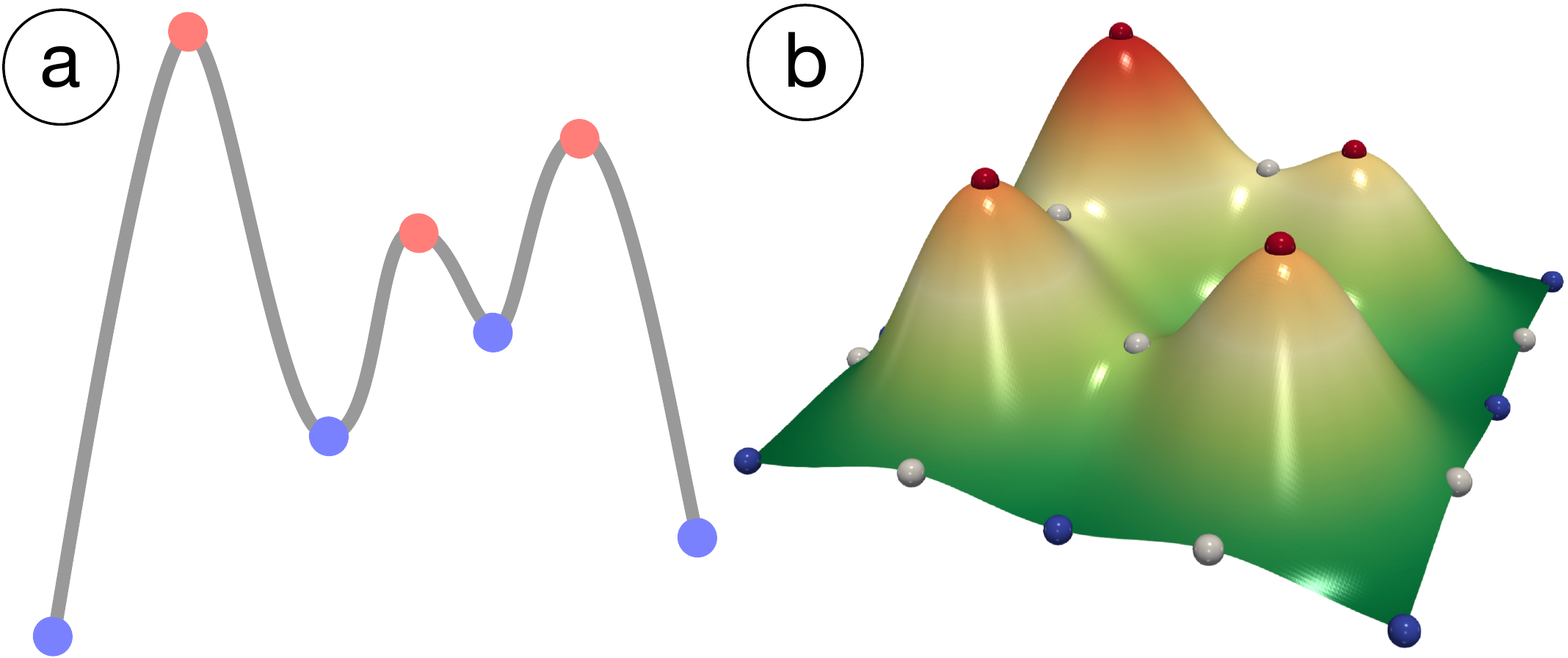}
    \caption{Morse functions with (a) a 1-dimensional and (b) a 2-dimensional domain, respectively.} 
    \label{fig:Morse-functions}
\end{figure}

\para{Morse functions.} Let $\Mspace$ be a smooth manifold and $f: \Mspace \to \Rspace$ a smooth function on $\Mspace$. 
A point $x \in \Mspace$ is a \emph{critical point} of $f$ if and only if the partial derivatives at $x$ are zero; otherwise, it is a \emph{regular point}. 
The image of a critical point is a \emph{critical value} of $f$. 
A critical point $x$ is \emph{non-degenerate} if the Hessian (the matrix of second derivatives) at $x$ is non-singular. 
 $f$ is a \emph{Morse function} if all its critical points are non-degenerate and have distinct function values. 
 \autoref{fig:Morse-functions} gives two examples of Morse functions with a 1- and a 2-dimensional domain, respectively. Critical points are always displayed as red (for local maxima), blue (for local minima), and white (for saddles) circles or spheres. 
 
\para{Morse theory.}  
For a Morse function $f: \Mspace \to \Rspace$, let $\Mspace_t := f^{-1}(-\infty, t] = \{ x \in \Mspace \mid f(x) \leq t\}$ denote sublevel sets of $f$. 
A basic result of Morse theory states that almost all functions are Morse functions. Technically speaking, the set of Morse functions forms an open dense subset of the space of smooth functions. 
In practice, a non-Morse function can be made into a Morse function by resolving degenerate conditions via the simulation of simplicity~\cite{EdelsbrunnerMucke1990}.
We assume all functions discussed in this paper to be Morse. 

The Morse lemma states that a function looks extremely simple near a non-degenerate critical point. 
Two fundamental theorems of Morse theory study how sublevel sets of a function changes topologically {\wrt} its critical points. 
A number of theoretical properties relevant to topological descriptors described in this section can be traced back to these two fundamental theorems. 
We refer interested readers to \cite[Theorems 3.1 and 3.2]{Milnor1963} in their original forms. 
In a nutshell, these theorems describe if and when the topology of  sublevel sets $\Mspace_t$ change as $t$ varies, in particular, when $t$ passes a critical value. 
Topological descriptors such as persistent diagrams and merge trees are related with one another via theorems of Morse theory as both are defined over the sublevel sets of a function.

In practice, we rarely find smooth functions. Instead, we are given samples of such functions, represented as a function on a point cloud sample of $\Mspace$. 
Oftentimes, we impose a combinatorial structure (\ie,~a simplicial complex $K$) on the sample as an approximation of $\Mspace$.  
Let $K$ be a simplicial complex with real values specified on its vertices; $|K|$ represents its \emph{underlying space}.  
We obtain a piecewise linear (PL) function $f: |K| \to \Rspace$ using linear extension over the simplices, where $f(x) = \sum_i b_i(x) f(u_i)$ ($u_i$ are vertices of $K$ and $b_i(x)$ are the barycentric coordinates of $x$)~\cite[page 135]{EdelsbrunnerHarer2010}. 
We can then apply Morse-theoretical ideas to this PL approximation.  
This application is justifiable according to the \emph{Simplicial Approximation Theorem}~\cite[page 56]{EdelsbrunnerHarer2010}, which states that every continuous function on  a triangulable topological space can be approximated by a PL function.

As described in subsequent sections, in some instances, features that form parts of topological descriptors are used in the comparative measures, in particular, critical points and their attributes, level sets (contours, or isosurfaces) defined as $f^{-1}(t)$ for some $t \in \Rspace$.  

\subsection{Persistence Diagrams and Barcodes}
Persistent homology is a widely used tool for TDA and visualization. 
Algebraically, it takes the form of a persistence module~\cite{ChazalSilvaGlisse2016}.
In this paper, we are mostly concerned with persistence homology that arises from \emph{sublevel set filtrations} of Morse functions. 
We refer the reader to~\cite{EdelsbrunnerHarer2010, CarlssonSilva2010, BendichEdelsbrunnerMorozov2013} for different ways to study persistent homology. 

\begin{figure}[!ht]
    \centering
    \includegraphics[width=0.98\columnwidth]{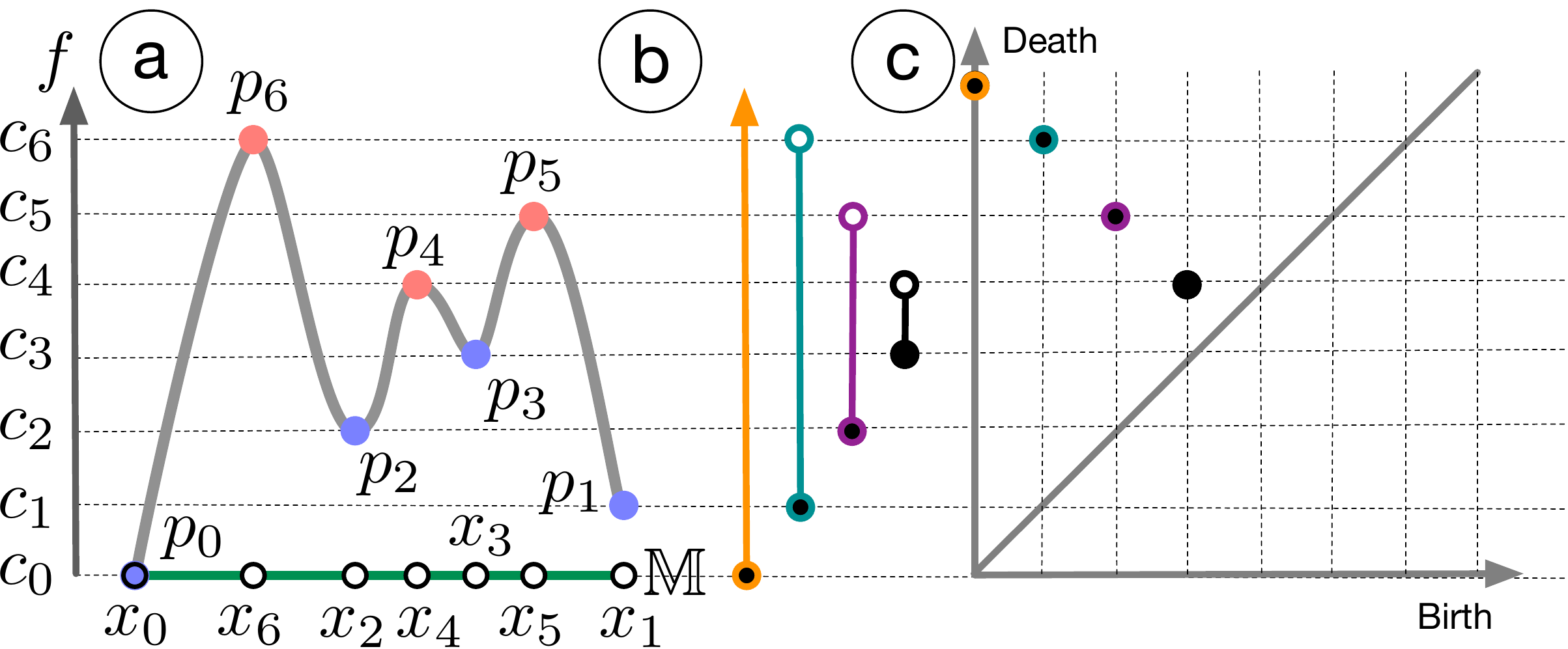}
    \vspace{-2mm}
    \caption{(a) The graph of $f: \Mspace \to \Rspace$, where each point $p_i = (x_i, c_i)$ for $c_i = f(x_i)$; together with (b) the 0-dimensional barcode and (c) 0-dimensional persistence diagram of $f$ based on its sublevel set filtration.} 
    \label{fig:dgm}
     \vspace{-4mm}
\end{figure}

\para{Persistence diagrams.} 
Let $f: \Mspace \to \Rspace$ be Morse and $\Mspace_{t}:= f^{-1}(\infty, t]$ its sublevel sets. 
Assuming $\Mspace$ is also compact, then a Morse function $f$ on a compact manifold contains finitely many critical points (as a consequence of the Morse lemma). 
Let $n$ be the (finite) number of critical values of $f$. 
Let $a_0 < \cdots < a_{n}$ be a sequence of regular values of $f$ such that each interval $(a_i, a_{i+1})$ contains exactly one critical value of $f$. 
A sublevel set filtration of $f$ is a sequence of sublevel sets connected by inclusions, 
$$\Mspace_{a_0} \to \Mspace_{a_1} \to \dots \to \Mspace_{a_n}.$$ 
Persistent homology studies the topological changes of sublevel sets by applying $k$-dimensional homology ($k \geq 0$) to this sequence, 
$$\Hgroup_k(\Mspace_{a_0}) \to \Hgroup_k(\Mspace_{a_1}) \to \dots \to \Hgroup_k(\Mspace_{a_n}).$$ 
Given a topological space $\Xspace$, the $0$-, $1$-, and $2$-dimensional homology groups, denoted as $\Hgroup_0(\Xspace)$, $\Hgroup_1(\Xspace)$, and $\Hgroup_2(\Xspace)$, respectively, capture the connected components, tunnels, and voids of $\Xspace$. 
We give an example of $0$-dimensional persistence homology based on the sublevel set filtration of a 1-dimensional Morse function in~\autoref{fig:dgm}. 

Formally, a $k$-dimensional persistence diagram $\D$ is the disjoint union of a multi-set of off-diagonal points $\{(b, d) \mid b \neq d, b, d \in \Rspace_{\geq 0}\}$ on the Euclidean plane $\bar{\Rspace}^2$ (where $\bar{\Rspace} = \Rspace \cup \{-\infty, +\infty\}$) and the diagonal $\Delta = \{(b,b) \ | \ b \in \Rspace_{\geq 0}\}$ counted with infinite multiplicity.
As illustrated in~\autoref{fig:dgm}a, let $c_i$ denote the critical values of a Morse function $f$ restricted to an interval $\Mspace \subset \Rspace$, $f: \Mspace \to \Rspace$, where $c_0 < c_1 < \dots < c_6$.   
Let $x_i$ denote the critical points of $f$.  Assume $f$ is Morse, then $c_i = f(x_i)$. For simplicity, we set $c_0 = 0$, $c_1 = 1$, and $c_i = i $, etc. 
Let $a_0 < a_1 < \dots < a_7$ be a sequence of regular values of $f$ such that each interval $(a_i, a_{i+1})$ contains exactly one critical value $c_i$.
The $0$-dimensional persistent homology captures how connected   components in the sublevel sets $\Mspace_{t}$ changes as $t$ varies from $a_0$ to $a_7$.  
At $t = a_0 < 0$, $\Mspace_{t} = \emptyset$. 
At $t = 0$, a single (connected) component appears in the sublevel set $\Mspace_{t}$ containing the global minimum $x_0$, we call this a \emph{birth} event at $\Mspace_{0}$.
At $t = 1, 2$, and $3$, a 2nd, 3rd, and 4th component appears in $\Mspace_{t}$ containing local minima $x_1$, $x_2$, and $x_3$, respectively. 
At $t = 4$, the component containing $x_3$ merges with the component containing $x_2$ as per the \emph{Elder Rule}~\cite[Page. 150]{EdelsbrunnerHarer2010}, referred to as a \emph{death} event: the component containing $x_3$ disappears (dies) while the component containing $x_2$ remains. 
At $t = 5$, the component containing $x_2$ merges with the component containing $x_1$ and dies. 
At $t = 6$, the component containing $x_1$ merges with the component containing $x_0$ and dies. 
Persistent homology pairs the birth and death events either as a set of intervals (called \emph{barcode}), or a multi-set of points in the plane (called \emph{persistence diagram}). 

\para{Barcodes.} A barcode is shown in~\autoref{fig:dgm}b. 
The component containing $x_0$ never dies, giving rise to a bar $[0, \infty)$ in the barcode that begins at $0$ and goes to $\infty$.
The component containing $x_1$ is born at $t=1$ and dies at $t=6$, which corresponds to a bar $[1,6)$. 
Similarly, the birth and death events of components containing $x_2$ and $x_3$ give rise to two additional bars $[2, 5)$ and $[3,4)$, respectively.
The \emph{persistence} of a bar $[b,d)$ in a barcode is defined to be $|d-b|$, which captures the life span of a component in the filtration. 
A persistence diagram is shown in~\autoref{fig:dgm}c, where each bar $[b,d)$ is mapped to a point $(b, d)$ on the plane.

\begin{figure}[!ht]
    \centering
    \includegraphics[width=0.98\columnwidth]{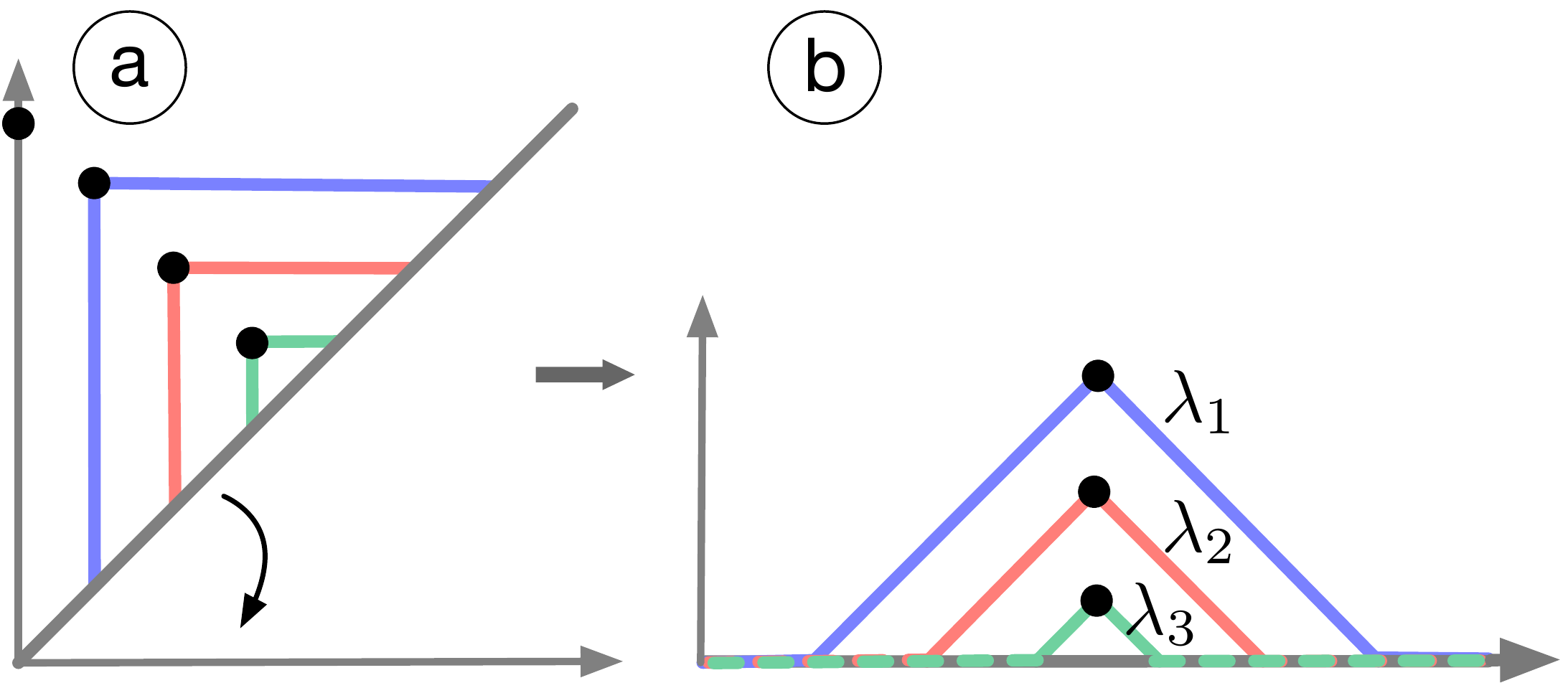}
    \caption{Rotating a persistence diagram in (a) to create a functional representation -- a persistence landscape in (b).}
    \vspace{-4mm} 
    \label{fig:persistence-landscape}
\end{figure}

\para{Other variants} exist, mostly derived from persistence diagrams or barcodes. 
The \emph{persistence landscape}~\cite{Bubenik2015} is a function-based representation of a persistence diagram. 
It maps a persistence diagram into a function space, which allows it to be easily integrated with tools from statistics and machine learning~\cite{BubenikDlotko2017,Bubenik2020}. 
Formally, for a birth-death pair $(b, d)$ in a persistence diagram, assuming $b$ and $d$ are finite, we define a piecewise-linear function $f_{(b,d)} : \Rspace \to [0, \infty]$ as 
\[
    f_{(b,d)}= 
\begin{cases}
    0,& \text{if } x \notin (b,d)\\
    x-b,              & \text{if  } x \in (b, \frac{b+d}{2}]\\
    -x+d,              & \text{if  } x \in [\frac{b+d}{2}, d)\\
\end{cases}.
\]
The \emph{persistence landscape} of the birth-death pairs $\{(b_i , d_i )\}_{i=1}^n$ in a persistence diagram is the sequence of functions $\lambda_k: \Rspace \to [0, \infty]$, where $\lambda_k(x)$ is the $k$-th largest value of $\{f_{(b_i,d_i)}(x)\}_{i=1}^n$ (for $k = 0, 1, 2, \dots$).  
$\lambda_k(x)=0$ if the $k$-th largest value does not exist. 
In other words, the \emph{persistence landscape} is a function $\lambda : \Nspace \times  \Rspace \to [0, \infty]$, where $\lambda(k,t) = \lambda_k(t)$~\cite{BubenikDlotko2017}. 
Intuitively, consider the points with finite birth and death times in a persistence diagram~(\autoref{fig:persistence-landscape}a). We construct a persistence landscape in~\autoref{fig:persistence-landscape}b by rotating the points by $45^{\circ}$ and building three linear functions, $\lambda_1$~(blue), $\lambda_2$~(red), and $\lambda_3$~(green), with these points.  

Another descriptor widely used in machine learning is the \emph{persistence image}~\cite{AdamsEmersonKirby2017}. 
It is a vector-based representation of a persistence diagram.
It can be informally considered as a heat map, which is generated from a weighted sum of Gaussian centered at each point $(b,p)$, where $b$ is the birth and $p=d-b$ is the persistence of a point in the persistence diagram. 

\emph{Betti curves} also summarize the information of persistent homology (\textit{e.g.},~\cite{Robins2002, GameiroMischaikowKalies2004, RieckSadloLeitte2020b, ChungLawson2020}). 
Recall the $k$-th Betti number is informally the number of $k$-dimensional holes (homology) of a topological space. 
For a filtration parameter $t$, the Betti curves at $t$ are the Betti numbers of the associated complex. 
Betti curves are arguably the simplest function-based representation of a persistence diagram (\cf, the persistence landscape).   
Turner~\etal~\cite{TurnerMukherjeeBoyer2014} introduced a summary statistic from persistence diagram, called the persistent homology transform~(PHT), to model surfaces in $\Rspace^3$ and shapes in $\Rspace^2$. 
Li~\etal~\cite{LiWangAscoli2017} proposed another persistence-based feature vectorization of a persistence diagram using a 1-dimensional  density function to compare neuronal trees; their feature vectorizations can be considered as a 1-dimensional version of the persistent images~\cite{AdamsEmersonKirby2017}.
Rieck~\etal~\cite{RieckSadloLeitte2017a} developed an inter-level set persistence hierarchy~(ISPH) to capture the spatial relationship between features in persistence diagram.

\subsection{Merge Trees, Contour Trees, and Reeb Graphs}
\label{Subsec:RG}

Topological descriptors such as merge trees, contour trees, and Reeb graphs capture topological changes of (sub)level sets of scalar fields, which are real-valued smooth functions.

\begin{figure}[!b]
    \centering
    \includegraphics[width=0.98\columnwidth]{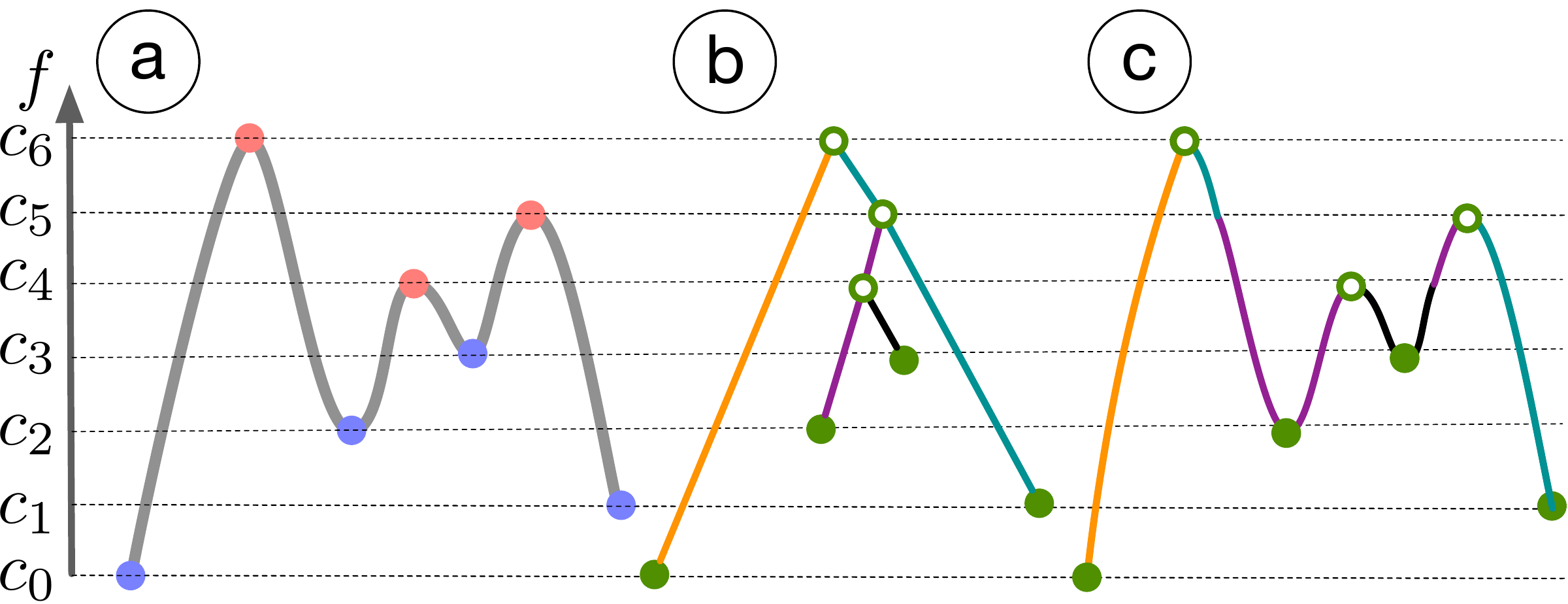}
    \caption{(a) The graph of a 1-dimensional Morse function $f$ restricted to an interval, $f: \Mspace \to \Rspace$; (b) the merge tree of $f$ shown abstractly, where branches are colored based on its branch decomposition; (c) the graph of $f$ is colored based on the branch decomposition in (b).} 
    \label{fig:mt}
\end{figure}

\para{Merge trees.} Given a Morse function $f: \Mspace \to \Rspace$ defined on a connected  domain $\Mspace$, a merge tree records the connectivity of its sublevel sets. 
Two points $x, y \in \Mspace$ are \emph{equivalent} ({\wrt}~$f$), $x \sim y$, if they have the same function value, that is, $f(x) = f(y) = t$, and if they belong to the same connected component of the sublevel set $\Mspace_t$, for some $t \in \Rspace$.
A \emph{merge tree} is the quotient space $\Mspace/{\sim}$ obtained by gluing together points in $\Mspace$ that are equivalent under the relation $\sim$. 
It keeps track of the evolution of connected components in $\Mspace_t$ as $t$ increases; see~\autoref{fig:mt} for an example. 
In the abstract view of a merge tree in \autoref{fig:mt}b, each leaf corresponds to a local minimum of $f$ that represents the birth of a connected component; each internal node corresponds to the merging of components; and the root represents the entire space as a single component. 
\autoref{fig:mt}b also visualizes the branches of the merge tree based on its branch decomposition. 
The connection between a merge tree and the barcode is apparent, \cf~\autoref{fig:dgm}(b) and~\autoref{fig:mt}(b-c), where a merge tree decomposes into a barcode following a branch decomposition process;  and bars in a barcode can be used to assemble a (non-unique) merge tree following a gluing process.
See~\cite{CatanzaroCurryFasy2020, Curry2017, KanariGarinHess2020} for references for the relation between a merge tree and a barcode.  
Note that the notions of join and split trees~\cite{CarrSnoeyinkAxen2003} are the two forms of merge trees; a join tree is the merge tree of $f$ and a split tree is the merge tree of $-f$. 

\begin{figure}[!ht]	
    \centering
    \includegraphics[width=0.98\columnwidth]{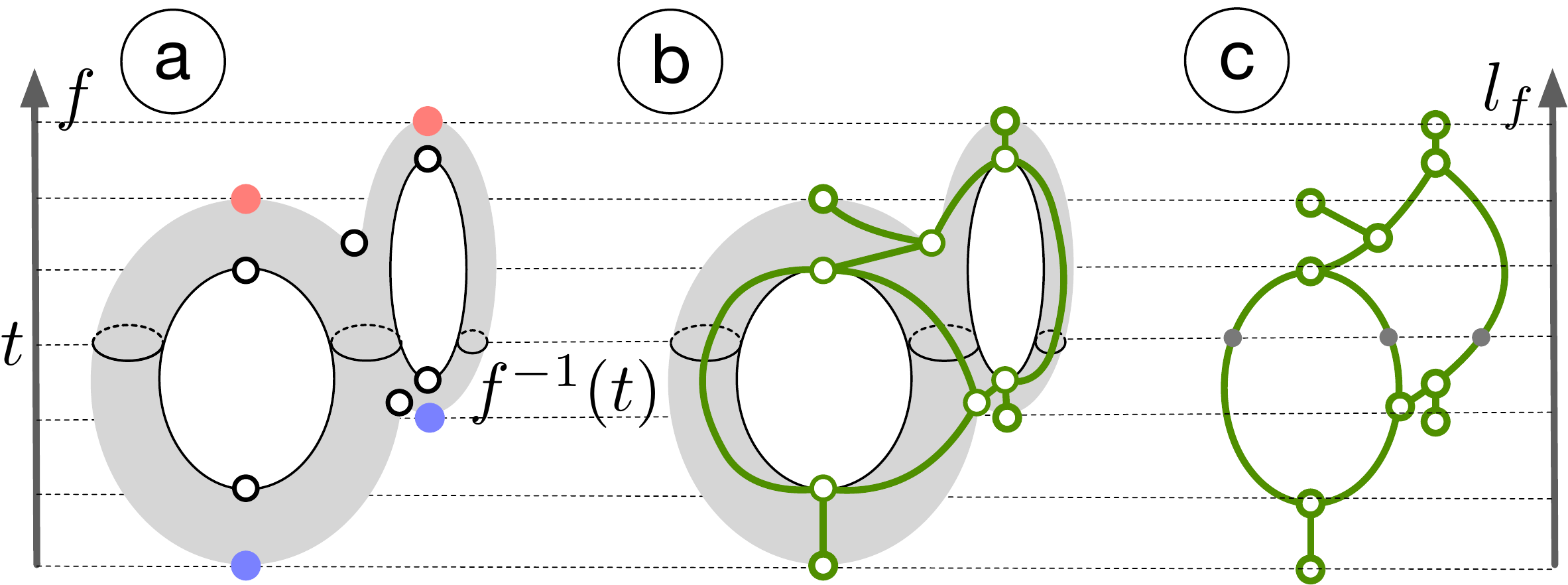}
    \caption{(a) A height function $f: \Mspace \to \Rspace$ defined on a double torus, (b) its Reeb graph embedded in the domain $\Mspace$, and (c) its Reeb graph shown in an abstract view. If the Reeb graph in (c) is further equipped with a function $l_f$ defined on its vertices, where $l_f$ is the restriction of $f$ to $V$, then we obtain a labeled Reeb graph.}
    \label{fig:Reeb}
\end{figure}

\para{Reeb graphs and contour trees.} A Reeb graph, on the other hand, relies on equivalence relations among points in the \emph{level sets} of a Morse function $f: \Mspace \to \Rspace$. 
Two points  $x, y \in \Mspace$ are \emph{equivalent}, $x \sim y$, if $f(x) = f(y) = t$, and if they belong to the same connected component of the level set $f^{-1}(t)$, for some $t \in \Rspace$. 
The \emph{Reeb graph} $\G_f:=\Mspace/{\sim}$ is the quotient space obtained by identifying equivalent points; see~\autoref{fig:Reeb}. 
Nodes in the Reeb graph have a one-to-one correspondence with the  critical points of $f$, while arcs connect the nodes. 
A point on an arc represents a connected component of a level set (i.e., a \emph{contour}) in $\Mspace$.  
Intuitively, as $t$ increases within the range of $f$, a Reeb graph captures the topological changes in the level sets of $f$, in particular, the appearances, disappearances, splitting, and merging among the connected components (contours) of $f^{-1}(t)$; 
see~\cite[section VI.4]{EdelsbrunnerHarer2010} for a formal treatment.
Bauer~\etal~\cite{BauerDiFabioLandi2016} worked with the notion of a \emph{labeled Reeb graph}, where the vertices of $\G_f$ are labeled by the function $l_f: V(\G_f) \to \Rspace$ induced by restricting $f: \Mspace \to \Rspace$ to its critical points. Then, $(\G_f, l_f)$ is the labeled Reeb graph of the data $(\Mspace,f)$, see~\autoref{fig:Reeb}c. 

\begin{figure}[!ht]	
    \centering
    \includegraphics[width=0.98\columnwidth]{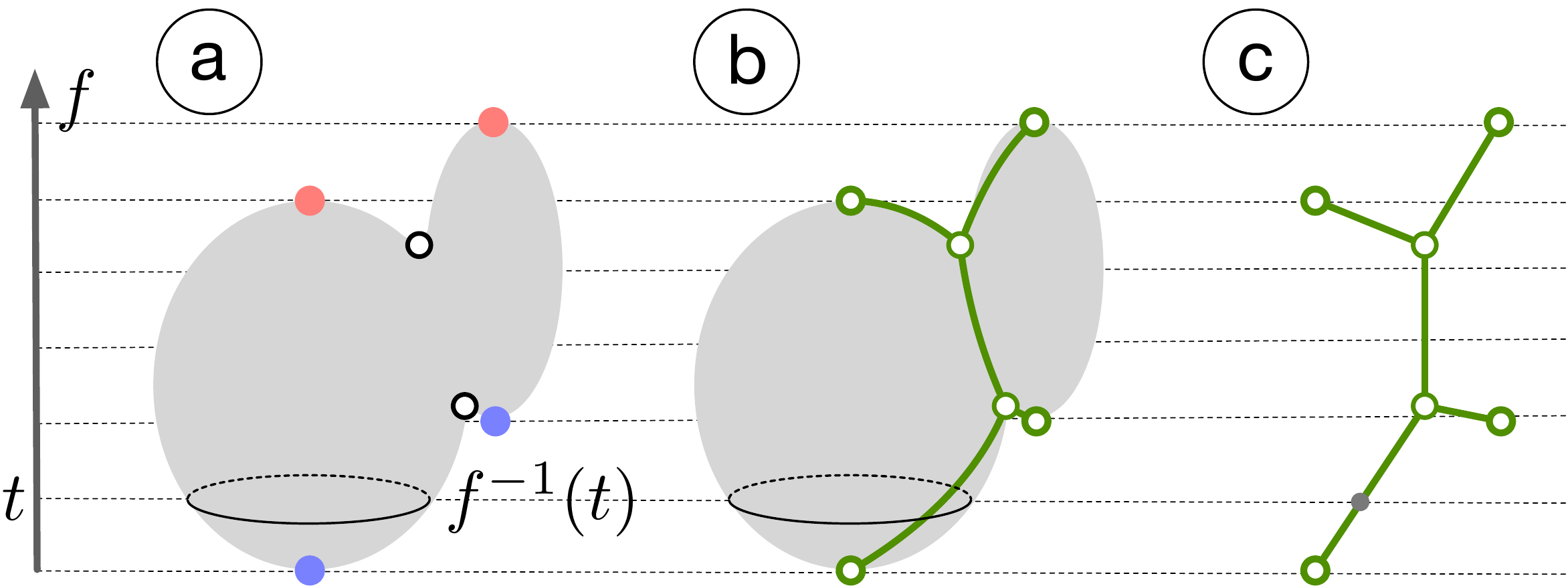}
    \caption{(a) A height function $f: \Mspace \to \Rspace$ defined on the surface of two (solid) balls glued together; (b) its contour tree embedded in the domain $\Mspace$; and (c) its contour tree shown in an abstract view.} 
    \label{fig:contour-tree}
\end{figure}

A contour tree is a special type of Reeb graph when the domain $\Mspace$ is simply connected. Then, $\Mspace/{\sim}$ gives rise to a tree; see~\autoref{fig:contour-tree} for an example involving a ``deformed'' spherical domain.   
The main difference between a contour tree and a merge tree is that the former captures the connectivity among level sets, while the latter encodes the connectivity among sublevel sets of a Morse function.

\para{Mapper constructions and mapper graphs.}
Given a point cloud $\Xspace \subset \Rspace^d$, we construct the \emph{nerve of a covering}. 
Let $I$ be an index set. A \emph{cover} of $\Xspace$ is defined as a set of open sets in $\Rspace^d$, $\Ucal = \{U_i\}_{i \in I}$ such that $\Xspace \subset \cup_{i \in I} U_i$. 
The \emph{nerve} complex of $\Ucal$ is a simplicial complex, $\Ncal(U) := \{J \subset I \mid \cap_{j \in J} U_j \neq \emptyset \}$. 
The 1-dimensional nerve of $\Ucal$, denoted as $\Ncal_1(\Ucal)$, is a graph.   
Each node $i \in I$ in $\Ncal_1(\Ucal)$ represents a cover element $U_i$, and there is an edge between $i,j \in I$ if $U_i \cap U_j \neq \emptyset$.  

Given a real-valued function $f: \Xspace \to \Rspace$, we start with a finite cover of $f(\Xspace) \subset \Rspace$ using intervals, that is, a cover $\Vcal = \{V_k\}$ such that $f(\Xspace) \subseteq \cup_{k} V_k$. 
We obtain a cover $\Ucal$ of $\Xspace$ by considering the clusters induced by points in $f^{-1}(V_k)$ for each $V_k$ as cover elements. 
The nerve of $\Ucal$ is a simplicial complex, and is referred to as the \emph{mapper} (or mapper construction) of $f$.
The 1-dimensional nerve of $\Ucal$, $\Ncal_1(\Ucal)$, is the \emph{mapper graph} of $(\Xspace, f)$. 

\begin{figure}[!ht]
    \centering
    \includegraphics[width=0.98\columnwidth]{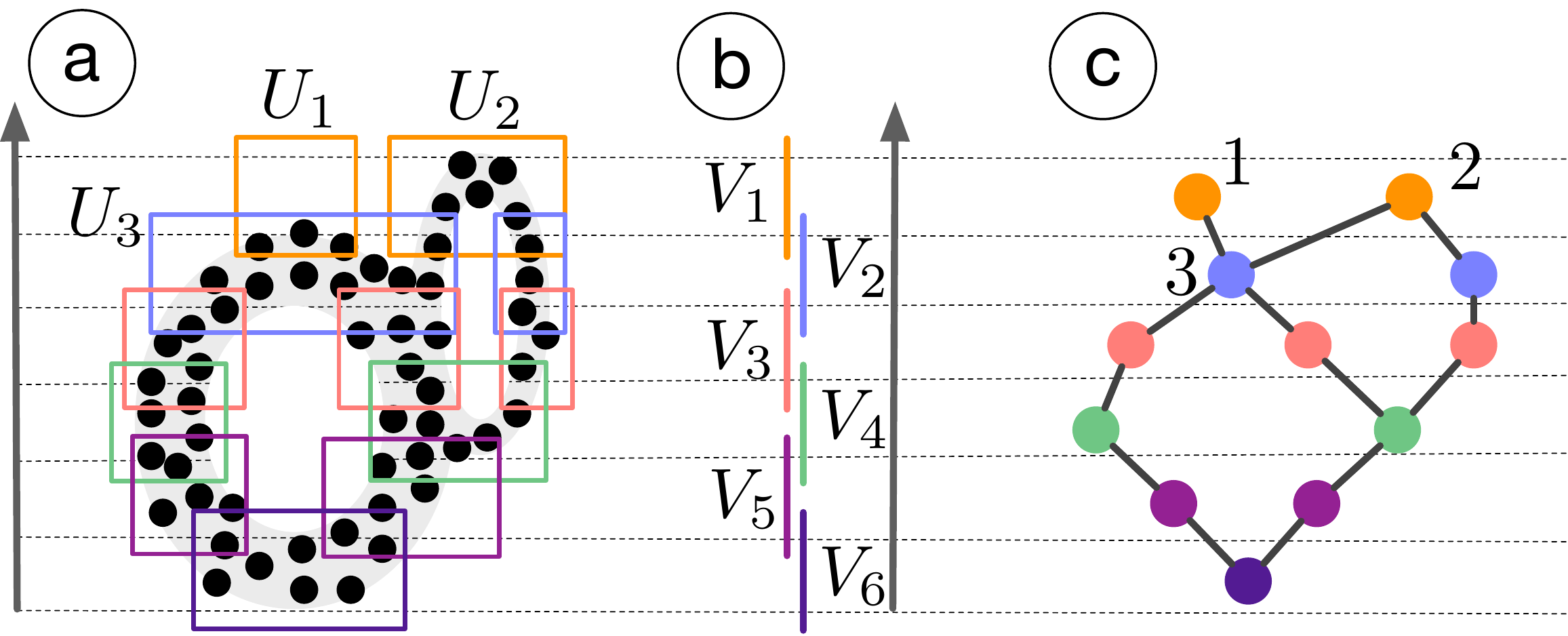}
    \caption{(a) A height function $f: \Xspace \to \Rspace$ defined on a point cloud sampled from a double annulus together with a cover, (b) the cover of $f(\Xspace)$ with intervals, and (c) its mapper graph.} 
    \label{fig:mapper-graph}
\end{figure}

Take as an example a 2-dimensional point cloud $\Xspace$ sampled from a double annulus that is equipped with a height function in~\autoref{fig:mapper-graph}a.  
Six intervals form a cover $\Vcal = \{V_1, V_2, \cdots, V_6\}$ of the image of $f$, that is, $f(\Xspace) \subset \bigcup_k V_k$ ($1 \leq k \leq 6$) in~\autoref{fig:mapper-graph}b. 
For each $k$, $f^{-1}(V_k)$ induces some clusters of points that are subsets of $\Xspace$; each cluster forms cover elements of $\Xspace$.
For instance, $f^{-1}(V_1)$ induces two clusters of points that are enclosed by the orange cover elements $U_1$ and $U_2$, and $f^{-1}(V_2)$ induces two clusters enclosed by the blue cover elements, one of which is $U_3$. 
The mapper graph shows that there is an edge between node $1$ and node $2$ in~\autoref{fig:mapper-graph}c since $U_1 \cap U_3 \neq \emptyset$. 

\para{Other contour-based topological descriptors} have been studied in recent years.  
Zhang {\etal} introduced the \emph{dual contour tree}, which is constructed from the contour tree of a volume by dividing its functional range into segments such that the connected contour tree edges within a segment become a node in the dual tree~\cite{ZhangBajajBaker2004}.  
The dual contour tree shares many resemblances with the mapper graph; see~\autoref{sec:single-fields} and~\autoref{sec:ensembles} for its applications in visualization.  
 
A \emph{branch decomposition tree}~(BDT) is derived from a contour tree~\cite{PascucciMcLaughlinScorzelli2004} or a merge tree~\cite{SaikiaSeidelWeinkauf2014}. 
A BDT represents the branch decomposition of a tree, with the nodes representing the branches and the edges representing their hierarchy. 
Saikia \etal~\cite{SaikiaSeidelWeinkauf2014} further introduced an \emph{extended branch decomposition graph}~(eBDG), which represents a forest of BDTs, where each of the BDTs is computed from a subtree of the merge tree.  

In addition to mapper graphs, Reeb graphs have several variants, many of which have not been utilized in scientific visualization. 
The \emph{$\alpha$-Reeb graph}~\cite{ChazalSun2014} defines the equivalence relation between points using open intervals of length at most $\alpha$. 
The \emph{extended Reeb graph}~\cite{BarraBiasotti2014} uses cover elements from a partition of the domain without overlaps.
The \emph{enhanced mapper graph}~\cite{BrownBobrowskiMunch2021} considers inverse images of intersections among the cover elements and encodes function values on its vertices and edges.  
Several variants of mapper constructions exist, as discussed in~\autoref{sec:multivariate-descriptors}.

\subsection{Morse and Morse-Smale Complexes}
\label{subsec:MS}

\begin{figure}[!ht]
    \centering
    \includegraphics[width=0.98\columnwidth]{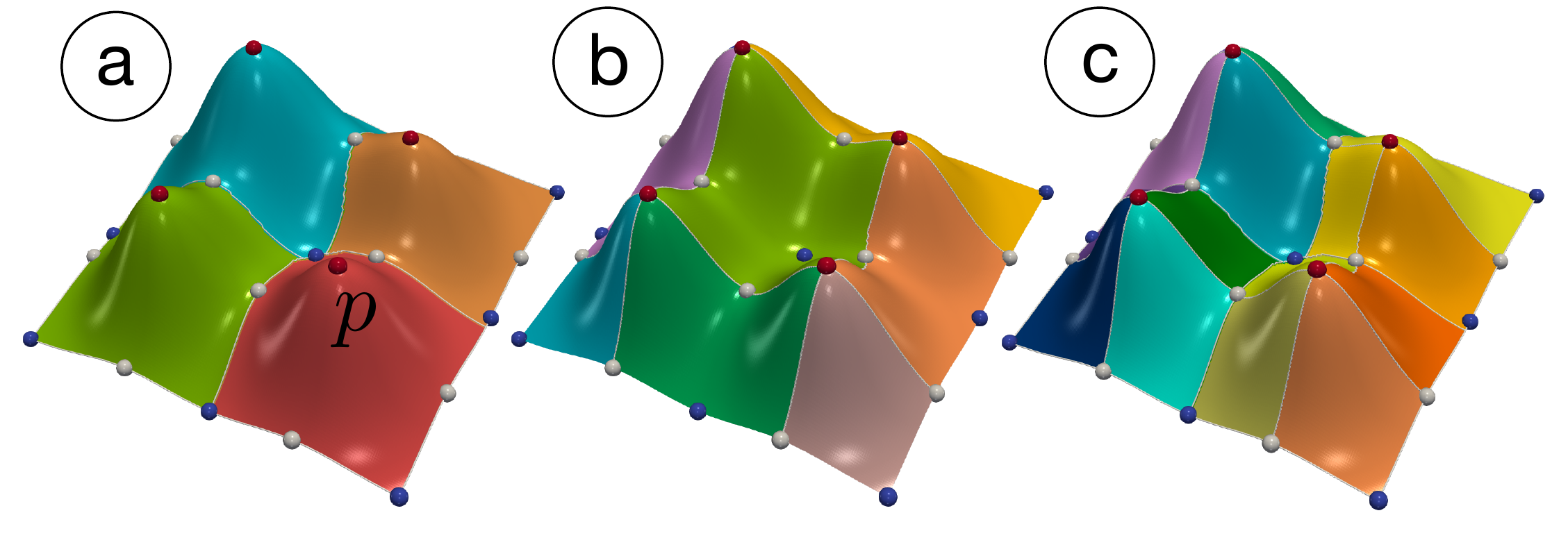}
    \caption{Given the 2-dimensional function $f$ from~\autoref{fig:Morse-functions}, (a) shows the Morse complex of $f$ (with stable manifolds), (b) shows the Morse complex of $-f$ (with unstable manifolds), and (c) is the Morse-Smale complex of $f$.} 
    \label{fig:msc}
\end{figure}

Let $f: \Mspace \to \Rspace$ be a Morse function,  $\grad{f}$ its gradient. 
At a regular point $x$, an \emph{integral line} is a maximal path whose tangent vectors agree with the gradient \cite{EdelsbrunnerHarerZomorodian2001}. 
An integral line begins and ends at critical points.   
The \emph{stable manifold} surrounding a critical point  $p$ includes $p$ itself and all regular points whose integral lines end at $p$. 
This is also referred to as the \emph{descending manifold} of $p$ since $f(p) \geq f(x)$ for all points $x$ in the stable manifold of $p$~\cite[Page 131]{EdelsbrunnerHarer2010}.
For instance, the stable manifold of the local maximum $p$ in~\autoref{fig:msc}a corresponds to the red ``bump''. 
The \emph{unstable manifold} (\emph{ascending manifold}) of a critical point   $p$ is the point itself together with all regular points whose integral lines originate at $p$~\cite[Chap.~VI, page 131]{EdelsbrunnerHarer2010}, see \autoref{fig:msc}b. 
Symmetrically, an unstable manifold (\emph{ascending manifold}) of $p$ in $f$ is a stable manifold of $p$ in $-f$. 
A Morse function $f$ is a \emph{Morse-Smale function} if the stable and unstable manifolds intersect transversally. 

Given a Morse-Smale function $f$ defined on a 2-dimensional domain, its stable manifolds surrounding local maxima decompose the domain into $2$-cells  (colored regions in~\autoref{fig:msc}a), whereas integral lines connecting the critical points are the $1$-cells, and critical points are the $0$-cells.  
These cells form a complex called a \emph{Morse complex} of $f$.  
Intersecting the stable and unstable manifolds of $f$ (equivalently, intersecting the Morse complex of $f$ and $-f$) gives rise to a refinement of the two complexes called the \emph{Morse-Smale complex} (MSC) of $f$, see~\autoref{fig:msc}c. 
Its $0$-cells are the critical points, and its $1$- and $2$-cells are the components of the unions of integral lines with a common origin and a common destination~\cite[Chap.~VI, page 134]{EdelsbrunnerHarer2010}.  
3D Morse and Morse-Smale complexes of $f: \Mspace \subset \Rspace^3 \to \Rspace$ are defined similarly based on the gradient behavior of points in its domain~\cite{EdelsbrunnerHarerNatarajan2003}. 
These complexes can be approximated in high dimensions for data analysis and regression~\cite{GerberBremerPascucci2010, GerberPotter2012}.
Persistent homology can be used to simplify a MSC~\cite{EdelsbrunnerHarerZomorodian2001}; see~\cite{GuntherReininghausSeidel2014} for a discussion. 

\para{Subsets of Morse-Smale Complexes.}
An \emph{extremum graph}, introduced by Correa \etal~\cite{CorreaLindstromBremer2011}, is a sparse subset of the MSC. 
It connects critical points along steepest ascending (or descending) lines, which join adjacent extrema~\cite{CorreaLindstromBremer2011}. 
It is designed to retain (some) important structural information of a MSC without visual clutter from the entire complex. 
A maximum graph contains maximum-saddle connections, whereas a minimum graph contains minimum-saddle connections. 
Thomas and Natarajan~\cite{ThomasNatarajan2013} augmented the extremum graph with topological and geometric information to facilitate the efficient detection of geometric symmetry in the electron microscopy data.  

Feng \etal~\cite{FengHuangJu2013} introduced \emph{feature graphs} to represent non-rigidly deformed surfaces.  
A feature graph is derived from the MSC of the Auto Diffusion Function (ADF), a solution to the heat equation. 
Nodes in a feature graph are critical points of a persistence-simplified MSC, which are connected by integral lines. 
Thus, a feature graph is the 1-dimensional skeleton of a simplified MSC. 

\subsection{Topological Descriptors of Multivariate Functions}
\label{sec:multivariate-descriptors}

We briefly describe topological descriptors of multivariate functions, although they are not the focus of this paper. 
We specifically focus on these multivariate descriptors as many of them are the direct extensions of their univariate counterparts. 
Given a multivariate function $F = (f_1, f_2, \cdots, f_k): \Mspace \to \Rspace^k$ ($k \geq 2$), we have three types of descriptors, those based on (a) the gradient behaviors of components $f_i$ (Jacobi sets), (b) the contours of $F$  (Reeb spaces, multivariate mapper constructions), and (c) multi-parameter persistent homology. 

\para{Reeb spaces, multivariate mapper constructions, joint contour nets.} 
Reeb spaces~\cite{EdelsbrunnerHarerPatel2008} are high-dimensional analogs of Reeb graphs. 
Given a multivariate function $F: \Mspace \to \Rspace^k$, the \emph{Reeb space} is  the quotient space obtained by identifying equivalent points, that is, $\Mspace/{\sim}$, where $x \sim y$ if $F(x) = F(y) = t \in \Rspace^k$ and $x$ and $y$ belong to the same connected component of the pre-image of $t$. 

Following the mapper construction for a scalar field (\autoref{fig:mapper-graph}), the filter function $f$ may be generalized to be a multivariate function, that is, $F: \Mspace \to \Rspace^k$ ($k \geq 2$). For instance, when $k=2$, the corresponding cover elements of $F(\Mspace) \subset \Rspace^2$ become rectangles. 
We call this the multivariate mapper construction in this paper to differentiate it from its univariate (scalar field) version.

Since a mapper graph is considered as a discrete approximations of a Reeb graph, the mapper construction for multivariate data $F: \Xspace \to \Rspace^k$ is a discrete approximation of the Reeb space~\cite{MunchWang2016}.  
There are other variants of such approximations, noticeably the \emph{joint contour nets} (JCNs)~\cite{CarrDuke2014}. 
The JCN applies  quantizations to the cover elements by rounding the function values. 
The \emph{multi-scale mapper}~\cite{DeyMemoliWang2016} is a sequence of mapper constructions connected by linear maps by varying the granularity of the cover elements.
The \emph{multi-nerve mapper}~\cite{CarriereOudot2018} computes the multi-nerve~\cite{VerdiereGinotGoaoc2012} of a cover. 
For comparing time-varying and multi-fields (see~\autoref{sec:time-varying-fields}), Agarwal \etal~\cite{AgarwalRamarmurthiChattopadhyay2020} introduced a multi-resolution Reeb Space (MRS), which is approximated as a series of JCNs at various levels of discretization.
 
\para{Jacobi sets.} 
The relation between two Morse functions $f,g: \Mspace \to \Rspace$ can be studied in terms of their Jacobi set~\cite{EdelsbrunnerHarer2004}, $J(f,g)$.
The Jacobi set is the collection of points in $\Mspace$ where the gradients of the functions align, that is, for some $\lambda \in \Rspace$,
\[
J(f,g) = \{x\in \Mspace \mid \nabla {f(x)}+\lambda \nabla {g(x)}=0{\mbox{ or }}\lambda \nabla {f(x)}+\nabla {g(x)}=0\}.
\] 

The Jacobi set has been used to derive local and global comparison measures of multiple scalar functions~\cite{EdelsbrunnerHarerNatarajan2004}. 
Several techniques have been developed for its topological simplification~\cite{NagarajNatarajan2011,BhatiaWangNorgard2015}. 
A relevant concept is \emph{Pareto sets}~\cite{HuettenbergerGarth2015}.  

\para{Multi-parameter persistence} is an active area of research, where previous results surrounding the indecomposables of multi-parameter persistence modules have been largely theoretical (see~\cite{CarlssonZomorodian2009,Lesnick2012} for relevant  readings).  
Multi-parameter versions of barcodes and their variants are actively researched, see recent results on multi-parameter persistence landscapes~\cite{Vipond2020} and persistence images~\cite{CarriereBlumberg2020} respectively.
Noticeably, the software RIVET~\cite{rivet} computes barcodes from ``slices" from 2-dimensional persistence modules.

%% file: sec-comparative-measures.tex
\section{Comparative Measures for Topological Descriptors}
\label{sec:comparative-measures}

Comparing scalar fields using their topological descriptors is an important tool in the study of scientific data. 
Defining and computing these comparative measures give rise to interesting problems both in theory and in practice. 
In this section, we review various definitions of comparative measures for  topological descriptors before discussing their applications in visualization in~\autoref{sec:single-fields}, \autoref{sec:time-varying-fields}, and~\autoref{sec:ensembles}. 
We defer the discussion on their mathematical and computational properties to~\autoref{sec:properties}. 
We give formal definitions in the forms of equations for some of the well-known comparative measures. We give informal descriptions for their variants. 
We defer detailed discussions to later sections for comparative measures designed specifically for visualization tasks, which oftentimes are coupled with heuristics and/or data-dependent modifications.   

Before diving into the technical descriptions of these comparative measures, we would like to discuss the different origins and motivations behind these developments. 
For instance, comparative measures for persistence diagrams, such as the bottleneck and Wasserstein distances, are related to optimal transport~\cite{Villani2003}. Functional distortion distances for Reeb graphs are the continuous version and a constant factor approximation of the extended Gromov-Hausdorff distances, a classic tool from the study of metric spaces; while interleaving distances originate from the algebraic study of persistence modules. 
Kernels for persistence diagrams interface with kernel methods for machine learning. 
The persistence scale-space kernel takes inspirations from the scale-space theory in signal processing, while persistence Fisher kernel is derived from information theory. 
Each comparative measure enjoys a set of desirable properties (\autoref{sec:properties}) and is suited for a specific collection of analysis and visualization tasks (\autoref{sec:single-fields}, \autoref{sec:time-varying-fields}, and~\autoref{sec:ensembles}), which motivated its development in the first place.

In the following sections, $\D$ represents a persistence diagram and its variants (persistence landscape and persistence image), $\T$ represents a tree-based descriptor, $\G$ represents a graph-based descriptor, and $\M$ represent complex-based descriptors, including Morse and Morse-Smale complexes.  
We emphasize the function as labels when a comparative measure  explicitly encodes information from the function (\textit{e.g.,} $\T_f$, $\T_g$), and we use numeric labels (\textit{e.g.,} $\D_1$, $\D_2$) otherwise. 

\subsection{Comparing Persistence Diagrams and Their Variants}
\label{sec:compare-diagrams}

We review classic distances between persistence diagrams, namely, bottleneck and $p$-Wasserstein distances, as well as distances between their variants, such as $p$-landscape distances. 
We also include kernels defined on persistence diagrams that interface with machine learning. 

\para{Bottleneck and Wasserstein distances.}
To compare persistence diagrams, the bottleneck distance~\cite{Cohen-SteinerEdelsbrunnerHarer2007,EdelsbrunnerHarer2008} and the Wasserstein distance~\cite{Cohen-SteinerEdelsbrunnerHarer2010} are well established and widely used, for instance, in similarity estimation~\cite{HajijZhangLiu2020} and machine learning tasks~\cite{Bubenik2015,ZhaoWang2019}. 

\begin{definition}{\cite[Bottleneck distance]{EdelsbrunnerHarer2008}}
\label{eq:bottleneck}
Given two persistence diagrams $\D_1$, $\D_2$ and a bijection $\eta: \D_1 \to \D_2$, the \emph{bottleneck distance} between $\D_1$ and $\D_2$ is defined as
\begin{align}
    d_{\infty}(\D_1,\D_2) &= \adjustlimits \inf_{\eta: \D_1 \to \D_2} \sup_{x \in \D_1} || x - \eta(x) ||_\infty. 
\end{align} 
\end{definition}

\begin{definition}{\cite[$p$-Wasserstein distance]{Cohen-SteinerEdelsbrunnerHarer2010}}
\label{eq:wasserstein}
The \emph{$p$-Wasserstein distance} is defined as
\begin{equation}
    d_{p}(\D_1,\D_2)=\left[\adjustlimits \inf_{\eta: \D_1 \to \D_2} \sum_{x \in \D_1} || x - \eta(x) ||_\infty^p \right]^{\frac{1}{p}}
\label{equation:p-Wasserstein}   
\end{equation} 
\end{definition}
While~\autoref{equation:p-Wasserstein} is a typical notion in the literature, Turner~\etal~\cite{TurnerMileykoMukherjee2014} discuss a more general formulation by introducing a second parameter (i.e., $q$)  to~\autoref{equation:p-Wasserstein} that specifies the degree of the point-wise norm; that is, by replacing $L^\infty$ norm  in~\autoref{equation:p-Wasserstein} with a $L^q$ norm; where $q=2$ in~\cite{TurnerMileykoMukherjee2014}.

\para{Kernels for persistence diagrams.}
Since persistence diagrams do not have the structure of an inner product space (\ie~Hilbert space), various kernels have been introduced to interface persistence diagrams with kernel-based machine learning models such as kernel support vector machines (SVMs).  
An intuitive way to think about kernels for SVMs is that kernels are similarity functions for a pair of objects. 
A number of kernels exist for persistence diagrams, such as the persistence scale-space kernel~\cite{ReininghausHuberBauer2015}, the persistence weighted Gaussian kernel~\cite{KusanoFukumizuHiraoka2017}, the sliced Wasserstein kernel~\cite{CarriereCuturiOudot2017}, and the persistence Fisher kernel~\cite{LeYamada2018}, denoted as $K_S$, $K_G$, $K_W$ and $K_F$, respectively.  

Let $\D_1$ and $\D_2$ denote two $k$-dimensional persistence diagrams.  
The \emph{persistence scale-space kernel}~\cite{ReininghausHuberBauer2015} $K_S$ is defined as
\begin{align}
\label{eq:scale-space-kernel}
  K_S(\D_1, \D_2,\sigma) = \frac{1}{8\pi\sigma} \sum_{p \in \D_1, q \in \D_2} e^{\frac{\|p-q\|}{8\sigma}} - e^{\frac{\|p-\overline{q}\|}{8\sigma}},
\end{align}
where $\forall \ q = (b,d) \in \D_2$, we define $\overline{q} = (d,b)$, that is, $\overline{q}$ is a reflection of $q$ along the diagonal $\Delta$; $\sigma$ is bandwidth of the Gaussian kernel. 

The \emph{persistence weighted Gaussian kernel} (PWGK)~\cite{KusanoFukumizuHiraoka2017} $K_G$ is defined as
\begin{align}
\label{eq:weighted-gaussian-kernel}
  K_G(\D_1, \D_2,\sigma) = \sum_{p \in \D_1, q \in \D_2} w(p)w(q)e^{-\frac{\|p-q\|^2}{2\sigma^2}},
\end{align}
where $w(p)$ is the weight assigned to the point $p$.
Kusano \etal~\cite{KusanoFukumizuHiraoka2017} suggest $w(p) = \arctan(C(d-b)^t)$ as the weight for $p = (b,d)$, where $C$ is a positive constant for practical purposes, and $t$ is assumed to be greater than the dimension of the underlying space.

Given a unit vector $\theta$ in $\Rspace^2$, let $L(\theta) = \{\lambda\theta \ | \ \lambda \in \Rspace\}$ denote the line and $\pi(\theta,p)$ denote the orthogonal projection of point $p$ on the line $L(\theta)$. To compute the \emph{sliced Wasserstein kernel}~\cite{CarriereCuturiOudot2017}, we first augment persistence diagram $\D_1$ with the orthogonal projection of points in $\D_2$ onto the diagonal (denoted as $\D_1^\Delta$) and vice versa (denoted as $\D_2^\Delta$) to obtain two new sets $\D_1^*$ and $\D_2^*$. That is, $\D_1^* = \D_1 \cup \D^\Delta_2$ and $\D_2^* = \D_2 \cup \D^\Delta_1$. 
The \emph{sliced Wasserstein distance} between these two sets is approximated as
\begin{align}
\label{eq:sliced-wasserstein}
&  SW(\D_1^*, \D_2^*,M) = \frac{1}{\pi}\sum_{j=1}^{M} \|V(\D_1^*, \theta_j) - V(\D_2^*, \theta_j)\|_1,
\end{align}
where $M$ is the number of directions, $\theta_j = j\pi/M - \pi/2$ and $V(\D_1^*, \theta_j)$ is the vector of dot products $<p, \theta_j>$ of all points $p \in \D_1^*$. The \emph{sliced Wasserstein kernel} is then computed as
\begin{align}
\label{eq:sliced-wasserstein-kernel}
  K_W(\D_1^*, \D_2^*,M) = e^{\frac{-SW(\D_1^*, \D_2^*,M)}{2\sigma^2}},
\end{align}

Given an $k$-dimensional persistence diagram $\D$ and a bandwidth $\sigma > 0$, we can define a smooth, normalized measure
\begin{equation}
\begin{aligned}
\label{eq:rhoD}
    \rho_{\D} &= \left[\frac{1}{Z}\sum_{u\in \D}N(x;u,\sigma I)\right]_{x\in \Theta}
\end{aligned}
\end{equation}
over a given set $\Theta$, where $I$ is the identity matrix, $N$ is a Gaussian function, and $Z = \int_{\Theta} \sum_{u\in A}N(x;u,\sigma I)\textrm{d}x$. Note that if $\Theta$ is the entire Euclidean space $\Rspace^{2}$, then $\rho_{\D}$ is a probability distribution similar to the case of persistence images~\cite{AdamsEmersonKirby2017}. 
Given two $k$-dimensional persistence diagrams $\D_1$ and $\D_2$, we obtain two new sets $\D_1^*$ and $\D_2^*$ by augmenting $\D_1$ with the orthogonal projection of points of $\D_2$ on the diagonal and vice versa. For these two sets, the persistence Fisher kernel~\cite{LeYamada2018} is defined as
\begin{equation}
\begin{aligned}
\label{eq:persistence-fisher-kernel}
    K_{F}(\D_1, \D_2) &= e^{-t d_{F}(\D_1^*, \D_2^*)}
\end{aligned}
\end{equation}
where $t > 0$ is a scalar parameter and $d_{F}$ is the Fisher information metric defined as follows: 
\begin{equation}
\begin{aligned}
\label{eq:the-Fisher-information-metric-between-points}
    d_{F}(\D_1^*, \D_2^*) &= \arccos\left(\int\sqrt{\rho_{\D_1^*}(x) \ \rho_{\D_2^*}(x)}\textrm{d}x\right)
\end{aligned}
\end{equation}

\para{Comparing variants of persistence diagrams.} 
Both persistence landscapes and persistence images (as well as the persistence-based feature vectorizations~\cite{LiWangAscoli2017})  can be used in machine learning algorithms such as SVMs under a Euclidean metric ({\eg}, $L^2$ or $L^p$). 

\begin{definition}~\cite[$p$-landscape distance]{Bubenik2015}
\label{eq:landscape}
If $\lambda_1$ and $\lambda_2$ are the persistence landscapes corresponding to persistence diagrams $\D_1$ and $\D_2$, the \emph{$p$-landscape distance} is  
\begin{align}
\Lambda_{p}(\D_1, \D_2) = || \lambda_1 - \lambda_2||_p. 
\end{align}
\end{definition}

Rieck \etal~\cite{RieckSadloLeitte2020b} defined a family of distances for Betti curves (also called the \emph{persistence indicator functions}), as well as corresponding kernels in order to use Betti curve in machine learning algorithms. 
Zhao and Wang~\cite{ZhaoWang2019} introduced a weighted-kernel for persistence images~(WKPI), its induced distance, and a metric-learning framework to learn the weights (and kernel) from labeled data. 
The persistent homology transform~(PHT) introduced by Turner~\etal~\cite{TurnerMukherjeeBoyer2014} comes with a distance measure, referred to as the \emph{PHT distance}, which captures similarity between shapes in shape classification. 
The inter-level set persistence hierarchies~(ISPHs) ~\cite{RieckSadloLeitte2017a,RieckSadloLeitte2020b} are directed trees, whose similarity can be measured by the edit distance (see~\autoref{sec:compare-reeb}).

\subsection{Comparing Reeb Graphs and Their Variants}
\label{sec:compare-reeb}

A number of metrics have been proposed for Reeb graphs and their variants such as merge trees, including functional distortion distance~\cite{BauerGeWang2014, BauerMunchWang2015}, edit distance~\cite{BauerDiFabioLandi2016, BauerLandiMemoli2020, SridharamurthyMasoodKamakshidasan2020}, interleaving distance~\cite{ChazalCohen-SteinerGlisse2009, MorozovBeketayevWeber2013, SilvaMunchPatel2016, MunchStefanou2018}, distances based on branch decompositions and matching~\cite{BeketayevYeliussizovMorozov2014,SaikiaSeidelWeinkauf2014}, and metrics for phylogenetic trees~ \cite{CardonaMirRossello2013}. 

\para{Functional distortion distances.}
Inspired by the Gromov-Hausdorff~(GH) distance  for  measuring  metric  distortions, Bauer~\etal~\cite{BauerGeWang2014} introduced the \emph{function distortion distance} for Reeb graphs. 
Let $f$ and $g$ be two real-valued functions on topological spaces $\Xspace$ and $\Yspace$ (the technical requirements are tame functions), together with maps $\varphi: \Xspace \to \Yspace$ and $\psi: \Yspace \to \Xspace$. 
Let $\G_f$ and $\G_g$ be the two Reeb graphs. Define 
\begin{align}
   & C(\varphi, \psi)  = \{(x, \varphi(x)) \mid x \in \G_f\}  \cup \{(\psi(y), y) \mid y \in \G_g\}, \\
   & D(\varphi, \psi) = \sup_{(x,y),(x',y')
    \in C(\varphi, \psi)}\frac{1}{2}|d_f(x,x')-d_g(y,y')|.
\end{align}
$C(\varphi, \psi) $ captures the set of correspondences between $\G_f$ and $\G_g$ induced by maps $\varphi$ and $\psi$. 

\begin{definition}{\cite[Functional distortion distance]{BauerGeWang2014}}
\label{RG:FD}
The \emph{functional distortion distance} between two Reeb graphs, $d_{FD}(\G_f, \G_g)$ is defined to be
\begin{equation}
    d_{FD}(\G_f,\G_g)=\inf_{\varphi, \psi}\max\{D(\varphi, \psi), ||f-g\circ\varphi||_\infty, ||g-f\circ\psi||_\infty\}.
\label{eq:function-distortion}  
\end{equation}
\end{definition}  
Here, $\varphi$ and $\psi$ are all continuous maps between $\G_f$ and $ \G_g$.

\para{Edit distances.}
We begin with edit distances for trees, since contour trees and merge trees are inherently tree-based representations.  
Inspired by the edit distance from computational linguistics~\cite{RistadYianilos1998} that quantifies dissimilarities between strings, 
Zhang and Shasha~\cite{ZhangShasha1989} introduced edit distance for  ordered labeled trees by computing the minimum-cost of node operations (\ie~``relabel", ``delete", and ``insert") that transform one tree into another. 
Zhang \etal~\cite{ZhangStatmanShasha1992} extended edit distance to unordered labeled trees. 
Tree edit distances have been used in many applications~\cite{ZhangShasha1989,RameshRamakrishnan1992,KleinTirthapuraSharvit2000}, including comparing topological structures such as merge trees. 
Rieck \etal~\cite{RieckSadloLeitte2017a} proposed persistence hierarchies  to related points in persistence diagrams, where tree edit distance-based dissimilarity is used to compare these hierarchies. 
Sridharamurthy \etal~\cite{SridharamurthyMasoodKamakshidasan2020} extended constrained tree edit distance~\cite{Zhang1996} based on dynamic programming with suitable modifications applicable to merge trees and showed its implementation in a feature-driven analysis of scalar fields.

\begin{definition}~\cite[Edit distance between merge trees]{SridharamurthyMasoodKamakshidasan2020}
\label{eq:edit-distance-mt}
The \emph{edit distance} between merge trees $\T_1$ and $\T_2$ is defined as 
\begin{equation}
 d_E(\T_1,\T_2) = \min_S \{\gamma(S)\},   
\end{equation}
where $S$ is a tree edit operation sequence from $\T_1$ to $\T_2$ that include edit operations such as ``relabel", ``delete", and ``insert"; and $\gamma$ is a cost function that assigns a non-negative real number to each operation. 
\end{definition}

Recently, Lohfink \etal~\cite{LohfinkWetzelsLukasczyk2020} adapted the graph-theoretic notion of the \emph{tree alignment}, which is similar to the edit distance mapping and is used to jointly visualize the contour trees from members of an ensemble. 

Bauer \etal~\cite{BauerDiFabioLandi2016, BauerLandiMemoli2020} introduced an \emph{edit distance} between labeled graphs, and applied it to Reeb graphs. 

\begin{definition}{\cite[Edit distance between labeled Reeb graphs, Definition 3.8]{BauerDiFabioLandi2016}}
\label{eq:edit-distance-labeled-reeb}
The \emph{edit distance between labeled Reeb graphs} $(\G_f, l_f)$ and $(\G_g, l_g)$ is defined as 
\begin{equation}
 d_{EG}((\G_f, l_f),(\G_2, l_g)) = \inf_S \{\gamma(S)\},   
\end{equation}
where $S$ varies in a set of arbitrarily long sequences of edit operations necessary to transform $(\G_f, l_f)$ into $(\G_g, l_g)$, and $\gamma(S)$ is the cost of an edit sequence. 
\end{definition}

\para{Interleaving distances.} 
Algebraically, the interleaving distance arises from $\epsilon$-interleavings of persistence modules; see~\cite{ChazalCohen-SteinerGlisse2009} for technical details. 
For topological descriptors, Morozov \etal~\cite{MorozovBeketayevWeber2013} defined an analog as the  interleaving distance between merge trees. 
Inspired by $L^\infty$-cophenetic metric introduced by Cardona \etal~\cite{CardonaMirRossello2013}, Munch \etal~\cite{MunchStefanou2018} introduced an interleaving distance between labeled merge trees, which is the $L^\infty$-distance between their induced matrices. 
Gasparovic {\etal} further studied interleaving distance intrinsic properties for the space of labeled and unlabeled merge trees, and used it to construct metric 1-centers for collections of labeled merge trees~\cite{GasparovicMunchOudot2019,YanWangMunch2020}. 
We describe the interleaving distance between merge trees as defined in~\cite{GasparovicMunchOudot2019}, which was shown to be equivalent to the original in~\cite{TouliWang2019}. 

Given two merge trees $\T_f$ and $\T_g$ that arise from functions $f: \Xspace \to \Rspace$ and $g: \Yspace \to \Rspace$, a \emph{$\delta$-good map} $\alpha: (\T_f, f) \to (\T_g, g)$ is a continuous map on the metric trees such that the following properties hold:
\begin{itemize}
    \item[I.] $\forall x \in |\T_f|$,  $g(\alpha(x))-f(x)=\delta$, where $|\T_f|$ denote the support (\ie~underlying space) of the tree; 
    \item [II.] $\forall w \in$ Im$(\alpha)$ with $x':= LCA(\alpha^{-1}(w)), f(x')-f(u) \leq 2\delta $ for all $u \in \alpha^{-1}(w)$;
    \item [III.] $\forall w \notin$ Im$(\alpha)$, depth$(w) \leq 2\delta$. 
\end{itemize}
$LCA(v, w)$ denotes the \emph{lowest common ancestor} of $v$ and $w$ in a tree. Intuitively, the $\delta$-goodness means that the points that are mapped from $\T_f$ to $\T_g$ via $\alpha$ do not change their function values much.

\begin{definition}{\cite[Interleaving distance between merge trees]{GasparovicMunchOudot2019}}
\label{eq:interleaving}
The \emph{interleaving distance} between merge trees $\T_f$ and $\T_g$ is defined as 
\begin{equation}
    d_{I}(\T_f, \T_g) = \inf\{\delta \mid \exists \delta\mbox{-good }\alpha:\T_f \to \T_g\}
\end{equation}
\end{definition}

\begin{definition}{\cite[Interleaving distance between labeled merge trees]{MunchStefanou2018}}
\label{eq:labelled-interleaving} 
Given two labeled merge trees $(\T_f, \pi)$ and $(\T_g, \pi')$, where maps $\pi:[n]\to V(\T_f)$ and $\pi':[n]\to V(\T_g)$ assign labels $[n]:=[1,2,\dots,n]$ to the nodes of $\T_f$ and $\T_g$, their \emph{interleaving distance} is
\begin{align}
    d_{IL}((\T_f, \pi), (\T_g, \pi')) = ||M(\T_f, \pi)-M(\T_g, \pi')||_\infty.
\end{align}
We use $M(\T_f, \pi)$ to denote the \emph{induced matrix} of a labeled merge tree $(\T_f, \pi)$, which is the symmetric matrix $M \in R^{n\times n}$, and $M_{ij} = f(LCA(\pi(i),\pi(j)))$.
\end{definition}

Additionally, Silva \etal~\cite{SilvaMunchPatel2016} defined a sheaf-theoretic interleaving distance between a pair of Reeb graphs as the interleaving distance between their cosheaves, and proved that this distance is stable under perturbations of the input data; see~\cite{SilvaMunchPatel2016} for technical details. 

\para{Distances based on branch decompositions or subtrees.}
Beketayev~\etal~\cite{BeketayevYeliussizovMorozov2014} defined a distance $d_{BR}$ between merge trees based on branch decompositions. 
They considered all branch decompositions of merge trees and found a minimum cost matching between them. 
\begin{definition}{\cite[Distance between merge trees based on branch decomposition]{BeketayevYeliussizovMorozov2014}}
\label{eq:branch}
Given two merge trees $\T_f$ and $\T_g$, and all of their possible branch decompositions $B_{\T_f}=\{R_1^f,\ldots,R_k^f\}$ and $B_{\T_g}=\{R_1^g,\ldots,R_k^g\}$, the distance between $\T_f$ and $\T_g$ can be defined as 
\begin{equation}
    d_{BR}(\T_f,\T_g)=\min_{R_i^f\in B_{\T_f}, R_j^g\in B_{\T_g}}( \epsilon_{\min}(R_i^f,R_j^g)),
\end{equation}
where $\epsilon_{\min}(R_i^f,R_j^g)$ is the lower bound for non-negative matching and removal costs of branch decompositions $R_i^f$ and $R_j^g$. 
\end{definition}

Saikia~\etal~\cite{SaikiaSeidelWeinkauf2014} defined a comparative measure for the \emph{extended branch decomposition graph}~(eBDG) based on minimizing the cost of matching between sequences of trees formed by branch decomposition of merge trees. 
In other words, they compared all subtrees of a merge tree.
Both the descriptor~(eBDG) and the comparative measure are computed by  dynamic programming. 
The authors also extended their work on eBDG to define a simple, histogram-based comparative measure for merge trees~\cite{SaikiaSeidelWeinkauf2015}. 
Instead of overlaying branch decomposition trees obtained from the subtrees, they described every subtree with a feature vector, referred to as a histogram~\cite{SaikiaSeidelWeinkauf2015}. 
To compare two histograms, they used the $L^2$-norm of the log-scaled bin values~\cite{SaikiaSeidelWeinkauf2015}. 
Subsequently, Saikia and Weinkauf~\cite{SaikiaWeinkauf2017} proposed a global similarity measure for feature tracking in time-varying fields. 
Their similarity measure is an extension of \cite{SaikiaSeidelWeinkauf2014} that involves a combination of spatial overlaps and histogram comparisons.

Thomas and Natarajan~\cite{ThomasNatarajan2011} defined a comparison measure based on constructing and comparing hierarchical descriptors of the subtrees of contour trees, and claimed that such a descriptor is stable in the presence of noise. 

\para{Comparative measures for variants of Reeb graphs.} 
A number of comparative measures are based on features or attributes derived from Reeb graphs and their variants.

Saggar~\etal~\cite{SaggarSpornsGonzalez-Castillo2018} used the mapper graph to study similarities among time-varying fMRI data. 
Each time frame of the fMRI data is interpreted as a point in a high-dimensional space; and two time frames are considered similar if they are connected in the mapper graph.
Hilaga~\etal~\cite{HilagaShinagawaKohmura2001} constructed a multi-resolutional Reeb graph~(MRG) based on geodesic distance, and designed a coarse-to-fine strategy to measure similarity between MRGs using the attributes of nodes in the MRGs. 
Biasotti~\etal~\cite{BiasottiMariniMortara2003} defined a similarity measure based on error tolerant graph isomorphism on extended Reeb graph~(ERG). 
Later, Barra and Biasotti~\cite{BarraBiasotti2013} developed a similarity measure for ERGs by applying a Gaussian kernel to vertex and edge attributes. 
Wu and Zhang~\cite{WuZhang2013} attached measures of similarities to contour tree branches for comparative analysis; such a measure is quantified based on contour overlaps. 

\para{Graph-based or tree-based comparative measures.} 
Finally, comparative measures developed in biology or graph theory may be applicable for topological descriptors. 
Cardona~\etal~\cite{CardonaMirRossello2013} defined a family of cophenetic metrics for comparing phylogenetic trees, which can be adopted as comparative measures for merge trees (\textit{e.g.,} ~\cite{MunchStefanou2018, GasparovicMunchOudot2019}). 
Tools developed for pairwise comparisons of graphs may be used for Reeb graphs and their variants; see surveys on graph distances~\cite{TantardiniIevaTajoli2019, WillsMeyer2020} and references therein. 
On the other hand, comparative measures for topological descriptors can be extended for general graphs as well. 
Dey~\etal~\cite{DeyShiWang2015} compared graphs via the persistence distortion distance. 
They compared a set of persistence diagrams constructed by defining scalar fields from various base points from the graphs. 
The sets are compared by Hausdorff distance, and individual persistence diagrams are compared by bottleneck distance.

\subsection{Comparing Morse and Morse-Smale Complexes}
A few papers have focused on comparative measures for Morse complexes and Morse-Smale complexes, most of which compare graphs derived from these complexes.  
This focus is not too surprising as a general form of stability for these complexes  appears to be elusive. 

\para{Comparing graphs derived from complexes.} 
Feng~\etal~\cite{FengHuangJu2013} studied the problem of computing feature correspondences between two non-rigidly deformed surfaces using feature graphs, which are 1D skeletons of simplified Morse-Smale complexes. 
Feature graphs are compared using a minimum-cost graph matching algorithm.
The authors observed (without proof) that such a feature graph is stable for surfaces differing by topology or by significant deformation. 

Thomas and Natarajan~\cite{ThomasNatarajan2013} focused on detecting symmetry in scalar fields using \emph{augmented extremum graphs}. They used the \emph{geodesic distances} between extrema, and the \emph{earth mover’s distance} between histograms of selected seed regions. 

In order to compare a pair of extremum graphs that may differ in the number of extrema and their adjunct relationships, Narayanan~\etal~\cite{NarayananThomasNatarajan2015} introduced the notion of a \emph{complete extremum graph}, which allows edges between all pairs of extrema in the graph.
They then defined a distance between extremum graphs based on computing the maximum distortion of the vertex sets and edge sets between the graphs. 
Specifically, Narayanan~{\etal} represented a complete extremum graph as an attributed graph. 
Each vertex $v \in V$ of the graph -- identical to a vertex from an extremum graph -- is assigned its persistence $p(v)$. 
Each edge $(u,v) \in E$ of the graph is assigned a cost $c(u,v)$ such that $c(u,v) \leq \min(p(u), p(v))$. 
The persistence of the global maximum is set to $1$. 
A scalar function is normalized to have a range of $[0,1]$ to ensure that $0 \leq p(v), c(u,v) \leq 1$.   
The distance between extrema graphs is defined based on these vertex and edge attributes~\cite{NarayananThomasNatarajan2015}. 
Let $\G_f=(V,E_V)$ denote a complete extremum graph of $f$ with vertex set $V$ and edge set $E_V$. 
Given two complete extremum graphs $\G_f=(V,E_V)$ and $\G_g=(U,E_U)$, a map $h: V \to U$ is called \emph{$\rho$-valid} for $\rho \in [0,1]$ if it is bijective and the edge distortion of corresponding edges is bounded by $\rho$. 
For a map $h$, let $D_{h}(\G_f,\G_g)$ denote the maximal distortion between the vertex set and edge set, that is, 
\begin{align}
     D_{h}(\G_f,\G_g) 
     &= \max_{v\in V} |p(v) - p(h(v)|  + \max_{(u,v)\in E} |c(u,v) - c(h(u,v))|. 
\end{align}

\begin{definition} \cite[Distance between extremum graphs]{NarayananThomasNatarajan2015}
For a fixed $\rho$, the distance $d_{\rho}$ between extremum graphs $\G_f$ and $\G_g$ is the minimum over all possible $\rho$-maps, 
\begin{align}
    d_{\rho}(\G_f,\G_g) & = \min_{h}\{D_{h} (\G_f, \G_g) \mid h \mbox{ is } \rho\mbox{-valid}\}.
\end{align}
\label{eq:extremum-graphs}
\end{definition}

%% file: sec-navigation.tex
\section{Navigating the State of the Art in Visualization}
\label{sec:navigation}

The comparative measures introduced in~\autoref{sec:comparative-measures}  have enabled a wide variety of visualization tasks. We categorize these tasks based on whether they are applied primarily to single scalar fields, time-varying scalar fields, or ensembles. The visualization tasks are described in~\autoref{sec:single-fields},~\autoref{sec:time-varying-fields}, and~\autoref{sec:ensembles}. 
The definitions of several comparative measures have been motivated for the most part by specific tasks associated with visualization or interactive exploration. 
We discuss these comparative measures with a focus on their roles in enabling the visualization tasks.~\autoref{table:navigate-descriptor-by-task} presents a guide for navigating the state-of-the-art described in these sections. 
We have also released the list of all references covered in this survey via SurVis~\cite{BeckKochWeiskopf2015}, the visual literature browser, which is available at \url{https://git.io/Jt2Hq}.

\begin{table*}[ht]
\begin{center}
\small
\resizebox{2.1\columnwidth}{!}{
\begin{tabular}{p{0.08\textwidth} p{0.08\textwidth} p{0.16\textwidth}p{0.08\textwidth}p{0.16\textwidth} p{0.08\textwidth}p{0.08\textwidth}p{0.08\textwidth}p{0.08\textwidth}p{0.08\textwidth}p{0.08\textwidth}}
    \toprule
    & \multicolumn{3}{c}{\textbf{Single scalar field}} & \multicolumn{3}{c}{\textbf{Time-varying scalar fields}} & \multicolumn{4}{c}{\textbf{Scalar field ensemble}} \\
    \cmidrule(lr){2-4} \cmidrule(lr){5-7} \cmidrule(lr){8-11}
     {\center Topological structure} & \multicolumn{1}{p{0.08\textwidth}}{\center Symmetry detection} & \multicolumn{1}{p{0.14\textwidth}}{\center Shape retrieval } & \multicolumn{1}{p{0.08\textwidth}}{\center Other tasks } & \multicolumn{1}{p{0.16\textwidth}}{\center Feature tracking } & \multicolumn{1}{p{0.08\textwidth}}{\center Global structure changes } & \multicolumn{1}{p{0.08\textwidth}}{\center Space-time structures } & \multicolumn{1}{p{0.08\textwidth}}{\center Clustering and classification } & \multicolumn{1}{p{0.08\textwidth}}{\center Summarization } & \multicolumn{1}{p{0.08\textwidth}}{\center Uncertainty visualization } & \multicolumn{1}{p{0.08\textwidth}}{\center Interactive exploration}\\
    \cmidrule(lr){1-1}\cmidrule(lr){2-4} \cmidrule(lr){5-7} \cmidrule(lr){8-11}
    Critical points/ \newline contours
      & 
      \cite{SchneiderWiebelCarr2008}
      
      \cite{ThomasNatarajan2014}
      & 
      & 
      & 
      \cite{ShamirBajajSohn2002} \cite{WeinkaufTheiselGelder2010}
      
      \cite{KastenReininghausHotz2011} \cite{ReininghausKotavaGunther2011}
      
      \cite{KastenHotzNoack2012} \cite{KastenZoufahlHege2012}
      
      \cite{ReininghausKastenWeinkauf2012} \cite{DoraiswamyNatarajanNanjundiah2013}
      
      \cite{LukasczykWeberMaciejewski2017} \cite{SolerPlainchaultConche2018}
      
      \cite{ValsangkarMonteiroNarayanan2019} \cite{EngelkeMasoodBeren2020}
      
      \cite{NilssonEngelkeFriederici2020}
      & 
      & 
      & \cite{FavelierFarajSumma2018} 
      & \cite{FavelierFarajSumma2018}
      & \cite{MihaiWestermann2014}
      
      \cite{GuntherSalmonTierny2014}
      & 
      \\
    \cmidrule(lr){1-1}\cmidrule(lr){2-4} \cmidrule(lr){5-7} \cmidrule(lr){8-11}
    Persistence \newline diagram
      & 
      & \cite{TurnerMukherjeeBoyer2014} 
      \cite{LiWangAscoli2017}
      
      \cite{ZhaoWang2019} \cite{HajijZhangLiu2020}
      & 
      & 
      & \cite{RieckSadloLeitte2017a}
      
      \cite{SolerPlainchaultConche2018}
      
      \cite{SolerPetitfrereDarche2019}
      & 
      & \cite{KontakVidalTierny2019}
      
      \cite{VidalBudinTierny2020}
      & 
      & 
      & 
      \\
    \cmidrule(lr){1-1}\cmidrule(lr){2-4} \cmidrule(lr){5-7} \cmidrule(lr){8-11}
    Merge tree
      & \cite{SaikiaSeidelWeinkauf2014}
      
      \cite{SaikiaSeidelWeinkauf2015}
      
      \cite{SridharamurthyMasoodKamakshidasan2020} 
      & \cite{SridharamurthyMasoodKamakshidasan2020}
      & \cite{BeketayevYeliussizovMorozov2014}
      & 
      \cite{SaikiaWeinkauf2017}
      
      & \cite{SaikiaSeidelWeinkauf2014}
      
      \cite{SridharamurthyMasoodKamakshidasan2020}
      
      \cite{LiPalandeWang2021}
      & 
      &  
      & \cite{YanWangMunch2020}
      & \cite{GuntherSalmonTierny2014}
      
      \cite{YanWangMunch2020}
      & \cite{PocoDoraiswamyTalbert2015} 
      
      \cite{YanWangMunch2020}
      \\
    \cmidrule(lr){1-1}\cmidrule(lr){2-4} \cmidrule(lr){5-7} \cmidrule(lr){8-11}
    Contour tree/ \newline Reeb graph
      & \cite{ThomasNatarajan2011}
      & \cite{HilagaShinagawaKohmura2001}  \cite{BiasottiMariniMortara2003}
      
      \cite{ZhangBajajBaker2004} \cite{BarraBiasotti2013}
      & 
      & \cite{EdelsbrunnerHarerMascarenhas2004}\cite{SohnBajaj2006}
      & 
      & \cite{BremerWeberPascucci2010}
      
      \cite{WeberBremerDay2011}
      & \cite{HilagaShinagawaKohmura2001}
      
      \cite{ZhangBajajBaker2004}
      & \cite{LohfinkWetzelsLukasczyk2020}
      & \cite{WuZhang2013}
      & 
      \\
    \cmidrule(lr){1-1}\cmidrule(lr){2-4} \cmidrule(lr){5-7} \cmidrule(lr){8-11}
      Morse-smale complex/ \newline Extremum graph
      & \cite{ThomasNatarajan2013}
      & \cite{FengHuangJu2013}
      & 
      & \cite{KastenReininghausHotz2011} \cite{ReininghausKastenWeinkauf2012}
      
      \cite{NarayananThomasNatarajan2015} \cite{KuhnEngelkeFlatken2017}
      
      \cite{SchnorrHelmrichChilds2019} \cite{SchnorrHelmrichDenker2020} 
      
      \cite{RieckSadloLeitte2020}
      & 
      \cite{NarayananThomasNatarajan2015}
      & 
      & 
      & 
      & 
      \cite{AthawaleMaljovecYan2020} 
      & 
      \\
    \cmidrule(lr){1-1}\cmidrule(lr){2-4} \cmidrule(lr){5-7} \cmidrule(lr){8-11}
    Other methods
      & 
      & \cite{AlliliCorriveau2007} \cite{DeyShiWang2015}
      & \cite{HuettenbergerHeineCarr2013}
      
      \cite{RieckLeitte2016}
      & \cite{AgarwalRamarmurthiChattopadhyay2020}
      & \cite{EdelsbrunnerHarer2004}
      
      \cite{EdelsbrunnerHarerNatarajan2004}
      
      \cite{NagarajNatarajanNajundiah2011}
      
      \cite{SaggarSpornsGonzalez-Castillo2018}
      & \cite{SaggarSpornsGonzalez-Castillo2018}
      & 
      & 
      & 
      & 
      \\
    \bottomrule
\end{tabular}
}
\end{center}
\caption{Navigating surveyed papers based on topological descriptors vs. visualization tasks.}
\label{table:navigate-descriptor-by-task}
\end{table*}

%% file: sec-classify-single-fields.tex
\section{Visualization Tasks for Single Fields}
\label{sec:single-fields}
We begin by considering comparative measures between single scalar fields. 
The comparison of single fields has many applications in scientific visualization, and in many cases serves as a building block in comparing time-varying scalar fields and ensembles. 
It plays an important role in tasks such as symmetry detection and shape matching. 
The former finds applications in the visual analysis of biomolecules, and the latter is essential for comparative visualization and computer vision. 
In the context of single fields, comparative measures may support the comparison between two different fields or between sub-structures within a single field ({\ie}, self-comparison). 

An important application of self-comparison is symmetry detection, discussed in~\autoref{sec:symmetry-detection}. 
Applications of comparisons between two fields include shape matching and retrieval, followed by matching shapes that are not represented by meshes, such as neuronal trees, see~\autoref{sec:shape-matching}.
We also discuss other visualization applications with single field comparisons in~\autoref{sec:single-other-tasks}, such as parameter tuning for ray casting algorithms, graphs, and social networks.  
For comparative measures already summarized in~\autoref{sec:comparative-measures}, we focus on their applications in specific visualization tasks. 
For other comparative measures designed mainly for visualization, we introduce the measures on a high-level before discussing their associated tasks.

\subsection{Symmetry Detection}
\label{sec:symmetry-detection}
\begin{figure}[!ht]
    \centering
    \includegraphics[width=0.99\columnwidth]{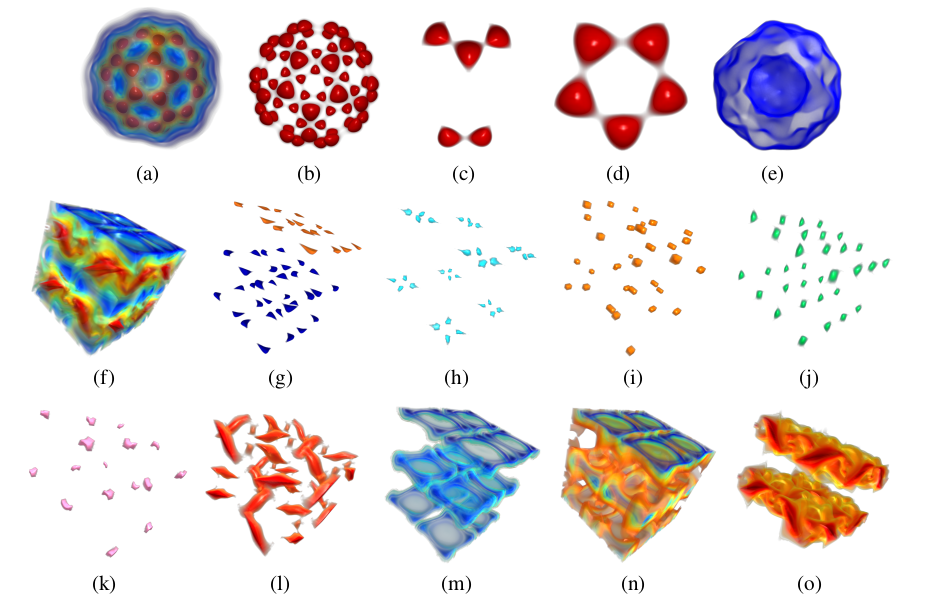}
    \caption{Symmetric regions identified within two datasets: Buckyball (top row) and vortex simulation (middle and bottom row). (a) and (f) are full volume renderings, whereas the rest correspond to symmetric regions. Image reproduced from Thomas and Natarajan~\cite{ThomasNatarajan2011}, cropped to show two datasets.} 
    \label{fig:symmetry}
\end{figure}

Symmetry detection refers to the identification of repeating structures within a single scalar field $f$. 
The repeats are identified based on a comparative measure, which is applied to compare a scalar field with itself. 
A typical pipeline would first construct a topological descriptor $\A$ from $f$, simplify 
$\A$ to remove noise, explicitly or implicitly enumerate sub-structures of $\A$, and compare pairs of these sub-structures. 
A refinement step may be incorporated to reduce the number of such comparisons, by selecting a specific sub-structure as a query or by applying spatial overlap criteria.  Symmetry detection is a first step in many visualization tasks such as query-based exploration, transfer function design, and linked volume editing, to name a few~\cite{ThomasNatarajan2014, MasoodThomasNatarajan2013}. 
Applications in molecular biology and allied fields have been demonstrated, where detecting repeating sub-structures in biomolecules is crucial to understand their shape and function. 
In~\autoref{fig:symmetry}, we see various symmetries identified within the Buckyball and vortex simulation datasets. 
In some cases, such as the vortex simulation data, the symmetries are not perceivable using a casual visual inspection.

\subsubsection{Merge Trees}
\label{sec:symmetry-detection-mt}
A rich set of techniques is available to detect symmetry by comparing merge trees and their variants, some of which are successful in discovering visually hidden symmetries. 
All papers described below showcase the utility of the method on \textsf{CryoEM} datasets, which consist of electron microscopy density maps of biomolecules. 
 
Saikia~\etal~\cite{SaikiaSeidelWeinkauf2014} computed the extended branch decomposition graph (eBDG) using a (simplified) merge tree $\T$ as input. 
An eBDG contains a union of all branch decomposition trees (sub-trees) of $\T$. 
To construct an eBDG, similarity scores are precomputed for all sub-trees against all other sub-trees using a combination of (normalized) volume and function differences. 
To detect self-similarity in a dataset, the eBDG is compared with itself.  
Since similarity scores are precomputed for all comparisons among sub-trees, given a region of interest, the method can report all similar sub-structures and hence support real-time exploration of the data. 
This approach also addresses the problem of seed selection~\cite{ThomasNatarajan2013} and supports slicing the 3D field and isolating symmetries that are not visually evident in 3D.

In a follow-up work, Saikia~\etal~\cite{SaikiaSeidelWeinkauf2015} provided an alternate method for symmetry detection. Each sub-tree is augmented with a feature vector, namely the histogram given by intensity distribution among  voxels of the sub-tree. 
This augmentation is computed together with the merge tree, and the resulting histograms (bin size $100$) are compared using an $L^2$ norm. 
The similarity scores are computed by comparing all pairs of sub-trees and a distance matrix is then constructed. A user can pick a voxel and then select the corresponding feature as the region of interest. Entries from the corresponding row in the distance matrix are picked, with an option to vary the distance threshold to refine the matches based on how close they are to the query.

Sridharamurthy~\etal~\cite{SridharamurthyMasoodKamakshidasan2020} used tree edit distance~(\cref{eq:edit-distance-mt}) to detect symmetric structures. After computing the merge tree of the scalar field, a set of sub-trees is selected based on persistence rank, and the edit distance is calculated by comparing all pairs of sub-trees to construct a distance matrix.  The method has limited utility as it does not support query-based similarity search, but can be used to detect symmetric structures by explicitly extracting sub-trees and comparing them. However, the method achieves results similar to those of Thomas and Natarajan~\cite{ThomasNatarajan2011}, who used the entire contour tree.

All three methods suffer from instabilities. 
Saikia~\etal~\cite{SaikiaSeidelWeinkauf2014} provided examples for false negatives and performed a perturbation analysis. 
The authors suggested a combination of volume and function differences as edge weights to alleviate the instability issue. 
Although histogram-based approaches~\cite{SaikiaSeidelWeinkauf2015} are robust for small perturbations, they cannot be a substitute for complicated branching. 
A case for which two trees with identical histograms was provided in the discussion. The number of bins used for the histogram can affect the final results. 
The authors suggested the possible use of non-linear binning in future work. 
Sridharamurthy~\etal~\cite{SridharamurthyMasoodKamakshidasan2020} achieved stability by merging saddle points into a multi-saddle based on an approach proposed earlier~\cite{ThomasNatarajan2011}. 
The merging is directed by a stabilization threshold that determines which critical points are merged. This approach works in practice, but with no theoretical guarantees.

\subsubsection{Contour Trees and Reeb graphs}
Thomas and Natarajan~\cite{ThomasNatarajan2011} defined a similarity measure for symmetry detection based on constructing and comparing hierarchical descriptors constructed from sub-trees of contour trees. 
The comparison is based on the max-weight matching of the hierarchical descriptors. The algorithm computes groups of symmetric regions and refines them in a post-processing step. 
The measure can be used to detect many kinds of symmetries that are not apparent via visual inspection. A stabilization parameter is used to handle instabilities. The method does not consider geometric information and thus cannot capture symmetries based on geometry. 
The method is used in applications such as symmetry-aware transfer function design and isosurface extraction.

In a subsequent work, Thomas \etal~\cite{ThomasNatarajan2014} exclusively used geometric information to extract symmetric structures in multiple scales. 
Schneider~\etal~\cite{SchneiderWiebelCarr2008} presented a similarity browser for comparing scalar fields, where similarity is defined as the relative overlap of the largest contours, and relevant contours are extracted by querying edges of a contour tree.

\subsubsection{Extremum Graphs and Morse Complexes}
Thomas and Natarajan~\cite{ThomasNatarajan2013} augmented a simplified extremum graph with edges that directly connect saddles, providing a good approximation of geodesic distance between pairs of extrema.  
Seed points are chosen from which the symmetric regions are grown using geodesic distance between extrema. Iteratively, the seeds are combined to form super-seeds, and symmetric regions are extracted via a region growing process. 
The method avoids computing matches between sub-trees or sub-structures and instead relies on geodesic distance; thus, it is robust in handling noise when compared to~\cite{ThomasNatarajan2011}. 
The method is used in applications such as proximity-aware volume visualization, linked volume editing and multi-mode volume rendering. Seed selection is a critical step, and symmetry detection depends on selection of a meaningful set of seeds. Seed selection and simplification depend on user-defined thresholds.

\subsection{Shape Matching and Retrieval}
\label{sec:shape-matching}

\begin{figure}[!ht]
    \centering
    \includegraphics[width=0.98\columnwidth]{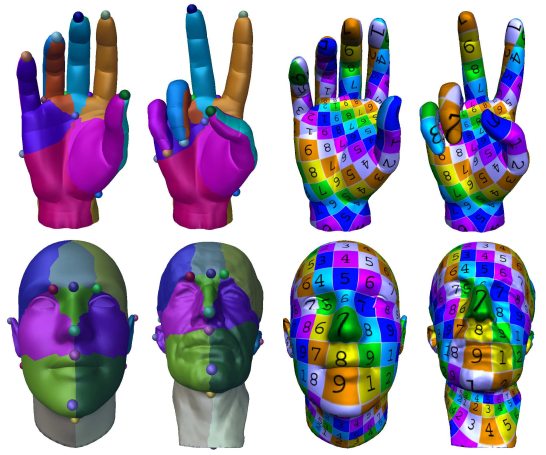}
    \caption{Shape-matching application: Point-and-patch correspondences followed by the texture transfer results of the hand and head data in different poses are shown. Image reproduced from Feng~\etal~\cite{FengHuangJu2013} with different alignment.} 
    \label{fig:match}
\end{figure}

Shape matching and retrieval is another important problem studied within the fields of computer graphics, computer vision, and visualization that can be addressed by comparative analysis of topological descriptors. 
Shape matching deals with comparing 3D shapes that are stored in the form of meshes. The solutions should be able to detect similarity/dissimilarity between shapes irrespective of view direction, scale, orientation, pose variation, and other transformations. The models often contain multiple attributes in addition to geometry and topology. Quantifying similarity is necessary to explore large databases of shapes. 
Biasotti~\etal~\cite{BiasottiCerriBronstein2014} surveyed existing methods from the perspective of maps between spaces. 
We restrict the discussion here to methods for which the comparative measure is based on topological descriptors. 
Although recent advances in learning-based methods do provide better results, topological descriptors provide scope for improvement as additional feature vectors.
Shape matching applications depend crucially on the choice of the scalar field/Morse function defined on the shapes. 
For example, the average geodesic distance from a randomly chosen set of vertices of the mesh is often used in the literature. Once the scalar field is computed, the rest of the pipeline is similar to symmetry detection. For each shape, a suitable topological descriptor is constructed, and the descriptors are compared to get a similarity score. The scores are used in a query-based system to match and retrieve shapes. \autoref{fig:match} shows one application of shape matching where correspondences of various parts of the shape mesh are found.

\subsubsection{Critical Points and Persistence diagrams}
\label{sec:shape-matching:Diagrams}
To study mesh similarity, Hajij~\etal~\cite{HajijZhangLiu2020} used persistence diagrams computed on the lower star filtration of the eigenfunctions of the Laplacian that store important geometric information. 
They alleviated the need for higher-dimensional persistence diagrams and showed that using a single eigenfunction itself has more discriminatory power than metric-based approaches. They used only the $0$-dimensional persistence diagrams, which are easy to compute. They compared the diagrams using bottleneck distance followed by 2D t-SNE projection of the distance matrix. They showcased their results on 60 meshes divided into 6 categories, available from Sumner and Popovi{\'c}~\cite{SumnerPopovic2004,SumnerPopovic2021}.
The results were shown for Fielder's vectors. They planned to combine signatures from multiple eigenfunctions and extend their method from triangulated meshes to point clouds and graphs.

Li {\etal}~\cite{LiWangAscoli2017} compared neuronal tree shapes by vectorizing neuron structures based on topological persistence, unlike traditional shape matching where the shapes are meshes. They proved that such a persistence-based signature is more effective in capturing the global and local structure than simple statistical summaries. They also proved, using a certain descriptor function, that a persistence-based signature contains more information than the classical Sholl analysis. The persistence diagram of the trees is computed with the descriptor function as the scalar field. The points in the diagram are then converted to a 1D density function by first assigning appropriate weights followed by converting the set of 1D points using a kernel estimate. The density function is then vectorized, which can be compared using standard norms like $L^1$ or $L^2$. Li {\etal} performed experiments on neuronal trees, used geodesic descriptor as the scalar field, showcased various tasks such as comparison and clustering, and analyzed results using classification accuracy. The space of the neurons was visualized. 
Future work involves building a database of descriptors and experiments using multiple descriptor functions.

Zhao and Wang~\cite{ZhaoWang2019} further used the Weighted Persistence Image Kernel~(WKPI) to compare neuronal trees and classify them in a similar fashion as Li {\etal}~\cite{LiWangAscoli2017}. They contrasted the classification accuracy using existing learning approaches and provided additional results for other graph data. 

Persistence homology transform~(PHT) was introduced by Turner {\etal}~\cite{TurnerMukherjeeBoyer2014}  as a topological descriptor of surfaces in 3D or curves in 2D. A unit vector induces a height field on the surface, for which a set of $d$ persistence diagrams can be computed. PHT is then defined as a map from all possible direction vectors (points on a unit sphere) to space of persistence diagrams. The authors proved that this transform is injective, and thus a metric defined on the space of persistence diagrams can be used to define a metric on the set of shapes. In practice, Wasserstein distance is used for comparing individual persistence diagrams. The distance is approximated by sampling the unit sphere to get a finite number of directions. The technique is demonstrated by applying it for classification and clustering of a set of shapes from MPEG-7 shape silhouette database~\cite{Sikora2001}. Seven class of objects with $20$ examples each totaling $1400$ objects are chosen, and $0$-th PHT is computed using $64$ evenly spaced directions. Then, distance computation considers various rotations and takes the minimum. The objects are then projected into 2D or 3D using multi-dimensional scaling. Turner {\etal} reported that the classes are well separated.

\subsubsection{Merge Trees}
Sridharamurthy~\etal~\cite{SridharamurthyMasoodKamakshidasan2020} used tree edit distance~(\cref{eq:edit-distance-mt}) to showcase shape matching using TOSCA non-rigid world dataset~\cite{BronsteinBronsteinKimmel2021}.
The shapes are in different poses and consist of both humanoid and non-humanoid shapes. Average geodesic distance field is calculated on the meshes, followed by a persistence simplification using a threshold of $1\%$ of the scalar field range. A distance matrix (DM) is computed by comparing all pairs of shapes. Each collection appears within a block of the DM, comprising low distance values irrespective of the pose. Some similarity across the blocks is also observed for humanoid shapes.

\subsubsection{Contour Trees and Reeb Graphs}
\label{sec:shape-matching:CTReebgraph}
Hilaga \etal~\cite{HilagaShinagawaKohmura2001} defined multi-resolution Reeb graphs (MRG) and a similarity measure to compare them. An approximation of geodesic distance based on Dijkstra's algorithm on the edge lengths is used as the scalar function upon which MRG is built. The similarity is computed by matching attributes defined on the MRGs. The MRGs are constructed on the shapes using a continuous scalar function, and a coarse-to-fine strategy is used to compare the graphs and compute the similarity. The similarity is used to find the best matches for a given query shape. The experimental setup of Hilaga {\etal} used 230 models collected from three sources: Viewpoint Models, 3DCAFE, and Stanford University models. They chose an object as the key and reported similar objects as retrieved by the method. The method depends on the resolution and two other parameters, range ($\mu_n$) and weight ($w$), which 
are typically set to $0.5$. They reported running times and mentioned that incorporating geometric information, extension to handle morphing, and application to pose estimation as potential future work.

Inspired by the work of Hilaga \etal~\cite{HilagaShinagawaKohmura2001}, Zhang \etal~\cite{ZhangBajajBaker2004} presented an algorithm to match volumetric functions based on multi-resolution dual contour trees. The matching is again based on weighted sum of attributes, normalized volume, function range, and Betti numbers of bounded contours. Electrostatic potential and electron density distributions within biomolecules (PDB data repository) are represented as scalar fields and compared. The method is robust with respect to both rigid body transformations and small perturbations to the scalar field. The paper reported the results of applying the matching algorithm on 242 protein chains assembled from different families. A clustering extension helps distinguish between different protein families. The authors proposed the use of sophisticated shape attributes and combined both electrostatic potential and electron density to improve the classification accuracy. They also described the scope for improving the tree matching algorithm.

Biasotti \etal~\cite{BiasottiMariniMortara2003}  defined a similarity measure based on error-tolerant graph isomorphism on an extended Reeb graph (ERG). An ERG is defined with respect to Euclidean distance from a point or with respect to integral geodesic distance. The measure depends on the choice of the function used to construct ERG and is shown to be a metric.
Barra and Biasotti~\cite{BarraBiasotti2013} used a kernel-based comparison of ERGs for shape retrieval. Kernels are defined on both vertex and edge attributes and computed by comparing all paths stemming from the two graphs being compared. 
Both papers discussed various applications on shape matching and retrieval contest (\textsc{shrec}) datasets, with detailed analysis of precision and recall.

\subsubsection{Extremum Graphs and Morse Complexes}
Feng \etal~\cite{FengHuangJu2013} studied the problem of computing corresponding features between two non-rigidly deformed surfaces, which is a key component in any shape-matching application. They compared their approach with other feature correspondence detection methods that involve geodesic and diffusion distances on various shapes with different values of weighting parameter. They also showcased the utility of their method in matching surfaces with non-zero genus, and showed examples of cross-parameterization and texture transfer.

\subsubsection{Other Descriptors}
Allili and Corriveau~\cite{AlliliCorriveau2007} provided a method to compare shapes by comparing the Morse shape descriptors (MSDs), which are topological descriptors defined on smooth manifolds. The MSD for an $n$-dimensional manifold is defined as a set of $(n+1)$ matrices with appropriate discretization of the scalar function. The similarity is measured as the weighted sum of distances between collections of MSD associated with contours. The main limitation is that the MSD is dependent on the Morse function used. No tools are available to select appropriate Morse functions for a given class of shapes. The descriptor allows multi-scale analysis due to discretization. The method is applied to 2D shapes only. 
The authors discussed that there is room to improve precision and recall, and to develop theoretical foundations to facilitate the choice of appropriate Morse functions.

Dey~\etal~\cite{DeyShiWang2015} used persistence distortion distance to compare surface meshes of different geometric models, some of them being the same models but in different poses. They used the geodesic distance from base points as the scalar field. The input is a $\delta$-sparse subsample of the 1-skeleton of the meshes constructed by randomized decimation. They proved that the error in the estimation of the distance is at most $12\delta$. Although the distance can be non-zero between a graph with itself (this depends on the choice of base points), the distances are small, confirming the stability.

\subsection{Other Visualization Tasks for Single Fields}
\label{sec:single-other-tasks}

Beketayev \etal~\cite{BeketayevYeliussizovMorozov2014} used distances  between merge trees (\cref{eq:branch}) to analyze the tuning of a ray tracing algorithm on a multi-core system as studied by Bethel \etal~\cite{BethelHowison2012}. 
Focusing on three parameters, work block width, height, and concurrency level, the authors recorded the performance for two datasets with the same parameter space but slightly different algorithms based on a ray selection method. The similarity of the datasets implies no significant difference is caused by the choice of ray selection method, which they confirmed by the small distance between them.

Rieck and Leitte~\cite{RieckLeitte2016} proposed a measure for comparing different clusterings of multivariate data and for studying the clustering quality. 
Given a multivariate dataset, the method first computes the Vietoris-Rips complex  and defines a suitable shape descriptor (a scalar field) on the point cloud. For a given clustering, the global version of the proposed measure is defined as the ratio of the total persistence sum of all clusters to the total persistence of the scalar field. This ratio captures the loss of features due to the clustering. They defined a local variant as well to capture the quality of the cluster. Although the measure is not a true comparative measure, it does enable a comparative analysis. The local measure is visualized as an attribute overlaid on the cluster, and the global measure for all clusterings is visualized as a network where similar clusterings are located close to each other. 

Topological descriptors have also been used for comparative analysis and visualization of discrete structures such as graphs and social networks~\cite{RieckFugacciLukasczyk2017,RieckLeitte2016b}.

%% file: sec-classify-time-varying.tex
\section{Visualization Tasks for Time-Varying Fields}
\label{sec:time-varying-fields}

Time plays a fundamental role in many processes. Prominent examples are physical models that simulate phenomena such as clouds formations in climate or weather modeling~\cite{DoraiswamyNatarajanNanjundiah2013,KuhnEngelkeFlatken2017,EngelkeMasoodBeren2020}, vortex shedding in flow simulations~\cite{KastenZoufahlHege2012,RieckSadloLeitte2020}, the simulation of combustion and burning structures~\cite{BremerWeberPascucci2010,NagarajNatarajanNajundiah2011}, or molecular dynamics simulations~\cite{SohnBajaj2006}. An example from medicine is time-varying measurements of the brain activity~\cite{SaggarSpornsGonzalez-Castillo2018}.
In all these examples, efficient analysis of the resulting dynamic data plays an increasing role. Key visualization and analysis tasks are the identification and tracking of features to understand the evolution of structural properties, find periodicity, or detect explicit events. 
Comparative measures play a primary role in this process. To this end, topological data analysis has proven to be a fundamental tool, and a large number of related publications are available, which will be discussed in this section along with topological descriptors being used.

In our context, the underlying assumption is that \emph{time} is a continuous variable.
However, typically, time-dependent data is available as a set of temporal snapshots. 
Generally, methods analyzing such data can be categorized depending on the treatment of the temporal dimension~\cite{Post2003}. The first possible approach is to analyze the data per time-slice and then compare the results. One can thereby find methods that explicitly track local features by solving an explicit correspondence problem~(\autoref{sec:tempFeatureTracking}) and methods that consider a global distance between the topological structures in one time-slice as a whole~(\autoref{sec:tempGlobalCompare}).
The third group of methods defines features as entities in the space-time domain where no explicit tracking is necessary~(\autoref{sec:spaceTimeFeatures}). 
In this report, we make an explicit distinction between time-varying fields and ensembles of scalar fields, where we consider ensembles to be a collection of scalar fields that arise from different parameter settings. Ensembles that are collections of time-varying fields are mainly discussed in the current section.

\begin{figure}[!t]
    \centering
    \includegraphics[width=0.95\columnwidth]{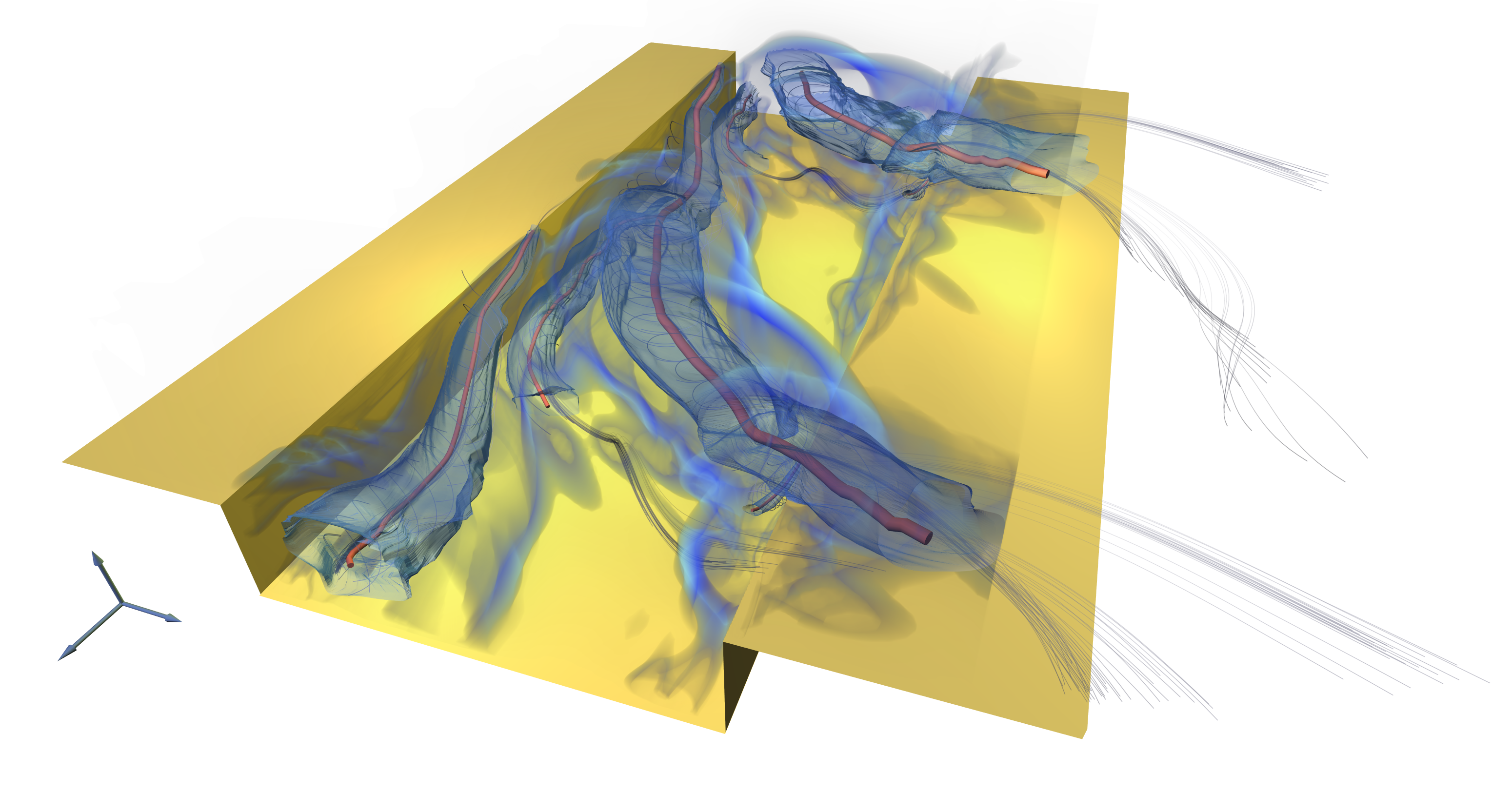}
    \caption{Time-dependent vortex core lines (red) and their associated vortex regions shown as transparent surfaces (blue) obtained from tracking minima and extremum structures in the acceleration magnitude filed of the 2D flow over a cavity. Volume rendering of the acceleration and a few path lines provide the context. Image reproduced from~\cite{KastenReininghausHotz2011}. } 
    \label{fig:cavity}
\end{figure}

\subsection{Feature Tracking and Event Detection}
\label{sec:tempFeatureTracking}

\begin{figure*}
    \centering
    \includegraphics[width=0.9\textwidth]{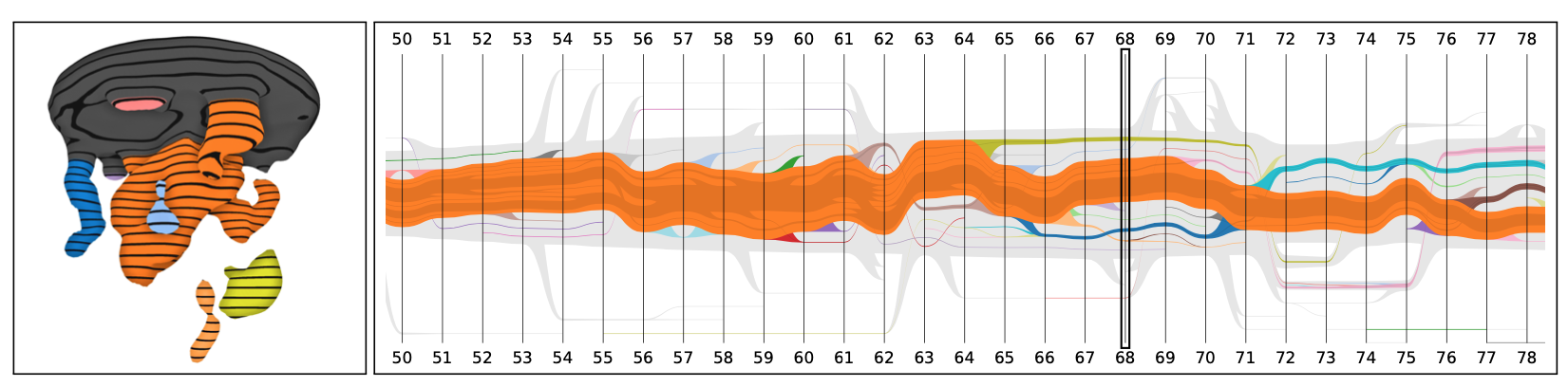}
    \caption{Nested tracking graph for one ensemble member of the viscous finger dataset. The graph illustrates the nesting hierarchy of contours across levels. The x-axis represents time and the y-axis is used to minimize edge crossings. The image shows the interface of the visual analytics framework consisting of a DVR window (left) and the interactive nested tracking graph (right). Image reproduced from~\cite{LukasczykWeberMaciejewski2017}.} 
    \label{fig:nestedTracking}
\end{figure*}

Tracking is mostly a two-step process in which topological features are extracted in each time slice and then matched solving a correspondence problem.
Therefore, feature tracking deals not only with the evolution of features but also with the identification of structural changes (events) between time steps, which are appearance (birth), disappearance (death), merging, or splitting of features. An optional simplification step using topological persistence is applied before computing the matching in the case of large data.

The most frequently used topological descriptors~(\autoref{sec:background}) are critical points~\cite{ReininghausKastenWeinkauf2012,SolerPlainchaultConche2018} or contours~\cite{ LukasczykWeberMaciejewski2017}, especially when it comes to real-world applications. 
However, some approaches have also been proposed for tracking the contour trees~\cite{SohnBajaj2006}, Reeb graphs~\cite{EdelsbrunnerHarerMascarenhas2004}, Morse cells~\cite{SchnorrHelmrichDenker2020}, or extremum graphs~\cite{KuhnEngelkeFlatken2017}.
Correspondence criteria use distance measures in the spatial domain or attribute space (\autoref{sec:comparative-measures}). 
These are often based on application-specific heuristics or feature overlap. Some methods also establish a correspondence building on an explicit temporal interpolation or optical flow~\cite{ValsangkarMonteiroNarayanan2019}. 
The respective papers are described in the following sections.

\begin{comment}
\begin{figure}[!b]
    \centering
    \includegraphics[width=0.9\columnwidth]{figs/abstract_merge_graph.png}
    \caption{Side by side comparison of abstract vortex merge graphs extracted using different scalar quantities: vorticity and the acceleration magnitude.~\cite{KastenZoufahlHege2012}} 
    \label{fig:vortex_merge_graph}
\end{figure}
\end{comment}

\subsubsection{Critical Point Tracking}
Weinkauf~\etal~\cite{WeinkaufTheiselGelder2010} described the dynamic behavior of critical points as streamlines of a higher-dimensional vector field in space-time, so-called \emph{feature flow fields}. 
The method relies on a continuous temporal interpolation of the field and applies numerical integration to generate the tracks. Originally, the method was applied to critical points in vector fields, but the concept is also applicable to critical points in scalar fields. 
Reininghaus~\etal~\cite{ReininghausKastenWeinkauf2012} tracked critical points in scalar fields to facilitate a discrete feature flow field. The feature tracks are gradient lines in the discrete field. 
Practically, feature tracking establishes a correspondence between extremal points in consecutive time steps using a forward and backward \emph{Morse matching}. This means an extremal point from time-step $t$ falls into the descending (resp. ascending) manifold of an extremal point in time-step $t+1$ and the other way round. 
The method explicitly keeps track of merge and split events. 
By tracking the persistence value over time, Reininghaus {\etal} introduced a temporal importance measure for the tracked features. The tracking itself is inherently local and fast. 
This method has been successfully applied to vortex tracking by Kasten~\etal~\cite{KastenHotzNoack2012,KastenZoufahlHege2012}, thus providing an abstract tracking graph.  
Reininghaus~\etal~\cite{ReininghausKotavaGunther2011} also used the approach to track critical points in the scale-space.
In an extension of this work, the descending manifolds of the tracked minima enclosed by a subset of the extremum structures are tracked and interpreted as vortex regions~\cite{KastenReininghausHotz2011} (\autoref{fig:cavity}). The correspondence of the extremum structure is inherited from the extremal points.
A similar approach was used by Engelke~\etal~\cite{EngelkeMasoodBeren2020,NilssonEngelkeFriederici2020} to track multi-centered cyclones defined as sets of critical points in a pressure field.

\begin{figure}[!b]
    \centering
    \includegraphics[width=0.95\columnwidth]{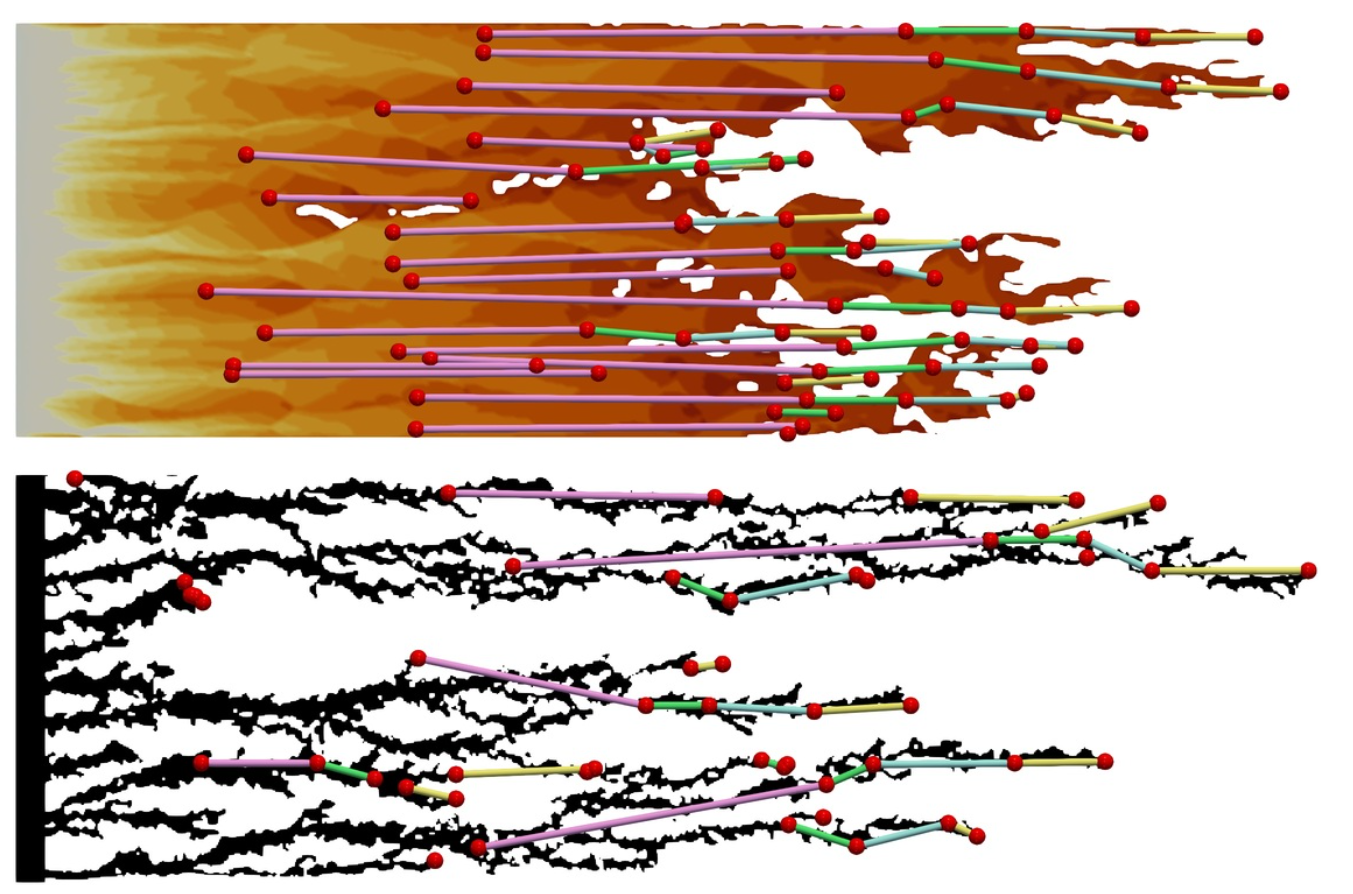}
    \caption{Critical point trajectories based on optimal matchings. The geometrical coherence of fingers allows to use a lifted version (that gives importance to the y-coordinate of fingers) of the Wasserstein metric to correctly track the evolution of persistence pairs. Image reproduced from~\cite{SolerPetitfrereDarche2019}.} 
    \label{fig:liftedWasserstein}
\end{figure}

Soler~\etal~\cite{SolerPlainchaultConche2018} introduced an extremal point tracking algorithm that solves the correspondence problem by minimizing the sum of the distances between matched pairs. The distance between two extrema is defined as a weighted sum of (i) the difference of the corresponding persistence pairs as used in the Wasserstein distance (\cref{eq:wasserstein}) and (ii) their geometric distance in the domain. The authors call the resulting metric {\em lifted Wasserstein metric}.
For an efficient computation of their tracking, they proposed a modification of the Kuhn-Munkres algorithm~\cite{Munkres1957}.
They demonstrated the usefulness of the approach on a few analytical datasets and examples from flow visualization (vortex tracking) (\autoref{fig:liftedWasserstein}).
As a future extension, they proposed to also include other attributes in the distance function assigned to the topological features. In a follow-up paper~\cite{SolerPetitfrereDarche2019}, the method was adapted to the analysis of the viscous finger by describing the dynamic process observed at the interface between fluids of different viscosity.

Valsangkar \etal~\cite{ValsangkarMonteiroNarayanan2019} used optical flow to study the temporal evolution of cyclones. Cyclonic centers are defined as local minima of the mean sea level pressure field. Candidate tracks are computed from an optical flow field, which then is clustered in a moving time window to obtain the final tracks. The tracks are visualized as a smooth curve interpolating the vertices of a track. The method supports the identification of merge and split events.

\subsubsection{Contour, Contour Trees and Merge Tree Tracking}
\label{sec:tempFeatureTracking:MTtracking}
Shamir~\etal~\cite{ShamirBajajSohn2002} proposed a progressive isosurface tracing algorithm that predicts the contour at time step $t+1$ based on the contour at time step $t$. 
In a later paper, Sohn and Bajaj~\cite{SohnBajaj2006} extended this approach by proposing a contour-tree based feature tracking method.
The method has some similarities with the methods introduced in \autoref{sec:compare-reeb}. Specific to this method, Sohn and Bajaj keep track of the contours (points on the contour tree) for all iso-values, which is summarized in a topology change graph~(TCG).
A temporal correspondence between contours is established in the case of \emph{significant} overlap over both their sublevel and superlevel sets.
To be able to track contours for iso-values, they facilitated the coherence of contours when changing iso-values. 
Additional ``critical points'' are added to the contour trees where two super level sets of the two trees ``collide'' and a new overlap region is created. The same process is done for sublevel sets. The significance value is a conceptual parameter in this approach. They demonstrated the utility of their method in tracking turbulent vortex structures and the binding of oxygen to hemoglobin molecule using time-dependent electron density maps.

Doraiswamy~\etal~\cite{DoraiswamyNatarajanNanjundiah2013} described a framework for the exploration of cloud systems at various scales in space and time based on infrared~(IR) brightness temperature images. 
The framework automatically extracts cloud clusters as contours for a given temperature threshold. To identify the threshold, the persistence diagram is used.
The movement of the cloud system is tracked using the optical flow between the pair of IR brightness temperature images. 

\begin{figure}[!b]
    \centering
    \includegraphics[width=0.98\columnwidth]{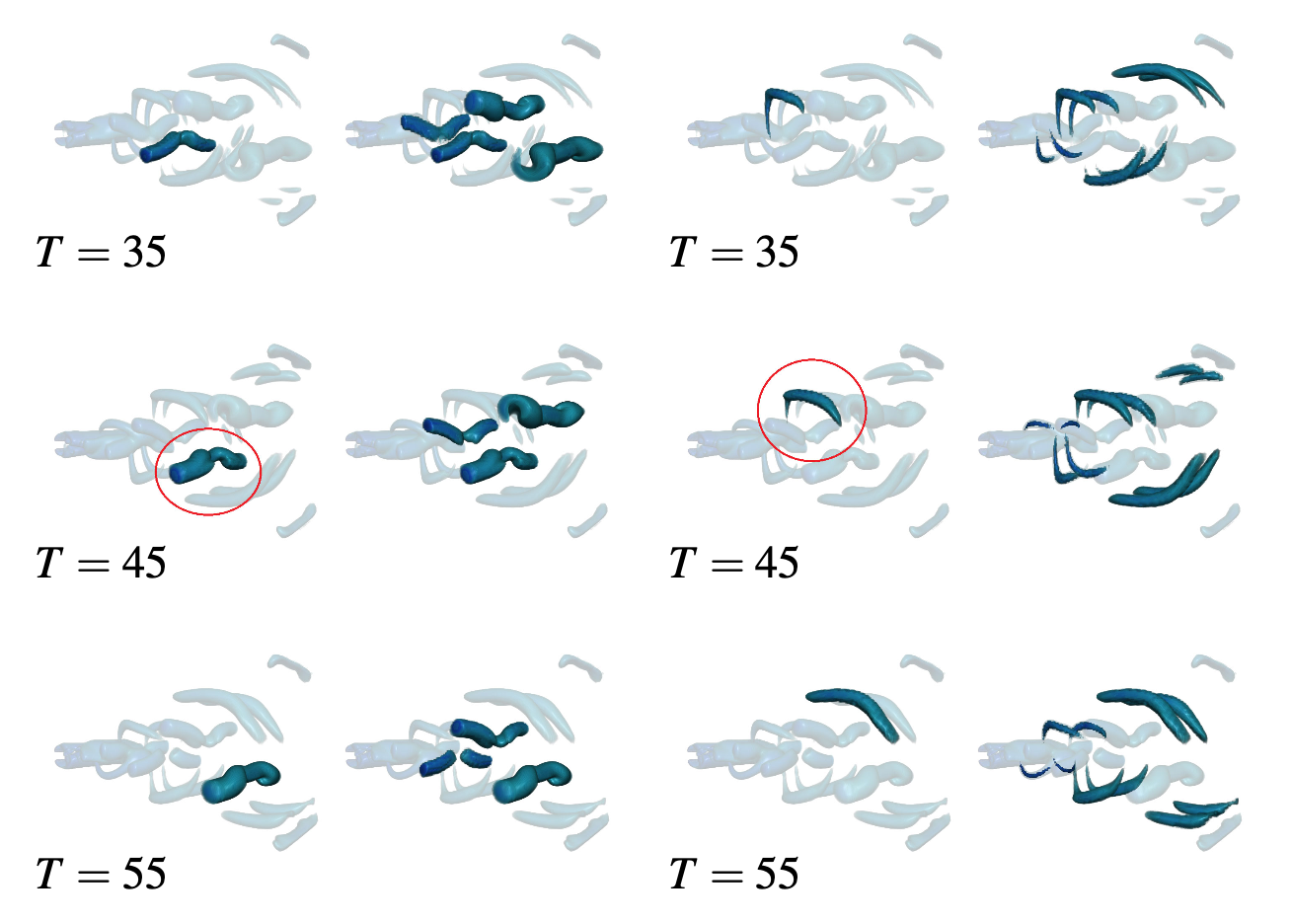}
    \caption{A primary and secondary vortex structure have been selected at $T = 45$ and tracked backwards and forwards in time. Their tracks have been used to find spatio-temporally similar structures in the entire dataset. Image reproduced from~\cite{SaikiaWeinkauf2017}.} 
    \label{fig:histogramTracking}
\end{figure}

Lukasczyk~\etal~\cite{LukasczykWeberMaciejewski2017} introduced nested tracking graphs as a hierarchical representation of feature tracks highlighting temporal events. Features are defined as contours and correspondence is established by contour overlap. They applied the approach to data from finite point set methods, computational fluid dynamics, and cosmology. The method has been integrated into a visual analytic approach~\cite{LukasczykAldrichSteptoe2017} for tracking of viscous fingers (\autoref{fig:nestedTracking}).

Saikia and Weinkauf~\cite{SaikiaWeinkauf2017} used subtrees of the merge tree for temporal feature tracking. 
Subtree similarity is based on a combination of spatial overlap and $\chi^2$-distance between local histograms. 
The result is a directed acyclic graph with nodes that represent features and the edges that record the similarity information. The final feature track is then computed as the shortest path in a graph, and thus considers not only the similarity between consecutive time steps but also global similarity. 
They applied their method to track primary and secondary vortex structures in the 3D time-dependent flow behind a cylinder (\autoref{fig:histogramTracking}) by considering the Okubo-Weiss criterion as the scalar field.

\begin{figure}[!ht]
    \centering
    \includegraphics[width=0.9\columnwidth]{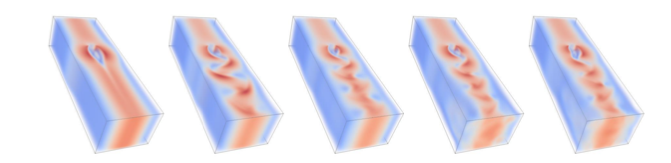}
    \includegraphics[width=0.6\columnwidth]{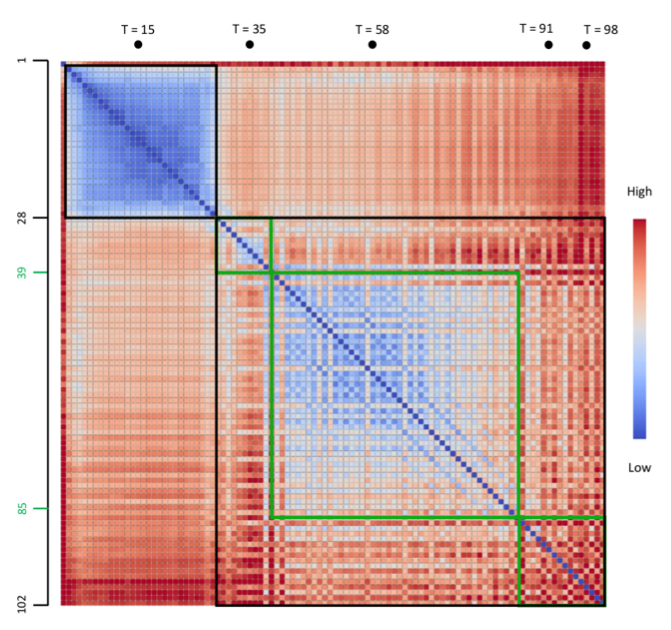}
    \caption{Tree edit distance matrix for all time steps of the 3D von K\'arm\'an vortex street. Columns corresponding to the selected time steps (images in the top row) are highlighted in the matrix. Interesting patterns are highlighted using black and green boxes. Image reproduced from~\cite{SridharamurthyMasoodKamakshidasan2020}.} 
    \label{fig:event}
\end{figure}

Time-varying Reeb graphs for the analysis of continuous space-time scalar data were proposed by Edelsbrunner~\etal~\cite{EdelsbrunnerHarerMascarenhas2004}. 
The nodes of the Reeb graphs for consecutive time steps, which correspond to the critical points of the scalar field, are connected using Jacobi curves in space-time~(\autoref{sec:background}).

\subsubsection{Extremum Graph and Morse Complex Tracking}
Schnorr~\etal~\cite{SchnorrHelmrichDenker2020} presented a two-step tracking approach for space-filling features (objects whose union covers the domain). 
The work was motivated by the analysis of dissipation elements, which are defined as Morse-Smale cells (\autoref{subsec:MS}). This problem is especially challenging since the number of overlapping cells in consecutive time steps can be very high. 
In a first step, the solution of a weighted (normalized volume overlap) bi-partite matching problem generates a 1:1 assignment of cells. This mapping is assumed to result in the most plausible matches. In a second step, additional edges are generated to allow for merge and split events by solving a maximum-weight independent-set problem.
Since the method is computationally expensive, the authors later published an algorithm that approximates the tracking achieving a substantial speed-up~\cite{SchnorrHelmrichChilds2019}.

Tracking of extremum structures~(\autoref{subsec:MS}) or skeletons is an unstable process since the resulting structures typically are sensitive to small variations in the data.
Kuhn~\etal~\cite{KuhnEngelkeFlatken2017} dealt with this challenge by proposing a space-time clustering approach for the extremum graphs from the individual time steps. The clusters are visualized as a distance field isosurface of the extremum graphs in space-time. 
The method has been developed in an application dealing with ash cloud tracking in a 2D domain.
Rieck~\etal~\cite{RieckSadloLeitte2020} analyzed the evolution of skeletons, which are extracted applying iterative thinning to binary images. To improve the temporal coherence, they introduce a novel persistence concept. They applied the method to the viscous fingers dataset.

Narayanan \etal~\cite{NarayananThomasNatarajan2015} used the distance $d_\rho$ (\autoref{eq:extremum-graphs}) to compare extremum graphs based on maximum common sub-graphs and to track turbulent vortex structures.

\subsection{Structural Change Detection}
\label{sec:tempGlobalCompare}
A global comparison of datasets derives a pair-wise distance measure between time steps, which often are displayed in a distance matrix or similarity plots over time~(\autoref{fig:event}). 
Typical visualization tasks are outlier, periodicity, or event detection. Periodicity detection refers to identifying structures that repeat over time. Event detection in this context refers to finding sudden changes in the overall structure of the data, which results in specific patterns in the distance matrix. 
Principally, all distance measures introduced in \autoref{sec:comparative-measures} can be used for this purpose. However, only a few of those have been used in visualization applications concerned with time-varying data.

Saikia \etal~\cite{SaikiaSeidelWeinkauf2014} used eBDG, which is computed from merge trees, and compared these trees using a dynamic programming algorithm to detect periodicity in the 2D Benard-von K\'arm\'an vortex street dataset. A similar application is showcased by Narayanan \etal~\cite{NarayananThomasNatarajan2015}.

Sridharamurthy~\etal~\cite{SridharamurthyMasoodKamakshidasan2020} utilized the tree edit distance in applications such as detecting periodicity (similar to ~\cite{SaikiaSeidelWeinkauf2014,NarayananThomasNatarajan2015}) and computing temporal summaries. 
They used a distance matrix to detect key events and summarized the Benard-von K\'arm\'an vortex street dataset as shown in \autoref{fig:event}.

Edelsbrunner~\etal~\cite{EdelsbrunnerHarerNatarajan2004} introduced a  global comparison measure for a set of $k \leq d$ scalar functions defined on a common $d$-dimensional manifold, and applied it to the study of time-varying functions. The comparison measure is defined as an integral of the norm of the wedge product of the $k$ derivatives. For the case of $k=d=2$, the measure is related to the Jacobi set of the two scalar functions (\autoref{sec:multivariate-descriptors}) and equals the integral of the persistence of all critical points of one function restricted to the level sets of the other.  The measure can be computed in linear time. It has been applied to combustion simulation data. A time plot of the comparison measure between the fuel and a variable representing the progress of the combustion helps identify the key stages of ignition, burning, and extinction. The paper also describes a local version of the measure obtained by restricting the computation to an infinitesimally small subdomain.  The measure is symmetric but does not satisfy the triangle inequality. The measure is also not scale-invariant and is therefore not suitable to compare the similarity of pairs of functions. An extension of the local measure to compare gradients of $k > d$ scalar functions was applied to climate science and combustion studies~\cite{NagarajNatarajanNajundiah2011}.

In a few applications, the comparison of persistence diagrams was also applied for the analysis of temporal data (\autoref{sec:compare-diagrams}).
Rieck~\etal~\cite{RieckSadloLeitte2017a} identified periodicity in time-varying temperature data from a climate application by comparing all pairs of time steps using inter-level set persistence hierarchies~(ISPHs) and visualizing the resulting distance matrix. 
In an application paper, Soler~\etal~\cite{SolerPetitfrereDarche2019} compared time-varying viscous finger datasets from ensemble simulation runs based on time-varying persistence diagrams. 
Hajij \etal~\cite{HajijWangScheidegger2018} employed the distances between persistence diagrams to visually detect structural changes in time-varying graphs. To quantify the structural difference between two time instances of a social network, they computed the bottleneck and Wasserstein distances between their persistence diagrams. The same framework is potentially applicable to the study of time-varying scalar fields.

Agarwal~\etal~\cite{AgarwalRamarmurthiChattopadhyay2020} proposed a similarity measure for the analysis of time-varying multi-fields. Each multi-field is represented as a Multi-resolution Reeb Space~(MRS) that is approximated as a series of joint contour nets~(JCNs) at various levels of data-range discretization. Between the nodes of adjacent resolutions, a parent-child relationship is introduced.
The computation of the similarity measure between two MRSs is a two-step process. 
In the first step, a list of matching pairs from the nodes of the respective MRSs is established, moving from the coarser to the finer resolution Reeb spaces. Two nodes are matched if they have the same resolution level, the parents have been matched, and they are topologically consistent.
In the second step, the similarity of the MRSs is computed as a sum of an attribute-based similarity between the nodes following ideas from Zhang~\etal~\cite{ZhangBajajBaker2004} and Hilaga~\etal~\cite{HilagaShinagawaKohmura2001} (\autoref{sec:compare-reeb}).
The visualizations show the abstract graphs and plots of the similarity measure over time. The paper focuses on the introduction of the method. Its use is demonstrated for time-varying multi-field data from computational physics. The similarity measure is not specific to the analysis of time-varying data. The correspondence principle could also be used for explicit feature tracking.

\begin{figure}[!hb]
    \centering
    \includegraphics[width=0.98\columnwidth]{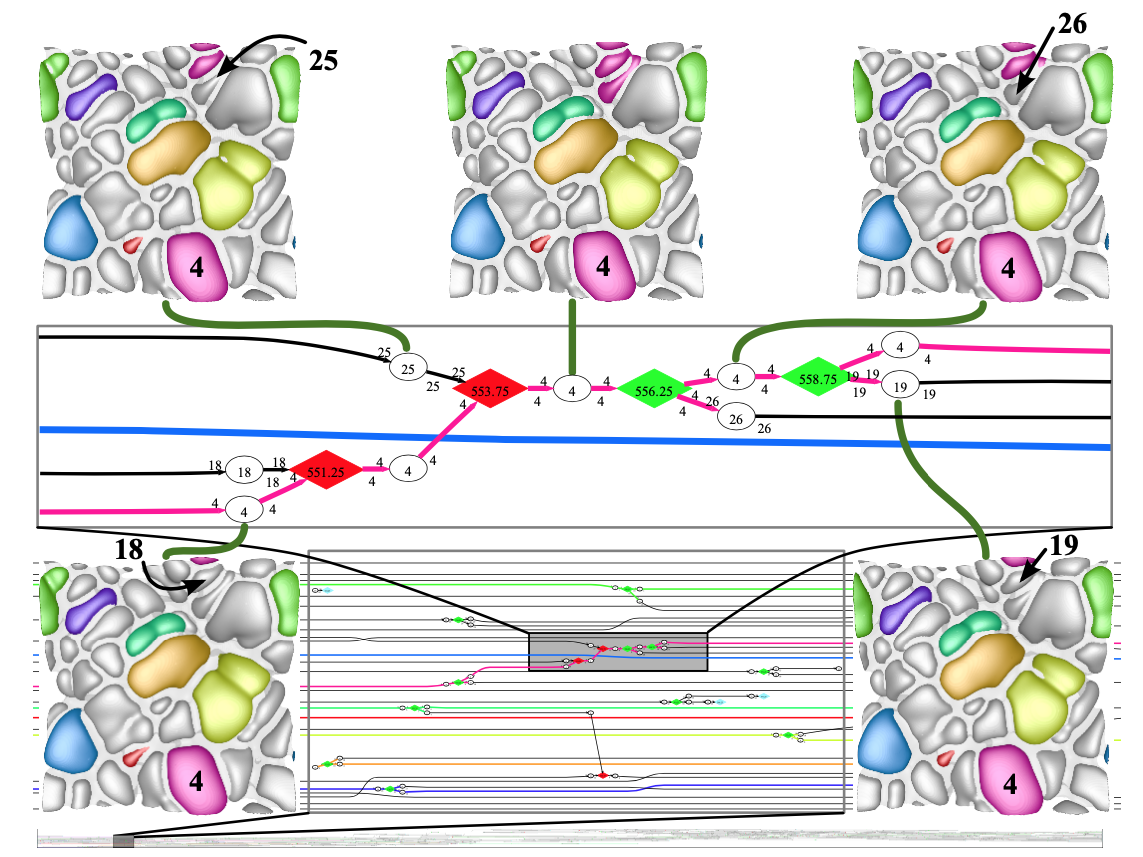}
    \caption{Reeb-graph based tracking graph of burning cells  with corresponding segmentations. Round nodes correspond to cells explicitly segmented by the Morse complex, diamonds to topological events between time steps. Red signifies a merge, green a split, and turquoise a birth/death event. Image reproduced from~\cite{BremerWeberPascucci2010}.} 
    \label{fig:burningCells}
\end{figure}

Given a set of merge trees $\Tspace = {\T_1, \dots, \T_k}$, Li \etal~\cite{LiPalandeWang2021} were interested in finding a basis set $\Sspace$ of merge trees such that each tree in $\Tspace$ can be approximately reconstructed from a linear combination of merge trees in $\Sspace$. 
A set of high-dimensional vectors can be sketched via matrix sketching techniques such as principal component analysis and column subset selection. 
Li {\etal} developed a framework for sketching a set of merge trees that combines the Gromov-Wasserstein framework~\cite{ChowdhuryNeedham2020} with techniques from matrix sketching. 
They demonstrated the applications of the framework in sketching merge trees that arise from time-varying data in scientific simulations ({\eg}, flow simulations, and the Red Sea eddy simulation from the IEEE Scientific Visualization Contest 2020), where their method is used to find good ensemble representatives and to identify outliers {\wrt} a chosen basis set. The same framework is potentially applicable for the ensembles of scalar fields discussed in~\autoref{sec:ensembles}.

\subsection{Space-Time Structures}
\label{sec:spaceTimeFeatures}
An alternative to feature tracking is to define topological structures directly in the space-time domain, in this case, the correspondence must not be established explicitly and no distance measures are required. 
Instead, one must assume an explicit temporal interpolation. 

Weber~\etal~\cite{WeberBremerDay2011} tracked subsets of isosurfaces using the Reeb graph. The work was motivated by the analysis of burning regions in simulated flames. Such regions are defined as parts of the temperature isosurfaces, where the fuel consumption rate is above a certain threshold. Boundary surfaces separating burning and non-burning regions are extracted in a two-phase contouring operation: a temperature iso-volume in 4D space-time is computed, from which the fuel-consumption-rate isosurface is constructed. Using time as a filter function, the Reeb graph of this boundary surface represents the evolution of burning regions. 
In a follow-up paper~\cite{BremerWeberPascucci2010}, this graph was embedded in an exploration framework and augmented with statistical attributes~(\autoref{fig:burningCells}).

Edelsbrunner~\etal~\cite{EdelsbrunnerHarer2004} proposed to compare individual time steps using Jacobi sets (see~\autoref{sec:background}). Although this approach provides a solid theoretical framework, the results are often very complex and hard to interpret when applied to real-world data. 

Saggar~\etal~\cite{SaggarSpornsGonzalez-Castillo2018} proposed a pipeline for the analysis of time-series of fMRI (functional magnetic resonance) data to explore the dynamical organization of the brain. Each timeframe is interpreted as a point in a high-dimensional space (number of pixels). After a dimension-reduction step using t-SNE~\cite{MaatenHinton2008}, mapper construction~(See \autoref{sec:background}) is applied to build a shape graph, where the nodes represent sets of timeframes. The results are visualized as an abstract graph augmented with aggregated node attributes linked to special visualizations. A distance between time frames can be defined via distances in the shape graph.
The authors stated that future work is needed to reduce the computational costs of the approach.

%% file: sec-classify-ensembles.tex
\section{Visualization Tasks for Ensembles}
\label{sec:ensembles}

\begin{figure}[!b]
    \centering
    \includegraphics[width=0.98\columnwidth]{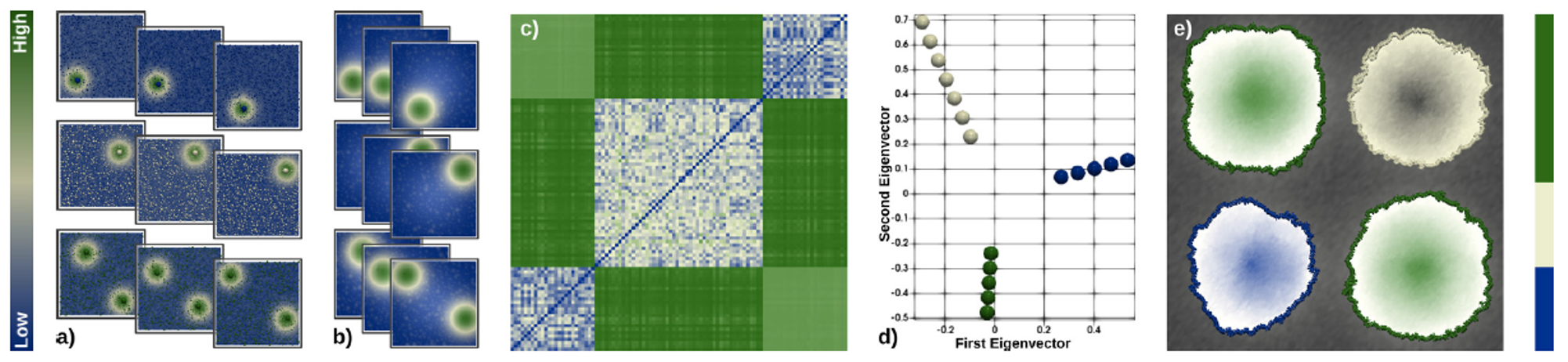}
    \caption{Persistence atlas for an ensemble of a synthetic scalar fields. (a) The input scalar fields. (b) Persistence maps. (c) Distance matrix for the persistence maps. (d) Projection of persistence maps to 2D. (e) The persistence atlas. Image reproduced from~\cite{FavelierFarajSumma2018}.}
    \label{fig:persistence-atlas}
\end{figure}

In this section, we focus on topological descriptors applied to scalar field comparisons for ensembles. 
An ensemble of scalar fields, in the context of this report, is a collection of scalar fields indexed on a set of parameters with no temporal relationship between them. Such a collection of scalar fields often results from multiple simulation runs obtained by varying the input parameters. 

The typical overall analysis task in an ensemble is to relate the observed features in the output scalar fields to the input parameter space. Topological descriptors play an important role and thus can reduce the task of understanding the space of scalar fields to a more tractable task of exploring the space of their corresponding topological descriptors. For example, instead of searching for patterns in the high-dimensional space of scalar fields, the job simplifies to finding patterns in the space of merge trees. Key tasks include clustering, classification, outlier detection, feature detection, summarization, and computation of structural statistics. 

\subsection{Clustering and Classification}
\label{sec:ensemble-clustering}
Applying and adapting techniques from pattern recognition and statistical analysis, such as clustering and classification to analyze ensembles, has seen active research in recent years. 
However, these ideas have been around for a long time. The work by Hilaga~\etal~\cite{HilagaShinagawaKohmura2001}, which used Reeb graph for shape matching and retrieval, is an early example of how topological approaches can work in the context of information retrieval, clustering, and classification. 
Another early topology-guided comparison for classification is the use of multi-resolution dual contour trees proposed by Zhang~\etal~\cite{ZhangBajajBaker2004}. They demonstrated applications to electrostatic potential and electron density fields for protein structures, resulting in a classification of proteins into categories. In both papers, a similarity matrix was used to capture distances between members of the ensemble; refer to~\autoref{sec:single-fields} for a more detailed discussion of these papers. 

Favelier~\etal~\cite{FavelierFarajSumma2018} described a framework called \emph{persistence atlas} that uses a \emph{persistence map} for clustering and trend variability analysis of ensembles. The persistence map is a derived scalar field that captures the spatial distribution and density of extrema in the input field weighted by their importance. Persistence is used as the importance measure. This derived map is then used as a signature of the scalar field. Eigenanalysis-based dimensionality reduction is applied to project the ensemble members to 2D where clusters are identified. Trend and spatial variability analyses are done per cluster by computing mandatory critical points~\cite{GuntherSalmonTierny2014}. The utility of the framework is demonstrated using simple synthetic data~(\autoref{fig:persistence-atlas}), multiple flow simulation ensembles, and ensembles originating from climate simulations. 

\begin{figure}[!ht]
    \centering
    \includegraphics[width=0.98\columnwidth]{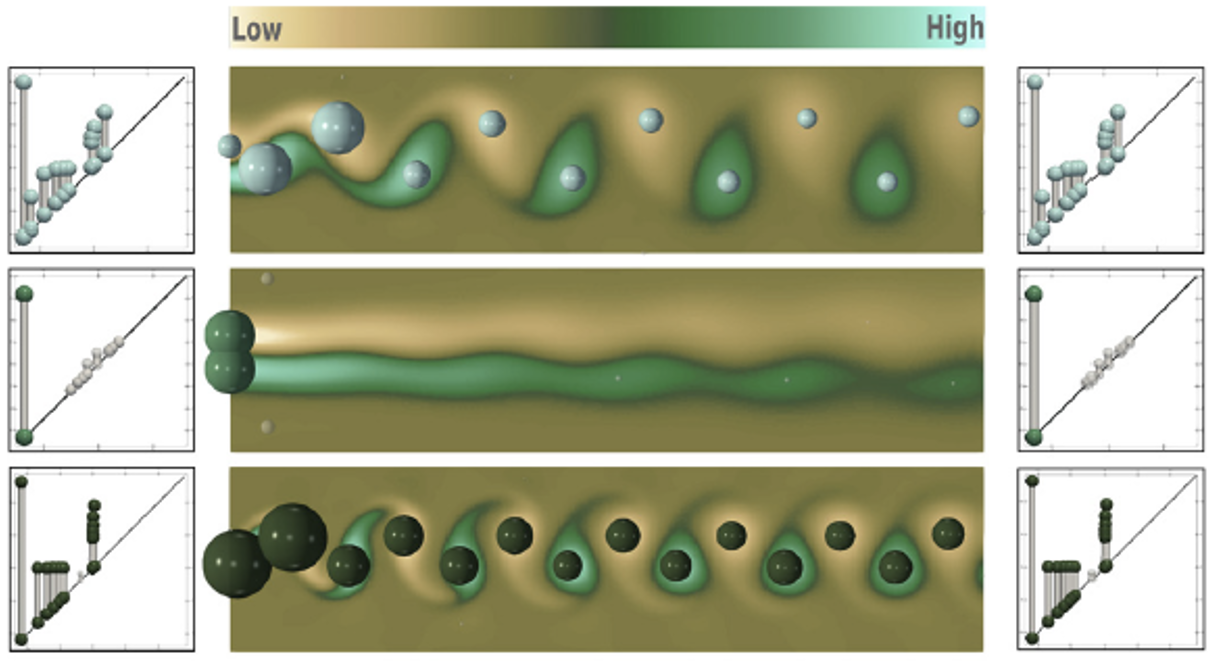}
    \caption{Three clusters identified in the flow data ensemble  correspond to different viscosity regimes. The Wasserstein persistence diagram barycenters computed using interruptible approach are shown on the left and the exact barycenters on the right. There is little difference between the diagrams, and the performance difference between these approaches is huge. Image reproduced from~\cite{VidalBudinTierny2020}, cropped to show three of the five clusters.}
    \label{fig:progressive-wasserstein}
\end{figure}

Vidal~\etal~\cite{VidalBudinTierny2020} pursued a more direct approach to clustering of persistence diagrams,  adapting the k-means algorithm based on Wasserstein distance. k-means is an iterative algorithm that proceeds in two steps. First, the point assignment step assigns each point to one cluster based on its distance to the cluster centroids. The second step updates the centroids. The mean persistence diagram of a set of persistence diagrams is computed as the Fr\'echet mean under the Wasserstein distance metric $d_{p=2}$. This requires repeated computation of pairwise distances between diagrams and the computation of Fr\'echet mean. If done naively, the computation is prohibitively costly and not time efficient for realistically sized data ensembles. The authors proposed a progressive algorithm for computing the Fr\'echet mean or discrete Wasserstein barycenters in a time-interruptible manner. This approach leads to improvements in runtime while producing similar quality results as shown via applications to ensembles studied earlier by Favelier~\etal~\cite{FavelierFarajSumma2018}. \autoref{fig:progressive-wasserstein} shows their results for a flow data ensemble.

The drawback of using k-means is the requirement of the number of expected clusters $k$ in the ensemble. Kontak \etal.~\cite{KontakVidalTierny2019} addressed this issue in their recent work in which they suggested finding the optimal number of clusters based on a minimization of established statistical score functions such as Akaike information criterion and Bayesian information criterion.

\subsection{Summarization}
\label{sec:ensemble-structure}
Another approach for providing an overview of a scalar field ensemble is to summarize the set of scalar fields within a single abstract structure. For this task, topological structures have recently been shown to be very useful. 

Lohfink~\etal~\cite{LohfinkWetzelsLukasczyk2020} described a technique for combining contour trees of multiple scalar fields of an ensemble in a single layout called fuzzy contour tree, 
 which provides a summary of the ensemble.
They use tree alignment, an idea similar to edit distance mappings, to identify similarities across multiple contour trees and to obtain a layout that can represent all trees simultaneously. 
A heuristic algorithm is used for  computing the tree alignments for a given similarity measure. 
The approach is applied to the visualization of two scalar field ensembles: 2D flow around a cylinder and a heated cylinder ensemble. The utility of this approach is demonstrated well for these examples. However, the scalability of the approach for feature-rich data is not clear. 
\begin{comment}
\begin{figure}[!hb]
    \centering
    \includegraphics[width=0.98\columnwidth]{figs/fuzzyCT.jpeg}
    \caption{A fuzzy contour tree highlights and align the common components from 2D heated cylinder ensembles~\cite{LohfinkWetzelsLukasczyk2020}.}
    \label{fig:fuzzyCT}
\end{figure}
\end{comment}

\begin{figure}[!h]
    \centering
    \includegraphics[width=0.98\columnwidth]{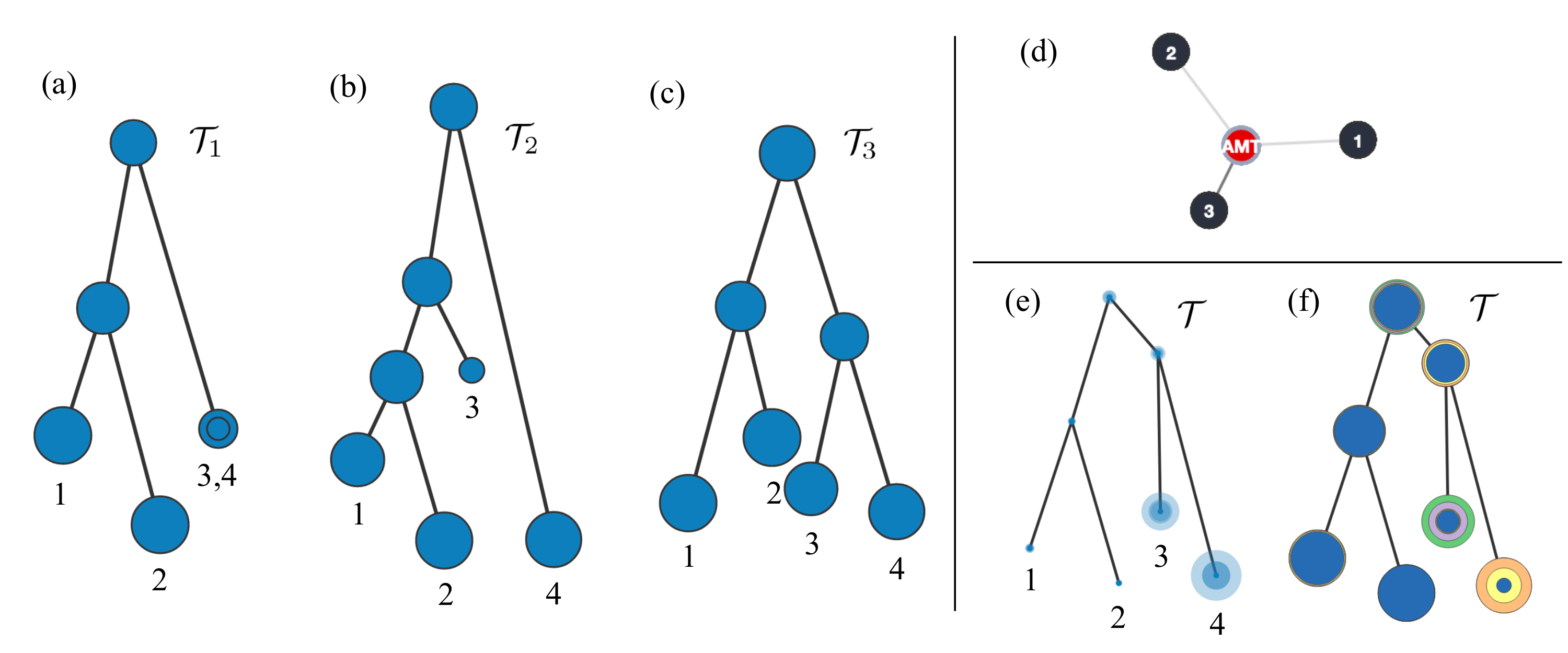}
    \caption{An example of a structural average of three input merge trees and a visualization of its uncertainty. Vertex consistencies for each ensemble member are encoded using the radius of the glyph in (a), (b), and (c). Graduated circular glyphs encode variational (e) and statistical (f) vertex consistencies for 1-center trees by applying graduated glyphs to minimum, first quartile, median, third quartile, and maximum across all ensemble members. A summary plot~(d) shows the interleaving distance between each input tree and the 1-center tree. Image generated using the visualization tool of~\cite{YanWangMunch2020}.}
    \label{fig:mean-merge-tree}
\end{figure}

Yan~\etal~\cite{YanWangMunch2020} proposed to use interleaving distances  (\cref{eq:labelled-interleaving}) to study a structural average of an ensemble of labeled merge trees. 
Such a structural average, referred to as a \emph{1-center tree}, minimizes its maximum distance to any other tree in an input ensemble.  
They used global and local structural consistency measures between the input and the 1-center tree to encode uncertainty; see \autoref{fig:mean-merge-tree} for an example. 
Different heuristics are used to compute the labeling between pairs of merge trees to compute the 1-center tree. 
Their work demonstrated the application of a nice theoretical framework to compute  structural averages of merge trees. 
In the supplementary material, the authors applied their framework to study neuron morphology. 
Their proposed consistency measure can be used to understand structural variations among an ensemble of neuron-cell-induced merge trees with respect to their 1-center, where such an ensemble arises from different reconstructions of the same neuron cell. 

Finally, we would like to mention that the persistence atlas~\cite{FavelierFarajSumma2018} framework discussed earlier also provides a summary of an ensemble. The mandatory critical point regions identified for each cluster can be embedded in the spatial domain. See~\autoref{fig:persistence-atlas}(e). However, this approach to summarization does not scale for large ensembles with complex features.

\subsection{Interactive Exploration}
\label{sec:ensemble-exploration}
Another significant aspect in the analysis of ensembles is to understand the relationship of the input parameters to the observed features in the scalar field instances, which is often explored fully through an interactive visualization tool. The work by Yan~\etal~\cite{YanWangMunch2020} provides some interactive exploration capabilities but it is restricted to exploring the space of merge trees.  

A more appropriate example in this context of parameter space exploration using topological comparison methods is the work by Poco~\etal~\cite{PocoDoraiswamyTalbert2015} exploring differences in species distribution models. They computed a locality-aware correspondence between similar extrema of two scalar fields that represent two species distribution models. The extrema of the two scalar fields are represented as a bipartite graph, and the correspondence is computed as a maximum weight matching. 
The weight of each edge is a product of two terms, a functional similarity term and a spatial distance term. 
After computing the above maximum topology matching between the extrema, the topological similarity is computed as the minimum amount of simplification required to obtain a perfect bipartite matching. Additionally, a functional similarity is computed from the perfect bipartite matching as the maximum function difference between the matched extrema pairs. 
Species distribution models are high-dimensional scalar functions, where the dimensions of the domain correspond to predictor variables of the model algorithm. The topological and functional similarity measures are integrated into a visual exploration tool with typical statistical plots and other domain-specific charts. The tool facilitates detailed inspection of selected ensemble members with a focus on identifying the regions of similarity and differences between the high dimensional scalar functions.

\subsection{Multi-field Data Comparison}
\label{ensemble-multi-field}
Scalar field ensembles can also be considered as a single multi-field. Thus, they may be analyzed using topological techniques for multi-field analysis. Here, we discuss a few results in this direction. Additionally, we also mention initial work on pairwise comparison of two multi-fields, which in turn can be used for the analysis of time-varying multi-field ensembles.

Huettenberger~\etal~\cite{HuettenbergerHeineCarr2013} extended the notion of extremal points to multi-fields based on Pareto optimality and Pareto dominance. The regions of consensus and disagreement among the ascending and descending manifolds of the extrema of the constituent scalar fields are identified. Pareto-extremal regions represent barriers across which all the constituent scalar fields cannot be jointly increased or decreased. These regions are joined by ascending and descending sets where the gradients of the scalar fields agree. Huettenberger~\etal  compared this approach against other topological methods for multi-field analysis like Jacobi sets and joint contour nets. The utility of the method is demonstrated on a simulated 2D flow with randomly perturbed input parameters, specifically to visualize the joint extremal structures of multiple realizations of the $\lambda_2$ field for vortex identification.

Liebmann and Scheuermann~\cite{LiebmannScheuermann2016} introduced a topological description that represents a set of Gaussian-distributed scalar fields as a whole. A set of singular patches serves as a counterpart to critical points. These patches are assigned a probability as an attribute. The patches are defined by a classification of the local scalar value configuration in a point’s neighborhood. Also, an adjacency relation between these patches is established, resulting in an abstract graph representation. In the proposed pipeline, the correlation matrix plays an important role in a first dimension reduction step (PCA). The analysis takes place in a high-dimensional space spanned by the largest eigenvectors of the correlation matrix. The graph is visualized using a force-directed layout to provide a first overview. Edge properties are encoded using lengths, thickness, and color. Furthermore, node properties, like patch probability and the topological type, are represented using glyphs. The spatial regions of the patches are shown with transparency, indicating the accumulated probability of all patches involved.

Both approaches discussed above are interesting ideas seeking to extend the well-established approaches of topological analysis for scalar fields to the largely unexplored domain of multi-field analysis. The wide popularity of contour trees and Reeb graphs in the visualization community resulted in a study of their generalization to multi-fields. The joint contour net~(JCN) proposed by Carr and Duke~\cite{CarrDuke2014} captures changes in specific subsets, called slabs, of the domain when varying multiple scalar fields simultaneously. 
They can be interpreted as a discrete approximation of the Reeb space~\cite{EdelsbrunnerHarerPatel2008}. 
The main idea is to apply a quantization of the values in the range of the multi-field resulting in a structure that is amenable to computation. JCNs are also closely related to Mapper~\cite{SinghMemoliCarlsson2007}.

\paragraph*{Comparison of multi-fields. }
An approach proposed by Agarwal~\etal~\cite{AgarwalChattopadhyayNatarajan2019} for comparing multi-fields suggested using fiber component distributions.
First, the JCN is computed for the given multi-field. Next, the frequency distribution is approximated by partitioning the range space into bins and counting the number of fiber components within each bin. 
A similarity score between two multi-fields is then simply computed as the $L^p$ distance between their corresponding fiber component frequency distributions.
The comparison measure finds application to computational chemistry, where it is used for identification of stable Pt-CO bonds using electron density multi-field data.
The comparison measure is also used for identifying nuclear scission events. 

In a follow-up work, Agarwal~\etal~\cite{AgarwalRamarmurthiChattopadhyay2020} extended the idea of comparison using multi-resolution Reeb graphs~\cite{HilagaShinagawaKohmura2001} and contour trees~\cite{ZhangBajajBaker2004} to multi-fields as well. 
They proposed multi-resolution Reeb space of a multi-field and computed a similarity score as a weighted average of the scores comparing node attributes, such as volume, range, number of connected components, and degree. 
Refer to \autoref{sec:tempGlobalCompare} for a more detailed discussion. This approach was also applied on the Pt-CO and the nuclear scission case studies with better results.

\subsection{Uncertainty Visualization}
\label{ensemble-uncertainty}
In the context of ensemble data analysis, the study and visualization of uncertainty~\cite{BonneauHegeJohnson2014} or variability in a set of scalar fields or their corresponding topological structures are crucial for drawing any meaningful conclusions from the data. We describe some selected approaches in this direction below.

Wu and Zhang~\cite{WuZhang2013} proposed an exploration framework for uncertain scalar fields. Their main goal was to show the variance in contours and topology. The core of the visualization and analysis is the contour tree of the field average considered as a pivot tree. In the visualization, this tree is augmented with uncertainty-variability glyphs. For every contour of the \emph{mean tree}, an average data-level uncertainty is attached to the tree branches in a ribbon-like fashion. For the contour comparison, they also introduced a branch correspondence concept, considering the branch with the largest overlap as most similar and computed an average difference of the contours. The abstract mean tree is overlaid with different variability measures and used to interact with the 3D spatial visualization. In 2D, the uncertainty glyphs are also visualized in the spatial representation. Two trees can be compared side-by-side or as an overlay. They showcase the framework via applications to uncertainty visualization for weather simulation and brain data.  
In the work of Yan~\etal~\cite{YanWangMunch2020}, a different notion of average merge tree is computed together with visual encodings of uncertainty information. 

Mihai and Westermann~\cite{MihaiWestermann2014} described an approach for finding a feature distribution across ensemble instances. The features they considered were defined by level sets and critical points. 
They proposed methods for classification of critical points with respect to different notions of stability. Two indicator functions, the gradient and Hessian, are computed at all vertices of the data. The first indicates the likelihood of the existence of a critical point, the second reveals the tendency of its type. Both indicator functions are interpreted as independent multivariate random variables at each grid point. Confidence intervals of the gradient are computed at each vertex and if it contains zero, then a glyph is rendered at the vertex for visualization. They demonstrated their approach using climate simulation data.

Gunther~\etal~\cite{GuntherSalmonTierny2014} introduced the notion of mandatory critical points that can be interpreted as the common topological denominator of uncertain scalar fields. These are regions and intervals where there is at least one critical point in any realization of the ensemble. Based on this concept, they also introduced the notion of mandatory merge trees. A simplification strategy  enables multi-scale visualizations. This approach was effectively demonstrated on two flow datasets, K\'arm\'an vortex street and heated cylinder, and two more ensemble datasets from astronomy and climate science. 
Athawale~\etal~\cite{AthawaleMaljovecYan2020} utilized mediatory critical points of~\cite{GuntherSalmonTierny2014} within their pipeline to explore uncertainty visualization of an ensemble of Morse complexes. They introduced three types of statistical summary maps – the probabilistic map, the significance map, and the survival map – to characterize the uncertain behaviors of gradient flows of Morse complexes.

%% file: sec-math-properties.tex
\begin{table*}[ht]
\centering
\resizebox{2.1\columnwidth}{!}{
\begin{tabular}{lcccccc}
\toprule
 \textbf{Measures} & \textbf{Citation} & \textbf{Notation} & \textbf{Metric} & \textbf{Stability} & \textbf{Discriminative} & \textbf{Complexity} \\
\bottomrule
\toprule
\multicolumn{7}{c}{Comparing persistence diagrams and their variants}
 \\ 
 \bottomrule
Bottleneck distance &\cite{Cohen-SteinerEdelsbrunnerHarer2007} &$d_\infty$ & \cellcolor{mygreen}extended peudometric &  \cellcolor{mygreen}Yes & baseline n/a & $O(n^{1.5}\log(n))$ \\
$p$-Wasserstein distance &\cite{Cohen-SteinerEdelsbrunnerHarer2010} &$d_p$ &\cellcolor{mygreen}extended peudometric & \cellcolor{mygreen}Yes & \cellcolor{mygreen} Yes & $O(n^{3})$ \\
$p$-landscape distance & \cite{Bubenik2015} &$\Lambda_{p}$  &\cellcolor{mygreen}metric & \cellcolor{mygreen} Yes & \cellcolor{babypink} No & $O(n^{2})$ \\ 
Persistence scale-space kernel  &\cite{ReininghausHuberBauer2015} & $K_S$ &n/a &\cellcolor{mygreen}Yes  &\cellcolor{myteal}unknown & $O(n^2)$ \\
Persistence weighted Gaussian kernel  &\cite{KusanoFukumizuHiraoka2017} & $K_G$ &n/a &\cellcolor{mygreen}Yes  &\cellcolor{myteal}unknown & $O(n^2)$ \\
sliced Wasserstein distance  &\cite{CarriereCuturiOudot2017} & $SW$ &\cellcolor{mygreen}extended peudometric &\cellcolor{mygreen}Yes  &\cellcolor{mygreen}Yes & $O(n^2\log(n))$ \\
Persistence Fisher kernel &\cite{LeYamada2018} & $K_F$ &n/a &\cellcolor{mygreen}Yes  &\cellcolor{myteal}unknown & $O(n^2)$ \\
Lifted Wasserstein  &\cite{SolerPlainchaultConche2018} & --- &\cellcolor{mypurple} conj. Metric & \cellcolor{myteal} unknown& \cellcolor{myteal} unknown & $O(n^3)$ \\
WKPI distance  &\cite{ZhaoWang2019} & --- &\cellcolor{mygreen}pseudometric &\cellcolor{mygreen} Yes  &\cellcolor{myteal}unknown & O(N) \\
PHT distance  &\cite{TurnerMukherjeeBoyer2014} & --- &\cellcolor{mygreen} metric &\cellcolor{mygreen} Yes &\cellcolor{myteal} unknown  & $O(n^3)$ \\
$L^p$ distance between persistence vectors& \cite{LiWangAscoli2017} & ---   & \cellcolor{mygreen}metric &\cellcolor{myteal}unknown &\cellcolor{myteal}unknown & $O(\max(n, l))$\\
\toprule
\multicolumn{7}{c}{Comparing Reeb graphs and their variants}
 \\ 
 \bottomrule
Functional distortion distance & \cite{BauerGeWang2014}& $d_{FD}$ &\cellcolor{mygreen}extended pseudometric & \cellcolor{mygreen} Yes & \cellcolor{mygreen} Yes & NP-hard \\
Edit distance between merge trees & \cite{SridharamurthyMasoodKamakshidasan2020}&$d_E$ &\cellcolor{mygreen}metric & \cellcolor{mypurple} We conj. No & \cellcolor{mypurple} conj. Yes & $O(n^{2})$ \\
Edit distance between labeled Reeb graphs & \cite{BauerDiFabioLandi2016} & $d_{EG}$ &\cellcolor{mygreen}extended pseudometric & \cellcolor{mygreen} Yes & \cellcolor{myteal} unknown & \cellcolor{mypurple} We conj. NP-hard \\
Interleaving distance between merge trees & \cite{MorozovBeketayevWeber2013}&$d_I$ &\cellcolor{mygreen}metric & \cellcolor{mygreen} Yes & \cellcolor{mygreen} Yes & NP-hard \\
Interleaving distance for labeled merge trees& \cite{GasparovicMunchOudot2019}& $d_{IL}$ &\cellcolor{mygreen}metric & \cellcolor{mygreen} Yes & 
\cellcolor{mypurple} We conj. Yes & $O(n^{2})$ \\
Interleaving distance between Reeb graphs & \cite{SilvaMunchPatel2016}& $d_{IG}$ &\cellcolor{mygreen}extended pseudometric & \cellcolor{mygreen}Yes & \cellcolor{myteal} unknown & NP-hard \\
Distance based on branch decompositions  &\cite{BeketayevYeliussizovMorozov2014} & $d_{BR}$ &\cellcolor{mypurple}We conj. Yes & \cellcolor{mypurple} We conj. No& \cellcolor{mypurple} We conj. Yes & $O(n^5\log(I_{\epsilon}))$ \\
Distance based on histograms for merge trees &\cite{SaikiaSeidelWeinkauf2015} & --- & \cellcolor{mygreen} metric & \cellcolor{myteal}unknown  & \cellcolor{myteal}unknown & $O(n^2 B)$\\
Distance based on tree alignment  &\cite{LohfinkWetzelsLukasczyk2020} &  --- &\cellcolor{myteal}unknown &\cellcolor{myteal}unknown &\cellcolor{myteal}unknown &$O(n^2)$ \\
Distance based on subtrees of contour trees  &\cite{ThomasNatarajan2011}&  --- &\cellcolor{myteal}unknown &\cellcolor{myteal}unknown &\cellcolor{myteal}unknown & $O(t^5 \log t)$ \\
Distance between extended Reeb graphs &\cite{BiasottiMariniMortara2003}&  --- &\cellcolor{mygreen}metric  &\cellcolor{myteal}unknown &\cellcolor{myteal}unknown & \cellcolor{myteal}unknown \\
Kernel between extended Reeb graphs &\cite{BarraBiasotti2013}&  --- &\cellcolor{myteal}unknown  &\cellcolor{myteal}unknown &\cellcolor{myteal}unknown & $O(\gamma^4_{max} + 2n^2 \log(n))$\\
\toprule
\multicolumn{7}{c}{Comparing Morse complexes, Morse-Smale complexes and their variants}
 \\ 
 \bottomrule
Distance between extremum graphs  &\cite{NarayananThomasNatarajan2015} & $d_{\rho}$ &\cellcolor{mygreen}metric &\cellcolor{myteal}unknown  &\cellcolor{myteal}unknown & NP-hard \\
Similarity between Morse-Smale complexes &\cite{FengHuangJu2013}&  --- &n/a &\cellcolor{myteal}unknown &\cellcolor{myteal}unknown & \cellcolor{myteal}unknown \\
Similarity between Morse-Smale cells &\cite{SchnorrHelmrichDenker2020}&  ---  &n/a &\cellcolor{myteal}unknown &\cellcolor{myteal}unknown & NP-hard \\
\toprule
\multicolumn{7}{c}{Other comparative measures}
 \\ 
 \bottomrule
Distance between Morse shape descriptor &\cite{AlliliCorriveau2007}&  --- &\cellcolor{myteal}unknown &\cellcolor{myteal}unknown &\cellcolor{myteal}unknown & $O(n^2)$ \\
Distance between persistence maps&\cite{FavelierFarajSumma2018}&  --- &\cellcolor{mygreen}metric  &\cellcolor{myteal}unknown &\cellcolor{myteal}unknown & $O(n^2M)$ \\
Distance between fiber component distributions &\cite{AgarwalChattopadhyayNatarajan2019}&  --- &\cellcolor{mygreen}metric  &\cellcolor{myteal}unknown &\cellcolor{myteal}unknown & \cellcolor{myteal}unknown
\end{tabular}}
\caption{Desirable properties for comparative measures surveyed in this paper.  
\textbf{Citation} indicates when a measure is first introduced. 
\textbf{Notation}: not all comparative measures are given a mathematical notation in this survey.
\textbf{Complexity}: $n$ is the number of critical points in a topological descriptor; $N$ in the number of pixels in a persistence image; $M$ denotes the number of vertices in the domain; $l$ is density vectorization parameter; $\gamma_{max}$ is the length of the maximal shortest path; $I_{\epsilon}$ is the search range for $\epsilon_{min}$; $B$ is the number of bins in a histogram; and $t$ is the number of branches in a branch decomposition.   
Colors encode yes (green), no (pink), conjecture (purple, abbreviated as conj.), n/a (not applicable, white), and unknown (blue). 
}
\label{table:properties}
\end{table*}

\section{Desirable Properties of Comparative Measures}
\label{sec:properties}

We now survey desirable properties of a comparative measure, denoted as $d(\A_1,\A_2)$, between a pair of topological descriptors $\A_1$ and $\A_2$. 
For each comparative measure $d$ reviewed in~\autoref{sec:comparative-measures},  we study its properties surrounding \emph{metricity}, \emph{stability}, \emph{discriminativity}, and \emph{computational complexity}. 
We give a systematic classification of surveyed comparative measures based on these properties, see~\autoref{table:properties}. 
There are a number of other properties that remain under-explored.  
For example, Bauer \etal~\cite{BauerLandiMemoli2020} considered a  distance for Reeb graphs to be \emph{universal} if it provides an  upper  bound  to  any  other  stable  distance. 
Carri\`{e}re and Oudot~\cite{CarriereOudot2017} considered a distance for Reeb graphs to be \emph{intrinsic} if it can be realized by a geodesic.
Gasparovic \etal~\cite{GasparovicMunchOudot2019} considered this property for distance between merge trees. 
It will be interesting to investigate these additional properties for a number of comparative measures in future works.

In this section, for discussions on computational complexity: $n$ and $m$ are the numbers of critical points in each topological descriptor (\emph{w.l.o.g.}, assume $m \leq n$); $N$ is the number of pixels in a persistence image; $M$ denotes the number of vertices in the domain; $l$ is density vectorization parameter; $\gamma_{max}$ is the length of the maximal shortest path; $I_{\epsilon}$ is the search range for $\epsilon_{min}$; $B$ is the number of bins in a histogram; and $t$ is the number of branches in a branch decomposition.  

\subsection{Comparative Measures for Persistence Diagrams}
We describe properties associated with the bottleneck distance $d_\infty$, the $p$-Wasserstein distance $d_p$, and the $p$-landscape distance $\Lambda_{p}$. 
We also discuss properties associated with a few other comparative measures for persistence diagrams. 

\para{Bottleneck distance.}
The space of persistence diagrams can be equipped with the bottleneck distance $d_\infty$~\cite{Cohen-SteinerEdelsbrunnerHarer2007}. 
Recall a persistence diagram is a multi-set of points in the extended plane $\overline{\Rspace}^2$. 
In the most general case, $d_\infty$ is an extended pseudometric~\cite[Page 51]{Oudot2017}. 
$d_\infty$ is a metric when the persistence diagrams are locally finite multi-sets in $\overline{\Rspace}^2$; see~\cite[Page 51]{Oudot2017} and~\cite[Theorem 4.10]{ChazalSilvaGlisse2016}. 
$d_\infty$ is proven to be stable~\cite{Cohen-SteinerEdelsbrunnerHarer2007} {\wrt} small perturbations of the input function.
\begin{theorem}[Stability of $d_{\infty}$~\cite{Cohen-SteinerEdelsbrunnerHarer2007}] 
Let $\Xspace$ be a triangulable space with continuous tame functions, $f, g: \Xspace \to \Rspace$. Then the persistence diagrams satisfy 
\begin{align}
d_\infty(\D_f, \D_g) \leq ||f-g||_\infty. 
\end{align}
\label{theorem:bottleneck-stability}
\end{theorem}
A topological space is \emph{triangulable} if there is a (finite) simplicial complex with homeomorphic underlying space~\cite{Cohen-SteinerEdelsbrunnerHarer2007}; and \emph{tameness} is a technical condition that ensures $f$ and $g$ to be well behaved. 
Many existing works use $d_{\infty}$ as a baseline comparative measure to determine the discriminativity of a newly introduced measure.  

The main challenge of computing $d_\infty$ between persistence diagrams is to find an optimal bijection $\eta$ between them. 
Such a task can be viewed as a bipartite graph matching problem, which has been studied for decades~\cite{HopcroftKarp1973,Bertsekas1979,Munkres1957}. 
Inspired by an auction algorithm~\cite{Bertsekas1979} and Hopcroft-Karp algorithm~\cite{HopcroftKarp1973}, Kerber {\etal}~\cite{KerberMorozovNigmetov2017} utilized the geometric structure of data into bijection matching and significantly improved the computation of both $d_\infty$ and $d_p$ in both runtime and memory consumption, where $d_\infty$ can be computed in $O(n^{1.5}\log n)$ time.

\para{$p$-Wasserstein distance.} 
$d_p$ is an extended pseudometric; it is a metric~\cite[Page 184]{EdelsbrunnerHarer2010} when the persistence diagrams are locally finite.
Furthermore, the set of persistence diagrams equipped with $d_p$ is shown to be a complete and  separable metric space~\cite{Cohen-SteinerEdelsbrunnerHarer2010}.
$d_p$ is also stable for a reasonably large class of functions~\cite{Cohen-SteinerEdelsbrunnerHarer2010}. 
\begin{theorem}[\cite{Cohen-SteinerEdelsbrunnerHarer2010}]
Let $f, g: \Xspace \to \Rspace$ be tame Lipschitz functions on a metric space whose triangulations grow polynomially with constant exponent $j$. 
Then, there are constants $C$ and $k > j$ no smaller than $1$ such that, for every $p \geq k$, 
\begin{align}
d_{p}(\D_f,\D_g) \leq C \cdot || f - g||_\infty^{1-\frac{k}{p}}. 
\end{align}
\end{theorem} 

As $p$ goes to $\infty$, $d_p$ approaches $d_\infty$ by defining the minimum of the maximum edge weight over all perfect matchings~\cite{KerberMorozovNigmetov2017}. 
Therefore, $d_p$ is more discriminative than $d_\infty$,
\begin{align}
d_\infty(\D_f,\D_f) \leq d_p(\D_f, \D_g). 
\end{align}

The exact computation of $d_p$ needs minimum bipartite matching, which requires $O(n^{3})$ effort. However, faster approximation algorithms are available for computing $d_p$ using the geometric techniques of Kerber \etal~\cite{KerberMorozovNigmetov2017}. 

\para{$p$-landscape distance.} $\Lambda_{p}$ is a  metric~\cite{BubenikDlotko2017}. 
It enjoys a form of stability under the same condition as for the stability of $d_p$ in~\cite{Cohen-SteinerEdelsbrunnerHarer2010}.
Specifically, the persistence diagram is stable {\wrt} the $p$-landscape distance if $p>k$ and $\Xspace$ has bounded degree-$k$ total persistence.
\begin{theorem}[\cite{Bubenik2015}, Theorem 16]
Let $\Xspace$ be a triangulable, compact metric space that implies bounded degree-$k$ total persistence for some real number $k \geq 1$, and let $f$ and $g$ be two tame Lipschitz functions. Then, for all $p \geq k$, 
\begin{align}
\Lambda_p(\D_f, \D_g)^p \leq C \cdot || f - g||_\infty^{p-k}. 
\end{align}
\end{theorem}
Here, the constant $C$ is related to the Lipschitz constants of $f$ and $g$. 
$\Lambda_{\infty}$ is also shown to be stable {\wrt} to the $L^\infty$ norm.
\begin{theorem}[\cite{Bubenik2015}, Theorem 12]
Given real-valued functions $f, g: \Xspace \to \Rspace$ on a topological space, 
\begin{align}
\Lambda_\infty(\D_f, \D_g) \leq ||f-g||_\infty.
\end{align}
\end{theorem}
$d_\infty$ is shown to be more discriminative than $\Lambda_{p}$ for $p=\infty$~\cite{Bubenik2015}.   
\begin{theorem}[\cite{Bubenik2015}, Theorem 13]
\begin{align}
\Lambda_{\infty}(\D_f, \D_g) \leq d_{\infty}(\D_f,\D_g) 
\end{align}
\end{theorem}
$\Lambda_p$ also contributes as a lower bound on the $p$-Wasserstein distance $d_{p}$~\cite[Corollary15]{Bubenik2015}. 
The above theorems suggest that $\Lambda_{p}$ is not discriminative {\wrt} the baseline $d_\infty$. 

Bubenik and D{\l}otko ~\cite{BubenikDlotko2017} developed algorithms to compute persistence landscapes and the landscape distance $\Lambda_p$. The former takes $O(n^2)$, where $n$ is the number of birth-death pairs, and the latter requires $O(P)$, where $P$ is the maximum number of critical points of two input persistence landscapes.
Since $P=O(n)$, we conclude the complexity is $O(n^2) + O(P) = O(n^2)$.

We now discuss properties associated with kernels on persistence diagrams.

\para{Persistence scale-space kernel.}
$K_S$ is 1-Wasserstein stable~\cite{ReininghausHuberBauer2015}, that is, it is upper bounded by $p$-Wasserstein distance when $p=1$ (denoted as $d_{p=1}$). 
Given two persistence diagrams (including points on the diagonal) $\D_1$ and $\D_2$, we have 
\begin{theorem}[\cite{ReininghausHuberBauer2015}, Theorem 2]
	\begin{align}
	||\Phi_\sigma(\D_1) - \Phi_\sigma(\D_2)||_{L_2(\Omega)} \leq \frac{1}{2\sqrt{\pi}\sigma} d_{p=1}(\D_1, \D_2),
	\end{align}
	where the $L_2$-valued 	
	 feature map $\Phi_\sigma: \Dcal \to L_2(\Omega)$ at scale $\sigma > 0$ of a persistence diagram $\D$ is defined as $\Phi_\sigma(\D)$, with $\Omega \subset \Rspace^2$ denoting the	closed half plane above the diagonal.
\end{theorem}
$K_S$ can be computed in $O(m \cdot n)$ time, where $m$ and $n$ denote the cardinality of the multi-sets $\D_1$ and $\D_2$ (not counting the diagonal), respectively; {\mywlog} assuming $m \leq n$, the complexity can be simplified as $O(n^2)$. 

\para{Persistence weighted Gaussian kernel.}
The distance defined by the reproducing kernel Hilbert spaces (RKHS) norm for the persistence weighted Gaussian kernel (PWGK) $K_G$ satisfies a form of stability~\cite{KusanoFukumizuHiraoka2017} {\wrt} the bottleneck distance. 
\begin{theorem}[\cite{KusanoFukumizuHiraoka2017}, Theorem 9]
	Let $\Mspace$ be a triangulable compact subspace in $\Rspace^d$, $X,Y \subset \Mspace$ be finite subsets, and $p > d + 1$. 
A persistence diagram $\D$ can be vectorized via the map, $E_{k_G}: \mu_\D^w \mapsto w(x)k_G( {\cdot}, x)$, where $w$ is a weight defined in~\autoref{eq:weighted-gaussian-kernel} and $\mu_\D:=\sum_{x \in \D}w(x)\delta_x$ for the Dirac delta measure at $x$. 
Now, given persistence diagrams $\D_X$ and $\D_Y$, and their vectorized representations $E_{k_G}(\mu_{\D_X}^w)$ and $E_{k_G}(\mu_{\D_Y}^w)$ of the RKHS,
	\begin{align}
	||E_{k_G}(\mu_{\D_X}^{w})-E_{k_G}(\mu_{\D_Y}^{w})||_{\Hcal_{k_G}}\leq L(k,w)d_\infty(\D_X, \D_Y),
	\end{align}
	where $||\cdot||_{\Hcal_{k_G}}$ represents the norm in RKHS, and $L(k,w)$ is a constant independent of $X$ and $Y$.
\end{theorem}
Given two persistence diagrams, the computation of $K_G$ involves $O(n^2)$ evaluations of $e^{-\frac{\|p-q\|^2}{2\sigma^2}}$, where $n$ is the number of points in the larger persistence diagram~\cite{KusanoFukumizuHiraoka2017}. Kusano {\etal} also investigated faster approximation algorithms using the random Fourier features~\cite{RahimiRecht2007}.  

\para{Sliced Wasserstein kernel and sliced Wasserstein distance.}
The sliced Wasserstein distance ${SW}$ (\autoref{eq:sliced-wasserstein}), which is designed for the sliced Wasserstein kernel $K_W$, is stable and discriminative as it preserves the metric between persistence diagrams~\cite{CarriereCuturiOudot2017}.
In particular, $SW$ is proved to be equivalent to $d_{p=1}$ (the $p$-Wasserstein distance for $p=1$), which implies that $SW$ is as discriminative as $d_{p=1}$. 
\begin{theorem}[\cite{CarriereCuturiOudot2017}, Theorem 3.3]
	Let persistence diagrams $\D_1$, $\D_2$ have cardinalities bounded by $N$, then
	\begin{align}
	\frac{d_{p=1}(\D_1, \D_2)}{2\times(1+2N(2N-1))} \leq SW(\D_1, \D_2,M) \leq 2\sqrt{2} d_{p=1}(\D_1, \D_2).
	\end{align}
\end{theorem}
The time required to compute $K_W$ is $O(n^2\log(n))$.

\para{Persistence Fisher kernel.}
The persistence Fisher kernel $K_{F}$ is proven to be stable~\cite{LeYamada2018}, because its induced squared distance $d_{K_F}^2$ is bounded by the Hilbert norm of the difference between two corresponding mappings $d_{F}$.  
\begin{theorem}[\cite{LeYamada2018}, Lemma 4.1]
	Let $\D_1$, $\D_2$ be two bounded and finite persistent diagrams, then
	\begin{align}
	d_{K_F}^2(\D_1, \D_2)\leq 2td_{F}(\D_1, \D_2),
	\end{align}
	where $t$ is a parameter of $K_F$ and $d_{K_F}^2$ is defined as 
	\begin{align}
	d_{K_F}^2(\D_1, \D_2) = K_F(\D_1, \D_1)+K_F(\D_2, \D_2)-2K_F(\D_1, \D_2)
	\end{align}
\end{theorem}
The time required to compute $K_F$ between a pair of persistence diagrams is $O(n^2)$. An accelerated version with fast Gauss transform can compute it in  $O(n)$ time~\cite{LeYamada2018}.

\para{Other comparative measures using persistence diagrams.}
Soler~\etal~\cite{SolerPlainchaultConche2018} proposed a lifted Wasserstein comparative measure for time-varying persistence diagrams. 
It is used to compare time-varying viscous finger datasets from ensemble simulation runs~\cite{SolerPetitfrereDarche2019}.  
The lifted Wasserstein measure is computed by a sparse persistence matching by augmenting points in the diagram with geometrical considerations, and the authors claimed (without proof) that the distance is a metric. 
They also claimed that it enhances geometrical stability of the feature tracking application in comparison to the Wasserstein distance.  
Its implementation uses an unbalanced Kuhn-Munkres algorithm for solving assignment problem, which takes $O(\min(m,n)^2\max(m,n))$, where $m$ and $n$ are the number of points in the persistence diagrams (which is roughly the number of critical points from the corresponding scalar fields). 
Assuming $m$ and $n$ are roughly the same, we simplify this to be $O(n^3)$, which is the running time of the Hungarian algorithm. 

The WKPI (weighted-kernel for persistence images) distance $d_{\omega}$~\cite{ZhaoWang2019} is a pseudometric induced by the inner product on a Hilbert space. It is proven to be stable {\wrt} small perturbation in persistence diagrams as measured by the $p$-th Wasserstein distance for $p=1$~\cite[Theorem 3.4]{ZhaoWang2019}. The runtime to compute the distance is $O(N)$, for $N$ being the number of pixels in each persistence image.

The persistent homology transform (PHT) defines a distance metric between shapes or surfaces~\cite{TurnerMukherjeeBoyer2014}. 
Computing the PHT of an object needs the persistence diagrams of the height function from various directions. 
The PHT distance averages the distance between diagrams of two input objects. 
Turner {\etal} proved that the map from a space of well-behaved shapes into the space of PHTs is an injective map~\cite{TurnerMukherjeeBoyer2014}. 
This map comes with a sense of stability as PHT of a finite simplicial complex is proved to be continuous~\cite[Lemma 2.1]{TurnerMukherjeeBoyer2014}.
Turner {\etal}~\cite{TurnerMukherjeeBoyer2014} used the Hungarian algorithm to compute the distances between two persistence diagrams in each of the directions. 
It takes $O((n+m)^3)$, where $n+m$ is the number of off-diagonal points in the two diagrams combined; assuming $m \approx n$, we simplify this to be $O(n^3)$. 
The $L^p$ distance between persistence vectors~\cite{LiWangAscoli2017} is by definition a metric (on the space of persistence vectors). 
Its computation takes $O(\max(n, l))$, where $n$ is the number of critical points and $l$ is the parameter used to vectorize the density function. 

Finally, there are a number of recent developments that aim to vectorize information from persistence diagrams to be interfaced with machine learning in a scalable way; most of which have not had direct applications in visualization.    
Noticeably, Hofer {\etal}~\cite{HoferKwittNiethammer2017} computed   parametrized projections of persistence diagrams that can be learned during neural network training. 
Moor {\etal}~\cite{MoorHornRieck2020} introduced a ``topological loss term'' for autoencoders that relate the topology of the data space with that of the latent space. 
The \emph{PersLay}~\cite{CarriereChazalIke2020} framework learns vectorizations of persistence diagrams by developing a differentiable layer for neural networks that processes information encoded persistence diagrams.

\subsection{Comparing Reeb graphs and their variants}
We discuss properties of functional distortion distance $d_{FD}$, edit distance $d_E$, interleaving distance $d_I$, 
as well as other measures for graph-based topological descriptors. 

\para{Functional distortion distance.}
$d_{FD}$ between a pair of Reeb graphs is an \emph{extended pseudometric}~\cite{BauerGeWang2014}. 
It is stable against perturbations of the input function~\cite{BauerGeWang2014}.
Let $\G_f$ and $\G_g$ denote the Reeb graphs for $(\Xspace,f)$ and $(\Xspace, g)$, respectively.
\begin{theorem}[\cite{BauerGeWang2014}, Theorem 4.1]
	Let $f, g: \Xspace \to \Rspace$  be tame functions whose Reeb quotient maps $\mu_f: \Xspace \to \G_f$ and $\mu_g: \Xspace \to \G_g$ have continuous sections. Then,
	\begin{align}
	d_{FD}(\G_f, \G_g) \leq ||f-g||_\infty.
	\end{align}
\end{theorem}
The \emph{tameness} is a technical condition requiring the functions to be well behaved. 
$d_{FD}$ between a pair of Reeb graphs is more discriminative than the bottleneck distance $d_\infty$ between persistence diagrams of the Reeb graphs~\cite{BauerGeWang2014}.
\begin{theorem}[\cite{BauerGeWang2014}, Theorem 4.2]
	\begin{align}
	d_\infty(\D_0(\G_f), \D_0(\G_g)) \leq d_{FD}(\G_f, \G_g). \\
	d_\infty(\D_0(\G_{-f}), \D_0(\G_{-g})) \leq d_{FD}(\G_f, \G_g).
	\end{align}
Here, $\G_{-f}$ and $\G_{-g}$ correspond to the Reeb graphs obtained by sweeping the range in the reverse direction. 	
\end{theorem}
The main cost of calculating $d_{FD}$ is to calculate the Gromov-Hausdorff (GH) distance $d_{GH}$ of input spaces. 
Schmiedl~\cite{Schmiedl2017} showed that $d_{GH}$ cannot be approximated within any reasonable bound in polynomial time. 

\para{Edit distance.}
$d_E$ was proven to be a metric by Sridharamurthy {\etal}~\cite{SridharamurthyMasoodKamakshidasan2020}.
The stability of $d_E$ is unknown. 
We conjecture that $d_E$ is not $L^{\infty}$-stable, that is, small changes in function values may cause pairing switches in the merge tree, resulting in a large increase in distance. 
$d_E$ is more discriminative than the bottleneck distance $d_\infty$ and the Wasserstein distance $d_p$ as conjectured by Sridharamurthy~\etal.   
The computation of $d_E$ takes $O(n^2)$ time for trees with bounded degree~\cite{SridharamurthyMasoodKamakshidasan2020}. 

\para{Edit distance between labeled Reeb graphs.}
$d_{EG}$ is shown to be an extended pseudometric~\cite{BauerDiFabioLandi2016, BauerLandiMemoli2020}.
It is also proven to be stable.
\begin{theorem}[\cite{BauerDiFabioLandi2016}, Corollary 4.2.]
Let $\Mspace$ be a connected, closed, orientable, smooth manifold of dimension 1 or 2. For every simple Morse functions $f, g: \Mspace \to \Rspace$, we have 
\begin{align}
d_{EG}((\G_f, l_f), (\G_g, l_g)) \leq ||f-g||_{\infty}.
\end{align}
\end{theorem}	
We conjecture that computing $d_{EG}$ is NP-hard since it is at least graph-isomorphism hard~\cite{BauerLandiMemoli2020}.

\para{Interleaving distance between merge trees.}
$d_I$ is proven to be a metric~\cite[Lemma 1]{MorozovBeketayevWeber2013}.  
$d_I$ is stable {\wrt} the largest difference between the two scalar functions~\cite{MorozovBeketayevWeber2013}.  
\begin{theorem}[\cite{MorozovBeketayevWeber2013}, Theorem 2]
	Given two merge trees $\T_f, \T_g$ defined by two scalar functions $f, g: \Xspace \to \Rspace$, then
	\begin{align}
	d_I(\T_f, \T_g) \leq  ||f-g||_\infty.
	\end{align}
\end{theorem}
$d_I$ is more discriminative than distances between persistence diagrams, such as the bottleneck distance $d_\infty$.
\begin{theorem}[\cite{MorozovBeketayevWeber2013}, Theorem 3]
	Given two tame functions $f, g: \Xspace \to \Rspace$, then
	\begin{align}
	 d_\infty(\D_f, \D_g) \leq d_I(\T_f, \T_g).
	\end{align}
Here, $\D_f$ and $\D_g$ are the $0$-dimensional persistence diagram of $f$ and $g$, respectively. 
\end{theorem}
Computing $d_{I}$ on pairs of merge trees or Reeb graphs is NP-hard~\cite{AgarwalFoxNath2018}. Later Touli and Wang~\cite{TouliWang2019} gave an FPT (fixed-parameter tractable) algorithm for computing $d_I$ for a pair of merge trees.
 
\para{Interleaving distance between labeled merge trees.}
$d_{IL}$ was shown to be a metric by Gasparovic {\etal}~\cite{GasparovicMunchOudot2019}. 
We conjecture that $d_{IL}$ is not $L^{\infty}$-stable {\wrt}  perturbations of the scalar field, since small changes in function values
may cause changes in node correspondences between a pair of merge tree.
But it has its own notion of stability~\cite{GasparovicMunchOudot2019}.
\begin{theorem}{\cite[Lemma 3.2]{GasparovicMunchOudot2019}}
	Given any pair of valid matrices $M_1, M_2$, and their associated merge trees $\T_1$ and $\T_2$, 
 	\begin{align}
 	d_{IL}(\T_1, \T_2) \leq ||M_1-M_2||_\infty.
 	\end{align}
\end{theorem}
We conjecture that $d_{IL}$ is more discriminative that $d_\infty$. 
Computing $d_{IL}$ between a pair of (leaf) labeled merge trees is polynomial in the number of leaves~\cite{MunchStefanou2018}. 
Considering leaf labeling strategies in~\cite{YanWangMunch2020}, the labeling step for $d_{IL}$ with tree mapping and Euclidean mapping takes $O(n^3)$ due to the use of the Hungarian algorithm. 
The distance computation for $d_{IL}$ takes $O(n^2)$. 

\para{Interleaving distance between Reeb graphs.}
$d_{IG}$ is an extended pseudometric~\cite[Proposition 4.5]{SilvaMunchPatel2016}.
It has $L^\infty$-stability. 
\begin{theorem}{\cite[Theorem 4.6 with simplified notations]{SilvaMunchPatel2016}}
Let $(\Xspace,f)$ and $(\Xspace,g)$ be the space that gives rise to the Reeb graphs $\G_f$ and $\G_g$, then 
\begin{align}
	d_{IG}(\G_f, \G_g) \leq ||f-g||_\infty,
\end{align}
where $\G_f$ represents the Reeb graph using a sheaf-theoretical language. 
\end{theorem}
The calculation of $d_{IG}$ is NP-hard~\cite{SilvaMunchPatel2016}.
 
\para{Distance based on branch decompositions.}  
For the distance $d_{BR}$ between merge trees based on branch decompositions~\cite{BeketayevYeliussizovMorozov2014}, whether $d_{BR}$ is a metric is unknown. We conjecture that it is a metric. 
Its stability was not investigated in the original work~\cite{BeketayevYeliussizovMorozov2014}. 
On the other hand, binary decisions made during the branch decomposition create instabilities {\wrt} the resulting branches~\cite{SaikiaSeidelWeinkauf2014}. The work in ~\cite{BeketayevYeliussizovMorozov2014} is believed (by Saikia \etal~\cite{SaikiaSeidelWeinkauf2014}) to alleviate such an issue by considering all possible branch decompositions. 
However, we still conjecture that $d_{BR}$ is likely unstable. 
Although no theoretical proof is offered, Beketayev \etal~\cite{BeketayevYeliussizovMorozov2014} used $d_\infty$ between persistence diagrams during their experiments as a baseline for the $d_B$ between merge trees, and observed that $d_{BR}$ is more discriminative than $d_\infty$.
The computational complexity of $d_{BR}$ depends on two quantities: first, the time complexity of a function, which for a predefined $\epsilon$, determines whether two branch decompositions match; and second, the number of iterations of binary search for $\epsilon_{min}$ -- the smallest $\epsilon$ for two branch decompositions to be $\epsilon$-similar. 
The former takes $O(n^2 m^2 (n+m))$, where $n$ and $m$ are the numbers of extrema in each merge tree respectively; assuming $n \approx m$, this is roughly $O(n^5)$.
Beketayev {\etal} claimed the latter to be ``moderate''; which we denote as $O(\log(I_{\epsilon}))$, $I_{\epsilon}$ being the search range.

For comparative measures involving histograms derived from merge trees, both $\chi^2$-distance~\cite{SaikiaWeinkauf2017} and $L^2$ distance~\cite{SaikiaSeidelWeinkauf2015} between histograms give rise to a metric. 
The running time including the construction of a distance matrix is 
$O(n^2 B)$, $B$ being the number of bins.
Saikia \etal~\cite{SaikiaSeidelWeinkauf2015} claimed that the merging of histograms works well under small perturbations in the data. 
The computation of the distance using $L^2$ distance of the log-scaled bin  values~\cite{SaikiaSeidelWeinkauf2015} takes $O(n^2B)$.
The $\chi^2$-histogram distance~\cite{SaikiaWeinkauf2017} can also be computed in polynomial time. 

The distance between a pair of extended Reeb graphs (ERGs)~\cite{BiasottiMariniMortara2003} is proven to be a metric. 
A kernel between extended Reeb graphs~\cite{BarraBiasotti2013} has a worst case running time of $O(\gamma^4_{max} + 2n^2 \log(n))$, where $\gamma_{max}$ is the length of the maximal shortest path.

\para{Other graph-based or tree-based comparative measures.} 
Some comparative measures are not well investigated regarding their mathematical properties, such as tree alignment distance~\cite{LohfinkWetzelsLukasczyk2020}, and similarity measure between subtrees of contour trees~\cite{ThomasNatarajan2011}. 
For two contour trees with bounded degree, the former can be calculated in $O(|T_1|\cdot |T_2|)$ (the cost of the alignment), which can be simplified to be $O(n^2)$ assuming the larger tree contains $n$ critical points.  
For the latter, a polynomial time algorithm is available with a worst case running time $O(t^5 \log t)$, where $t$ is the number of branches.
The eBDG~\cite{SaikiaSeidelWeinkauf2014} can be calculated in $O(N d \log d)$, where $N$ is the number of nodes in eBDG and $d$ is average branching factor. The comparison measure can be calculated in $O(N_1 N_2 \log_{l_1 + 1} (N_1 + 1)\log_{l_2 + 1} (N_2 + 1))$, where $N_1,N_2$ are number of nodes, and $l_1, l_2$ are average levels in the two trees.

Some comparative measures proposed for visualization have no analysis of mathematical properties or computational complexity, such as similarity measures derived from attributes~\cite{HilagaShinagawaKohmura2001,ZhangBajajBaker2004,SohnBajaj2006,SchneiderWiebelCarr2008, WuZhang2013, SaggarSpornsGonzalez-Castillo2018, AgarwalRamarmurthiChattopadhyay2020}.

\subsection{Comparing Morse and Morse-Smale Complexes}

\para{Distance between extremum graphs.} $d_{\rho}$ is proven to be a metric~\cite{NarayananThomasNatarajan2015}. 
Narayanan {\etal} mentioned studies on its stability and discriminativity as topics for future work. 
Computing $d_{\rho}$ involves weighted maximum clique enumeration, which has exponential time complexity $O(3^{n/3})$ ($n$ being the number of vertices, using the Bron-Kerbosch clique enumeration algorithm); hence, it is feasible for only small graphs~\cite{NarayananThomasNatarajan2015}.

\para{Feature correspondences with Morse-Smale cells.}
The feature correspondence framework using Morse-Smale complexes~\cite{FengHuangJu2013} contains no theoretical guarantee.
Feng {\etal}~gave empirical evidence of stability. However, the stability follows primarily because the Auto Diffusion Function (ADF) that they design is smooth and noise free.
Schnorr~\etal~\cite{SchnorrHelmrichDenker2020} introduced a feature tracking framework using dissipation elements~(DEs), which are by definition, equivalent to 3D Morse-Smale cells~\cite{GyulassyBremerGrout2014}. 
They determined features correspondences by solving a maximum-weight independent set problem, which is NP-hard in general. 
However, with certain assumptions, the authors reduced the problem to computing weighted, bipartite graph matching in practice, making the computation tractable.

\subsection{Other Comparative Measures}
Finally, we discuss known properties associated with other topological descriptors that arise specifically in visualization tasks. 
As described in~\autoref{sec:single-fields}, given a Morse function defined on a manifold, the Morse shape descriptor (MSD)~\cite{AlliliCorriveau2007} uses relative homology groups to encode the topology among pairs of sublevel sets of the function. 
The ranks of relative homology groups (represented as Betti numbers) across multiple levels encode ``complete topological information about the critical regions of the manifold as well as the extension of the regions where no topological change is observed''~\cite{AlliliCorriveau2007}. The distance between a pair of Morse shape descriptor (MSD) can be computed in $O(n^2)$ time ($n$ is the number of critical points in a Morse function, which is used for the discretization of level sets).

For topological descriptors used in ensemble visualization, as described in~\autoref{sec:ensembles}, a $L^2$ distance is defined between two persistence maps~\cite{FavelierFarajSumma2018}; its computation takes $O(n^2M)$, where $M$ denotes the number of vertices in the domain. 

In the study of time-varying and multi-field comparison (see~\autoref{sec:time-varying-fields}), the $L^p$ distance between two fiber component distributions is proven to be a metric~\cite{AgarwalChattopadhyayNatarajan2019}.
The mathematical properties associated with the similarity measures for multi-resolution Reeb spaces remain unknown due to partially heuristic matchings between nodes and attributions, in a way similar to the situation in~\cite{HilagaShinagawaKohmura2001,ZhangBajajBaker2004}.  

Poco \etal~\cite{PocoDoraiswamyTalbert2015} introduced a topology-based measure that computes a locality-aware correspondence between similar extrema of two scalar fields, to help ecologists compare species distribution models. The authors found that this similarity measure is stable under the influence of noise in practice. The algorithm has a complexity of $O(n^3)$ because of the maximum weight bipartite graph matching step.

\subsection{Open Source Implementations}
Efficient and robust open source software for computing topological structures and comparative measures are key to their broad adoption within application domains. 
A number of tools are available for computing persistence diagrams and/or  barcodes, together with their bottleneck and Wasserstein distances, including \textsf{GUDHI}~\cite{Gudhi2021}, \textsf{PHAT}~\cite{BauerKerberReininghaus2021}, \textsf{Dionysus}~\cite{Morozov2021}, \textsf{R-TDA}~\cite{FasyKimLecci2021},  \textsf{HERA}~\cite{KerberMorozovNigmetov2021}, \textsf{persim}~\cite{Persim2021}, \textsf{Perseus}~\cite{Perseus2021}, and \textsf{Ripser}~\cite{Bauer2021}. 
Their utilities extend beyond visualization, into applications in machine learning tasks. 
Many of these tools are implemented in C/C++ or Python, while a few provide Python/R wrappers. 
Otter \etal~\cite{OtterPorterTillmann2017} discussed software for persistence homology and also provided installation guides and use cases. Computation tools for merge trees, contour trees and Reeb graphs together with support for computing branch decomposition representations, symmetry detection, and feature tracking are available in software such as \textsf{Recon}~\cite{DoraiswamyNatarajan2021}, \textsf{contour-tree}~\cite{Doraiswamy2021}, \textsf{mtlib}~\cite{Saikia2021}, \textsf{AMT}~\cite{Yan2021}, and \textsf{SymmetryViewer}~\cite{ThomasNatarajan2021}. 
These software are implemented in C/C++, Python, or Java. 
Tools for computing Morse-Smale complexes alike include \textsf{mscomplex3d}~\cite{ShivashankarNatarajan2021}, \textsf{MSCEER}~\cite{Gyulassy2021}, \textsf{CompExtGraph}~\cite{Narayanan2021}.  
\textsf{Topology ToolKit (TTK)}~\cite{TiernyFavelierLevine2021} is a popular toolkit designed to work together with the visualization software \textsf{ParaView}~\cite{AhrensGeveciLaw2005}, that supports the computation and visualization of persistence diagrams,  merge trees, contour trees, Reeb graphs, and Morse-Smale complexes,  together with persistence based simplifications of  these descriptors. It also allows computation of bottleneck/Wasserstein distances between persistence diagrams and feature tracking via nested tracking graphs.

%% file: sec-future.tex
\section{Future Research Opportunities}
\label{sec:future}

After analyzing the collection of work discussed in this survey, we found that there are several research gaps and hence opportunities in the study of scalar field comparison with topological descriptors. We now discuss this topic in detail.

Looking at~\autoref{table:properties}, we immediately observe that many comparative measures are available with nice mathematical properties.~\autoref{table:navigate-descriptor-by-task}, on the other hand, shows that topological descriptors and their associated comparative measures have been used for a wide variety of visualization tasks, for which they seem to be especially well suited. 
On a closer look, we observe that the measures appearing in both tables do not match very well. Many mathematically sound comparative  measures have not found practical applications. 
On the other hand, some comparison-based visualization tasks have been developed using heuristics; the properties of the associated measures are not investigated comprehensively nor supported by theoretical guarantees, 
\emph{which leads to the question about the reason for this gap between theory and practice.}
We attempt to provide some answers below and point out respective research opportunities.

\subsection{Computational Efficiency and Stability}
 
The computation of metrics proposed for graph-based topological descriptors is NP-hard (\textit{e.g.,}~\cite{Bille2005, BjerkevikHavardBotnan2018, AgarwalFoxNath2018}), which makes them feasible for only small graphs. 
Much more work is needed to develop concepts, algorithms, and implementations applicable to large real-world data.

\para{Approximation algorithms.} 
One approach to efficiency is to develop approximation algorithms for computing the comparative  measures. 
This approach can be achieved by reducing the complexity of the input data using concepts proposed in mapper~\cite{SinghMemoliCarlsson2007} or cosheaf~\cite{SilvaMunchPatel2016, BrownBobrowskiMunch2021} {\wrt}~the Reeb graph, which are gaining popularity in computational topology.
Cavanna \etal~\cite{CavannaJahanseirSheehy2015} studied   sparse filtrations that selectively prune data points from the input and proved that such filtrations give good approximations to the barcodes.
Touli \etal~\cite{TouliWang2019} suggested an FPT (fixed-parameter tractable) algorithm that approximates the interleaving distance between merge trees.
However, for many of the distance measures introduced in \autoref{sec:comparative-measures},  no efficient algorithms are available. Here, we see big opportunities for the development of approximation algorithms by applying controlled relaxation of hard mathematical constraints to achieve better performance. 

\para{Heuristic matching strategies.} Many approaches tackle this challenge by providing heuristic graph matching strategies. This alternative to the use of classic algorithms for graph isomorphism  significantly reduces the time for comparing graph-based topological structures~\cite{ZhangBajajBaker2004,WuZhang2013,YanWangMunch2020,LohfinkWetzelsLukasczyk2020}. Finding appropriate heuristics while maintaining a mathematically sound framework is a challenging task. 
Important tasks that need to be addressed in the future include a thorough evaluation of mapping strategies by conducting computational experiments and theoretical analysis.

\para{New topological descriptors.} 
Some works introduced new topological descriptors that are easier to compute (\textit{e.g.,}~\cite{ZhangBajajBaker2004}). 
However, a detailed investigation of their mathematical properties has not been undertaken.
A general need for new topological descriptors remains, in particular, those that encode application-specific information for comparative analysis and visualization. 
Understanding the mathematical and computational properties associated with any new topological descriptor will help to increase its utilization in applications.  

\para{Scalable computation.} 
Scalable computations of comparative measures rely partially on scalable  computations of topological descriptors. 
Some efforts have been made to develop scalable computation of topological descriptors using careful engineering, parallel or distributed computation, for instance, for mapper constructions~\cite{HajijAssiriRosen2020,ZhouChalapathiRathore2020}, merge trees~\cite{MorozovWeber2013}, contour trees~\cite{MorozovWeber2014}, and Morse-Smale complexes~\cite{GyulassyBremerPascucci2019,SubhashPandeyNatarajan2020}, to name a few.  
A few recent efforts have adapted the comparative analysis to in situ environments~\cite{FriesenAlmgrenLukic2016, SolerPlainchaultConche2018}, 
which is an important research direction in dealing with large real-world data. 

\para{Stable measures for gradient-based descriptors.} 
Both~\autoref{table:properties} and ~\autoref{table:navigate-descriptor-by-task} show that few comparative measures have been developed for gradient-based topological descriptors such as Morse and Morse-Smale complexes and their variants. 
These descriptors are more sensitive to perturbations in the scalar fields, making it difficult to design effective comparative measures.
Furthermore, topological descriptors for multi-fields such as the Jacobi sets, Reeb spaces, and multivariate mappers are relatively less understood. Computing/comparing these descriptors remains a challenge.

\subsection{Application-Specific Topological Feature Descriptors}
Defining features of interest in an application context is a challenging problem in itself, involving a variety of considerations.  While topological descriptors provide a good abstraction, they may not capture all characteristics of the data. A domain-specific interpretation of the comparison measure is a further challenge.

\para{Augmentation of topological descriptors.} 
Some attempts have been made to define topological descriptors that can be further augmented with more geometric- or attribute-related information, such as the comparison of dual contour trees that establishes correspondences based on node attributes~\cite{ZhangBajajBaker2004}. An extension of this idea could be beneficial in many applications.

\para{Heuristic node mapping strategies.} 
Heuristic strategies for node mapping, as mentioned above, can lead to the possibility of integrating domain knowledge into the comparison process. The heuristic strategies often lead to a violation of the metric axioms and stability properties, and hence require a solid evaluation.

\para{Topological descriptors for multi-fields.} 
The design of comparative measures for multi-fields also has potential for application-specific feature comparison. However, this is also a challenging problem.  
The introduction of topological descriptors for multi-fields is still in its infancy, and the desirable properties are not well understood. 
Measures for their comparison also need further exploration.

\para{Topological descriptors for ensembles.} 
Ensemble simulations are ubiquitous and pose a specific challenge to all analysis methods, including topological ones.
So far, most methods are based on pairwise comparison of ensemble members (\autoref{sec:ensemble-exploration}), and often visualized as similarity matrices or topological summaries (\autoref{sec:ensemble-structure}). These methods usually assume a Gaussian distribution of the data, which is restrictive.
Few papers go beyond pairwise comparison and support clustering and outlier detection (\autoref{sec:ensemble-exploration}). Even in these cases, new comparative measures are required to complete the analysis.
The work in~\cite{LiPalandeWang2021} represents an interesting paradigm shift in the study of ensemble data, where matrix sketching (and, in general, techniques from randomized linear algebra) can be used to obtain ensemble representations and to detect outliers. 

\subsection{Integration in Visualization Pipelines}
We see potential for the integration of topological methods into interactive tools and visualization pipelines for all visualization tasks discussed in this report. This integration will support the exploration and comparison of scalar fields, and analysis of time-varying data and ensembles. 

The biggest hurdle to suc	cess in applications is the availability of the methods within open-source software such as \textsf{ParaView}~\cite{AhrensGeveciLaw2005} and the \textsf{Topology ToolKit}~\cite{TiernyFavelierLevine2018}, which allows visualization experts to integrate topological methods even if they are not expert developers of TDA  techniques.

%% file: sec-conclusion.tex
\section{Conclusion}
\label{sec:conclusion}

This state-of-the-art report presents a taxonomy of existing approaches that develop or utilize topological descriptors for the comparative analysis and visualization of scalar fields. In addition, a major contribution of this report is the collection of mathematical and computational properties for the various comparative measures of topological descriptors in the literature, which spans  applied topology, computational topology, topological data analysis, and  visualization. 
Some of the comparative measures described in this report have been developed and used in fields outside visualization, such as machine learning, computer vision, and computer graphics. 
Although the focus of this report is on visualization applications, we have included a few references to these other connections when appropriate. 

The development and deployment of visualization techniques and tools based on these comparative measures have impacted various application domains. We list below a set of application areas together with references to the description above that discusses specific visualization tasks.
\begin{itemize}
    \item \textbf{Structural Biology}: Topological analysis of biomolecular structures imaged using various microscopy techniques has benefited from the development of symmetry detection  (highlighting repeating substructures in biomolecules, \autoref{sec:symmetry-detection-mt}), shape matching  (protein classification, \autoref{sec:shape-matching:CTReebgraph}), and contour-tree-based tracking methods (bond tracking, \autoref{sec:tempFeatureTracking:MTtracking}). 
    \item \textbf{Climate Science}: Topological descriptors have played a key role in the development of cloud tracking methods (\autoref{sec:tempFeatureTracking}), identifying periodicity in surface temperatures (\autoref{sec:tempGlobalCompare}), and uncertainty visualization (\autoref{ensemble-uncertainty}).
    \item \textbf{Combustion Studies}: Pairwise comparison of physical quantities measured during a combustion simulation has resulted in an improved understanding of the different stages of combustion (\autoref{sec:tempGlobalCompare} and~\autoref{sec:spaceTimeFeatures}).
    \item \textbf{Neuroscience}: Space-time structures built on fMRI data are helpful in the study of the dynamic organization of the brain. Comparing neuronal trees helps us understand how the brain functions (\autoref{sec:spaceTimeFeatures} and \autoref{sec:shape-matching:Diagrams}).
    \item \textbf{Computational Physics and Chemistry}: Comparative  measures between multi-fields are helpful in the study of stable Pt-CO bonds and identification of nuclear scission points in simulation data (\autoref{ensemble-multi-field}).
    \item \textbf{Ecology}: Comparative measures based on TDA have been successfully used for exploration and better understanding of the species distribution models (\autoref{sec:ensemble-exploration}).
\end{itemize}

We believe that addressing the various research gaps outlined in~\autoref{sec:future} will enable further applications both in the above-mentioned and other areas of science and engineering. 

%% file: STAR-MT-refs.bbl
\newcommand{\etalchar}[1]{$^{#1}$}

%% file: sec-bio.tex
\section*{Short Biographies}

\paragraph*{Lin Yan} is a PhD student in the Scientific Computing and Imaging (SCI) Institute, University of Utah.
Her research interests are topological data analysis and visualization.  
Her recent work includes statistical analysis of merge trees, uncertainty visualization of Morse--Smale complexes, and topological signatures for vector fields. 
Email: lin.yan@utah.edu.

\paragraph*{Talha Bin Masood} is a Postdoctoral Fellow at Link\"oping University in Sweden. He received his Ph.D. in Computer Science from Indian Institute of Science, Bangalore. His research interests include scientific visualization, computational geometry, computational topology, and their applications to various scientific domains.
Email: talha.bin.masood@liu.se. 

\paragraph*{Raghavendra Sridharamurthy} is a PhD student in the Department of Computer Science and Automation at Indian Institute of Science, Bangalore. His interests include scientific visualization, computational topology and its applications. 
Email: raghavendrag@iisc.ac.in. 

\paragraph*{Farhan Rasheed} is PhD student at Link\"oping University in Sweden. He graduated from Heidelberg University with a degree in Scientific Computing. His research interests includes scientific visualization, topological data analysis, machine learning, and medical image computing. 
Email: farhan.rasheed@liu.se.

\paragraph*{Vijay Natarajan} is the Mindtree Chair Professor in the Department of Computer Science and Automation at Indian Institute of Science, Bangalore. He received his Ph.D. in Computer Science from Duke University. His research interests include scientific visualization, computational topology, and geometry processing. In current work, he is developing topological methods for time-varying and multi-field data visualization, and studying applications in biology, material science, and climate science. 
Email: vijayn@iisc.ac.in. 

\paragraph*{Ingrid Hotz} is currently a Professor in Scientific Visualization at the Link\"oping University in Sweden. She received her Ph.D. degree from the Computer Science Department at the University of Kaiserslautern, Germany. She worked as a postdoctoral researcher at IDAV at the University of California, Davis. Previous positions include leading a junior research group in visualization at the Zuse Institute Berlin and leading the scientific visualization group at the German Aerospace Center (DLR).   Her research interests lie in data analysis and scientific visualization, ranging from basic research questions to effective solutions to visualization problems in applications. This includes developing and applying concepts originating from different areas of computer sciences and mathematics, such as computer graphics, computer vision, dynamical systems, computational geometry, and combinatorial topology. 
Email: ingrid.hotz@liu.se. 

\paragraph*{Bei Wang} is an Assistant Professor at the School of Computing and a faculty member at the Scientific Computing and Imaging (SCI) Institute, University of Utah. She received her Ph.D. in Computer Science from Duke University. She is interested in the analysis and visualization of large and complex data. Her research interests include topological data analysis, data visualization, computational topology, computational geometry, machine learning, and data mining. Her recent work includes statistical analysis and uncertainty visualization of topological descriptors. 
Email: beiwang@sci.utah.edu. 